\documentclass[10pt,preprint]{aastex}
\usepackage{epsfig}
\usepackage{amsmath}
\usepackage{upgreek}
\usepackage{natbib}
\usepackage{graphicx}
\usepackage{longtable}
\usepackage{float}

\newcommand{\noprint}[1]{}

\begin{document}

\title{Herschel observations of EXtraordinary Sources: Analysis of the full Herschel/HIFI molecular line survey of Sagittarius B2(N)\textsuperscript{*}}

\author{
Justin L. Neill\altaffilmark{1}, 
Edwin A. Bergin\altaffilmark{1}, 
Dariusz C. Lis\altaffilmark{2}, 
Peter Schilke\altaffilmark{3}, 
Nathan R. Crockett\altaffilmark{1}, 
C\'{e}cile Favre\altaffilmark{1}, 
Martin Emprechtinger\altaffilmark{2}, 
Claudia Comito\altaffilmark{4},
Sheng-Li Qin\altaffilmark{3}, 
Dana E. Anderson\altaffilmark{1}, 
Andrew M. Burkhardt\altaffilmark{1}, 
Jo-Hsin Chen\altaffilmark{5}, 
Brent J.  Harris\altaffilmark{6},
Steven D. Lord\altaffilmark{7}, 
Brett A. McGuire\altaffilmark{8}, 
Trevor D. McNeill\altaffilmark{1},
Raquel R. Monje\altaffilmark{2}, 
Thomas G. Phillips\altaffilmark{2},
Amanda L. Steber\altaffilmark{6}, 
Tatiana Vasyunina\altaffilmark{6}, 
Shanshan Yu\altaffilmark{5}
}

\altaffiltext{1}{Department of Astronomy, University of Michigan, 500 Church Street, Ann Arbor, MI 48109, USA; jneill@umich.edu, ebergin@umich.edu}
\altaffiltext{2}{California Institute of Technology, Cahill Center for Astronomy and Astrophysics 301-17, Pasadena, CA 91125, USA}
\altaffiltext{3}{Physikalisches Institut, Universit\"{a}t zu K\"{o}ln, Z\"{u}lpicher Str. 77, 50937, K\"{o}ln, Germany}
\altaffiltext{4}{Max-Planck-Institut f\"{u}r Radioastronomie, Auf dem H\"{u}gel 69, 53121, Bonn, Germany}
\altaffiltext{5}{Jet Propulsion Laboratory, California Institute of Technology, 4800 Oak Grove Drive, Pasadena, CA 91109, USA}
\altaffiltext{6}{Department of Chemistry, University of Virginia, McCormick Road, Charlottesville, VA 22904, USA}
\altaffiltext{7}{National Herschel Science Center, California Institute of Technology, Pasadena, CA USA}
\altaffiltext{8}{Division of Chemistry and Chemical Engineering, California Institute of Technology, Pasadena, CA 91125, USA}

\begin{abstract}
A sensitive broadband molecular line survey of the Sagittarius B2(N) star-forming region has been obtained with the HIFI instrument on the \emph{Herschel} Space Observatory, offering the first high-spectral resolution look at this well-studied source in a wavelength region largely inaccessible from the ground (625--157 $\upmu$m).  From the roughly 8,000 spectral features in the survey, a total of 72 isotopologues arising from 44 different molecules have been identified, ranging from light hydrides to complex organics, and arising from a variety of environments from cold and diffuse to hot and dense gas.  We present an LTE model to the spectral signatures of each molecule, constraining the source sizes for hot core species with complementary SMA interferometric observations, and assuming that molecules with related functional group composition are cospatial.  For each molecule, a single model is given to fit all of the emission and absorption features of that species across the entire 480--1910 GHz spectral range, accounting for multiple temperature and velocity components when needed to describe the spectrum.  As with other HIFI surveys toward massive star forming regions, methanol is found to contribute more integrated line intensity to the spectrum than any other species. We discuss the molecular abundances derived for the hot core, where the local thermodynamic equilibrium approximation is generally found to describe the spectrum well, in comparison to abundances derived for the same molecules in the Orion KL region from a similar HIFI survey.  Notably, we find significantly higher abundances of amine- and amide-bearing molecules (CH$_3$NH$_2$, CH$_2$NH, and NH$_2$CHO) toward Sgr B2(N) than Orion KL, and lower abundances of some complex oxygen-bearing molecules (CH$_3$OCHO in particular).  In addition to information on the chemical composition of the hot core, the strong far-IR dust continuum allows a number of molecules to be detected in absorption in the Sgr B2(N) envelope for the first time at high spectral resolution, and we discuss the possible physical origin of the kinematic components observed in absorption.  Additionally, from the detection of new HOCO$^+$ transitions in absorption compared to published HCO$^+$ isotopic observations, we discuss constraints on the gas-phase CO$_2$ abundance and compare this to observations of the ice composition in the Galactic Center region, and to CO$_2$ abundance estimates toward other high-mass star-forming regions.  The reduced HIFI spectral scan and LTE model are made available to the public as a resource for future investigations of star forming regions in the submillimeter and far-infrared.
\end{abstract}

\noindent{\footnotesize{\textsuperscript{*}Herschel is an ESA space observatory with science instruments provided by European-led Principal Investigator consortia and with important participation from NASA.}}

\section{Introduction}

The Sagittarius B2 complex, located $\sim 100$ pc from the center of the Milky Way, is the most massive star-forming region in our Galaxy, with $M \sim 10^7 M_\odot$ \citep{lis89, lis90}.  Since the early days of molecular radio astronomy, this source has been the most fruitful one for investigating interstellar chemistry; of the approximately 175 molecules detected to date in the interestellar medium (ISM), over half have been detected in this source and a large fraction of these were first discovered there \citep[][and references therein]{menten11b}.  Sgr B2 extends over several arcminutes of the sky \citep[$\sim 50$ pc, assuming a distance from the Sun of 8 kpc, from][]{reid93, eisenhauer03, reid09} and contains three known massive cores, referred to as N, M, and S, for north, main, and south, respectively, aligned nearly north--south and separated across about $1.5'$ \citep{goldsmith87, gaume90}.  The N and M cores are the brightest regions within Sgr B2; of these two, the N core has the richer molecular spectrum from centimeter to far-infrared wavelengths, including the signatures of many of the most complex molecules detected in the ISM to date.  A number of studies have proposed that N is in an earlier evolutionary stage than M \citep{vogel87, miao95, kuan96, qin11}.  A number of unbiased spectral surveys of Sgr B2(N), from the centimter and millimeter \citep{cummins86, turner91, nummelin00, belloche08, belloche09, belloche13} to the far-infrared \citep{goicoechea04, polehampton07, etxaluze13}, have been performed in the last few decades, revealing the chemical richness of this source.  Sgr B2(N) is therefore a central source for studies of the physics and chemistry of massive star formation.

Sgr B2(N) contains molecular gas in a variety of physical states.  A central star forming complex with a diameter of  $\sim 5''$ (0.2 pc) contains multiple H II regions and hot gas and dust associated with multiple massive protostars \citep{gaume90, belloche08, qin11}.  This gas is warm ($T_\textnormal{kin} \ge 150$ K) and dense ($n$(H$_2$) $> 10^7$ cm$^{-3}$) and contains high abundances of complex, prebiotic molecules best studied in the millimeter wave region \citep{herbst09}.  There is also an extended, cooler, and less dense envelope that has a rich molecular content.  Observations with the Green Bank Telescope between 8--50 GHz have discovered a number of complex organic molecules with low rotational temperatures \citep{hollis04a, hollis04b, hollis06, remijan05, loomis13, zaleski13}, many in absorption against the nonthermal centimeter wave continuum of the H II regions \citep{hollis07}, suggesting an origin in moderate-density ($\sim 10^4$ cm$^{-3}$) gas.

Sgr B2 is part of the Central Molecular Zone (CMZ), which extends over $\sim 2^\circ \times 0.5^\circ$ (280 $\times$ 70 pc) \citep{morris96, jones12}.  The CMZ is unique within our galaxy as having diverse chemistry, including high abundances of oxygen-bearing complex molecules such as methanol, methyl formate, and glycolaldehyde \citep{requenatorres06, requenatorres08, yusefzadeh13}, but many of its clouds lack the protostellar objects that are normally needed to provide the energy to synthesize these molecules and liberate them from ice grains \citep{garrod08}.  An enhanced cosmic-ray ionization rate (of $10^{-15}-10^{-14}$ s$^{-1}$) has been inferred for the CMZ compared to the rest of the Galaxy from the detection of widespread warm and diffuse H$_3$$^+$ \citep{oka05, yusefzadeh13b}.  The Sgr A* black hole also has been suggested to be the source of a recent X-ray burst \citep{sunyaev93, ponti10, nobukawa11, ryu13} which could be powering ionization and energetic chemistry even in regions where the dust temperatures are $\sim 20-30$ K \citep{lis90, molinari11}.

Despite the numerous studies of the molecular content of Sgr B2(N), for the smaller molecules that are the ``building blocks'' for the rich chemical diversity seen in the ISM, many transitions lie in the submillimeter and far-infrared, where the Earth's atmosphere inhibits investigations with ground-based telescopes across much of the spectral bandwidth.  Additionally, for many other molecules that are observable from the ground (e.g. CO, HCN), high-excitation transitions that trace the most energetic gas in star-forming regions fall in this region.  The \emph{Herschel} Space Observatory \citep{pilbratt10} contains three instruments for spectroscopy in the far-infrared.  The Heterodyne Instrument for the Far Infrared (HIFI), in particular, is a high-resolution ($R > 5 \times 10^5$) instrument with nearly continuous spectral coverage from 480--1910 GHz \citep{degraauw10}, which enables the first high-spectral resolution investigations of star forming regions in this wavelength region.  

As part of the Herschel Observations of EXtra-ordinary Sources (HEXOS) key program, complete HIFI spectral surveys of five positions within the Sgr B2 and Orion molecular clouds have been obtained:  Orion KL, Orion S, the Orion Bar, Sgr B2(M), and Sgr B2(N) \citep{bergin10}.  In this manuscript, we present the HIFI spectral survey of Sgr B2(N).  For most of the 44 molecules detected to date in this survey a model has been developed, assuming local thermodynamic equilibrium (LTE) conditions, to the emission and absorption features observed over the spectral bandwidth.  Most molecules are detected with features from a wide range of energy states, which serve to constrain the excitation of each species.  Additionally, with the high spectral resolution of HIFI, the multiple kinematic components associated with Sgr B2(N) and its envelope, as well as those in the clouds along the line of sight, are spectrally resolved.  The fully reduced spectral scan, along with the LTE models presented in this paper, will be released to the public through the Herschel Science Center\footnote{http://herschel.esac.esa.int/UserProvidedDataProducts.shtml}.  As Sgr B2(N) is a template source for interstellar chemistry and high-mass star formation, we anticipate that this will be a valuable archival data product for future investigations.

The structure of the remainder of the paper is as follows.  In \S 2.1 we describe the HIFI observations of Sgr B2(N) and the data reduction procedure, and in \S 2.2 we briefly describe complementary SMA observations used to constrain the spatial extent of the hot core emission.  In \S 3 we describe the methods used to model the data set, and some statistics of the survey and the model fits.  \S 4 discusses some of the characteristics of the spectral fits to each molecule in the survey.  Following this, \S 5 provides a discussion of the overall results:  \S 5.1 presents the abundances of each molecule detected in the hot core, with a comparison of the results to those derived from the HEXOS/HIFI survey of the Orion KL region \citep{crockett13}.  In \S 5.2 we describe some of the properties of the Sgr B2 envelope, with a discussion of the observed kinematic components in \S 5.2.1, and an estimate of the CO$_2$/CO gas-phase ratio (from detections of their protonated ions, HCO$^+$ and HOCO$^+$) in \S 5.2.2, before concluding in \S 6.

\section{Observations and Data Reduction}

\subsection{HIFI}

The HIFI spectral survey of Sgr B2(N) was acquired over the period between 2010 September and 2011 April.  The frequency range, observation ID, date of observation, aperture efficiency \citep[taken from][] {roelfsema12}, and root-mean-square noise level for each band is given in Table 1.  The pointing center of the observations was $\alpha_\textnormal{J2000} = 17^\textnormal{h} 47^\textnormal{m} 19^\textnormal{s}.88$ and $\delta_\textnormal{J2000} = -28^\circ 22'18''.4$; a pointing error of $2''$ is assumed \citep{pilbratt10}.  The wide band spectrometer (WBS) was used, with a spectral resolution of 1.1 MHz.  Spectra were collected in dual beam switch mode, with reference beams $3'$ to the east and west of the science target.  Notably, therefore, some transitions with extended emission may have contamination in the off-position.  This will also be mentioned in \S 4 when appropriate.  The SPIRE line surface brightness maps shown in \cite{etxaluze13} indicate that some lines, particularly those of low-lying CO as well as the atomic fine structure transitions, may fall into this category, though the footprint of SPIRE is not large enough to include the HIFI off-position.  Spectra in both instrumental polarizations were acquired.

Details of the reduction of HIFI spectral surveys of other molecule-rich sources can be found elsewhere \citep{bergin10, crockett10, crockett13}, but we briefly summarize the procedure here.  The data reduction was performed using the Herschel Interactive Processing Environment \citep[HIPE,][]{ott10} version 8.  Because HIFI is a double-sideband spectrometer, a given spectral channel must be measured with multiple local oscillator (LO) settings in order to uniquely determine the sideband from which each transition arises.  For the Sgr B2(N) spectrum, each frequency was measured with eight unique LO settings in bands 1--5, and four in bands 6--7.

\begin{deluxetable}{c c c c c c l c}
\tablenum{1}
\tablewidth{0pt}
\tablecaption{Frequency ranges and observation parameters of the Sgr B2(N) spectral survey.}
\tablehead{Band & Frequency range & Observation ID & Observation date & $\eta_A$\tablenotemark{a} & Integration time & RMS\tablenotemark{b}  & $\theta_\textnormal{HPBW}$ \\
			&      (GHz)		&			&			&			& (s) & ($T_A$, K) & $('')$}
\startdata
1a & 479.5--561.1 & 1342205491 & 22 Sep 2010 & 0.68 & 9206 & 0.015 & 41 \\
1b & 554.5--636.5 & 1342206364 & 12 Oct 2010 & 0.67 & 7958 & 0.02 & 36 \\
2a & 626.1--726.0 & 1342204703 & 15 Sep 2010 & 0.67 & 13268 & 0.025 & 31 \\
2b & 714.1--801.2 & 1342204812 & 17 Sep 2010 & 0.67 & 11702 & 0.03 & 28 \\
3a & 799.1--860.0 & 1342204731 & 16 Sep 2010 & 0.67 & 6657 & 0.04 & 26 \\
3b & 858.1--961.0 & 1342204829 & 17 Sep 2010 & 0.67 & 14051 & 0.05 & 23 \\
4a & 949.1--1061.0 & 1342218198 & 07 Apr 2011 & 0.66 & 16130 & 0.09 & 21 \\
4b & 1050.3--1122.0 & 1342206370 & 12 Oct 2010 & 0.66 & 10521 & 0.06 & 20 \\
5a & 1108.2--1242.8 & 1342205855 & 05 Oct 2010 & 0.56 & 15017 & 0.2 & 18 \\
5b & 1227.2--1280.0 & 1342215934 & 12 Mar 2011& 0.56 & 5980 & 0.25 & 17 \\
6a & 1425.9--1535.0 & 1342204692 & 14 Sep 2010 & 0.64 & 16403 & 0.5 & 14 \\
6b & 1574.0--1702.8 & 1342206498 & 15 Oct 2010 & 0.64 & 13336 & 0.5 & 13 \\
7a & 1697.3--1797.9 & 1342216701 & 24 Mar 2011 & 0.63 & 10232 & 0.6 & 12 \\
7b & 1788.4--1906.5 & 1342206643 & 16 Oct 2010 & 0.62 & 11687 & 0.6 & 12
\enddata
\tablenotetext{a}{Aperture efficiencies taken from \cite{roelfsema12}.}
\tablenotetext{b}{Noise levels measured on spectra smoothed to a resolution of 1 km s$^{-1}$.}
\end{deluxetable}

The scans were first processed by the HIPE pipeline to Level 2 (i.e., suitable for scientific analysis).  Spurs were removed from the scans, either flagged automatically by HIPE or identified by eye.  A preliminary deconvolution (conversion of the double-sideband spectra into single-sideband, performed using the \emph{doDeconvolution} task in HIPE) was performed.  The continuum was then fit using a low-order polynomial, saved, and removed from the scans.  Because strong lines (defined here as $\Delta T > 10$ K) can introduce ghosts in the deconvolution procedure, these transitions were removed as well.  The \emph{doDeconvolution} task was then performed on the continuum-subtracted scans with the strong lines removed.  Deconvolution was also performed on the scans with the strong lines included, which were used to acquire information on the intensities of the strong lines themselves.  The strong lines were then inserted into the deconvolved scan.  The data were then exported to CLASS\footnote{http://www.iram.fr/IRAMFR/GILDAS}.    In all spectra presented here, the H and V polarizations were averaged together in order to improve the signal-to-noise ratio.  Two versions of the data are produced, one with the continuum removed and one with the continuum included; the latter data product was created by adding the fit continuum from the earlier step to the final deconvolved spectrum.  The spectra were corrected for the aperture efficiency from Table 1, and are presented on the $T_A$ scale.  In the figures presented here, the spectra were smoothed to a resolution of approximately 1 km s$^{-1}$.  The spectrum with the continuum removed was used for the modeling of emission lines, while the spectrum with the continuum included was used for absorption lines.  A full view of the HIFI spectrum is presented in Figure 1.  For this figure and all plots given on a frequency axis, a local standard of rest (LSR) source velocity of 64 km s$^{-1}$ is assumed.

\begin{figure}
\figurenum{1}
\centering
\includegraphics[width=6.0in]{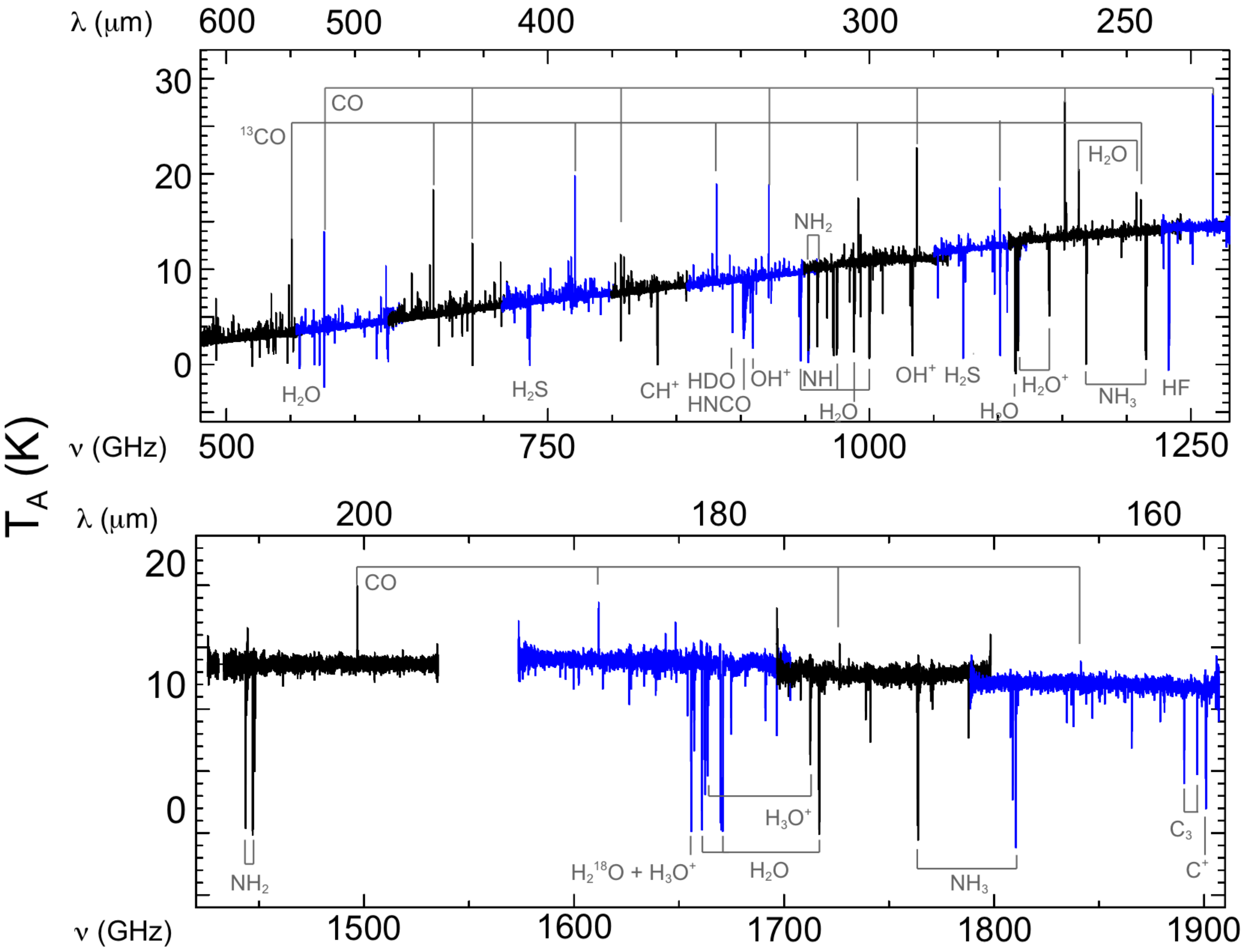}
\caption{Full view of the Sgr B2(N) HIFI survey.  For this figure, the spectrum has been smoothed to a spectral resolution of 8 km s$^{-1}$.  HIFI bands are divided into ``a'' and ``b'' bands (see Table 1); here we show ``a''  in black and ``b''  in blue.  Strong emission and absorption lines are identified.}
\end{figure}

Due to the high line density of the spectrum, particularly in the lower ($\nu < 1$ THz) bands of HIFI, baseline errors of up to 0.2 K are evident in the continuum-subtracted spectrum.  In many cases, these errors can be identified by eye, and in the figures presented here, the continuum level was adjusted line-by-line to correct for these errors.  However, in some regions with high line density, uncertainty of up to 0.2 K remains on the true baseline level.

\subsection{SMA}

Interferometric observations of Sgr B2(N), acquired with the Submillimeter Array (SMA), were used as auxiliary data in this work.  We use these observations to determine the source sizes for molecular emissions in the hot core.  An analysis of the continuum measured in these observations, along with details of the data acquisition and calibration procedures, can be found in \cite{qin11}. Briefly, data were acquired in both the compact configuration (with seven antennas in service, and projected baselines from 9--80 k$\lambda$) and in the very extended configuation (with eight antennas, and baselines from 27--590 k$\lambda$).  The measured spectral bandwidth was 342.2--346.2 and 354.2--358.2 GHz, with spectral resolution of 0.8125 MHz ($\sim 0.7$ km s$^{-1}$).

After calibration, the continuum was subtracted from the spectral data cube.  Channels presumed to be line-free were imaged and self-calibrated by the method presented in \cite{qin11}.  The continuum image has a spatial resolution of $0.4'' \times 0.24''$.  The spectral data cube was deconvolved using the CLEAN algorithm.  The spatial resolution for the molecular line images presented here is $0.62'' \times 0.60''$.  The absolute flux scale is estimated to have an uncertainty of 20\%.  It was estimated that roughly 60\% of the submillimeter continuum emission, particularly from the extended envelope, is resolved out in these observations \citep{qin11}.  Most of the SMA transitions presented here are of complex molecules, and generally in excited energy levels, so most of the emission likely arises from the hot core, to which the SMA observations are most sensitive.

\section{Data Analysis}

\subsection{Modeling emission lines}

The survey was modeled using the XCLASS program\footnote{https://www.astro.uni-koeln.de/projects/schilke/XCLASS}, which provides the functionality of the CLASS software with access to the CDMS \citep{muller01, muller05} and JPL \citep{pickett98} spectral catalogs.  This program uses the local thermodynamic equilibrium (LTE) assumption to model the spectrum.  For each molecule, we have developed a single model to describe the emission and absorption features of the molecule across the full HIFI spectral bandwidth.

For emission lines, we use the continuum-subtracted spectrum.  XCLASS therefore describes the spectrum by \citep{comito05, zernickel12}:

\begin{equation}
\Delta T_A(\nu) = T_\textnormal{line}(\nu)-T_\textnormal{cont}(\nu) = \sum_m \sum_c \eta_{bf} (\theta_{m,c},\nu) J(T_{\textnormal{ex}_{m,c}},\nu) \left(1-e^{-\tau(\nu)_{m,c}}\right) e^{-\tau_\textnormal{dust} (\nu)}
\end{equation}

\begin{equation}
\tau(\nu)_{m,c} = \sum_l \tau(\nu)_{l,m,c}
\end{equation}

\begin{equation}
\tau(\nu)_{l,m,c} = \frac{c^3}{8\pi\nu^3} A_{ul} N_\textnormal{tot}^{m,c} \frac{g_l e^{-E_l/kT_{\textnormal{ex}_{m,c}}}}{Q(m, T_{\textnormal{ex}_{m,c}})} \left(1-e^{-h\nu/kT_{\textnormal{ex}_{m,c}}}\right) \phi(\nu)_{m,c}
\end{equation}

\begin{equation}
\phi(\nu)_{m,c} = \frac{2\sqrt{\textnormal{ln}(2)}}{\sqrt{\pi}\Delta v_{m,c}} \textnormal{exp}\left(-\frac{\left(\nu-\nu_l\left(1-\frac{v_{\textnormal{LSR}_{m,c}}}{c}\right)\right)^2}{\left(\Delta {v_{m,c}\frac{\nu_l}{c}}\right)^2/4\,\textnormal{ln}\,2}\right)
\end{equation}

\noindent The beam filling factor, $\eta_{bf}$, is given by

\begin{equation}
\eta_{bf} (\theta_{m,c},\nu) = \frac{{\theta_{m,c}}^2}{{\theta_{m,c}}^2 + \theta_\textnormal{beam}(\nu)^2}
\end{equation}

\noindent where $\theta_\textnormal{beam} ('') = \frac{21200}{\nu (\textnormal{GHz})}$ for HIFI.  The spectrum is summed over the lines, $l$, for each component, $c$, of each molecule, $m$.  In these equations, $T_\textnormal{ex}$ represents the excitation temperature, $\tau(\nu)_{m,c}$ the line optical depth, $\tau_\textnormal{dust}(\nu)$ the optical depth due to dust, $N_\textnormal{tot}$ the molecular column density, and $\Delta v$ the full-width at half-maximum linewidth.  The source function is defined as $J(T,\nu) = \frac{h\nu}{k} \frac{1}{\textnormal{exp}(\frac{h\nu}{kT})-1}$.  The line frequencies $\nu_l$, Einstein A coefficients $A_{ul}$, state degeneracies $g_l$, lower-state energies $E_l$, and molecular partition functions $Q(T_\textnormal{ex})$ are derived from the CDMS and JPL catalogs.  In the treatment of emission lines, we assume intermixed, isothermal gas and dust; see Appendix 1 for a derivation of Eq. (1) in this case.  We also ignore the cosmic microwave background ($T_\textnormal{cmb}$ = 2.7 K), because $J(T_\textnormal{cmb})$ is negligible ($< 5$ mK) at HIFI frequencies.  This relatively simple model does not incorporate the effects of thermal gradients in the dust opacity, and also does not self-consistently incorporate the observed continuum brightness into the dust modeling.  More sophisticated radiative transfer models will be needed to move beyond these simplifications for a source with structure as complex as Sgr B2(N).

In most cases, particularly for molecules with only optically thin lines, the column density and source size cannot be independently derived in the modeling.  Therefore, the SMA observations are used to derive source sizes for the hot core.  However, we emphasize that we have not developed models to fit the intensities of the SMA transitions in this work.  Figure 2 shows single-transition integrated SMA flux maps of six molecules detected in the HIFI survey.  For these images, the emission has been integrated from \mbox{$\sim$60--68 km s$^{-1}$.}  At this spatial resolution, these images show that the molecular emission has very complex structure, and that different molecules have different spatial distributions.  In order to estimate the total emitting area of each molecule, we count the number of pixels with a flux within a factor of $e$ of the brightest pixel in the image, and multiply by the pixel area in arcsec$^2$.  This area is then converted into an equivalent spherical source size diameter by the following relation:

\begin{equation}
d = \sqrt{\frac{4A}{\pi}}
\end{equation}

\noindent Although this is a crude approximation, because of the complex structure of Sgr B2(N) as shown in the images in Figure 2, it is a reasonable measure of the total extent of the molecule's emission.  The abundances derived from the HIFI survey are therefore averaged over the regions in which each molecule is emissive.  For molecules where multiple transitions are detected in the SMA observations, each detected transition was analyzed in this way, and the average has been taken as the source size from the HIFI data.

Interpretation of the relative abundances of chemically related species, such as when discussing the chemistry (\S 5.1), requires that the same source size is used for modeling of each species---i.e., it must be assumed that the molecules emit from the same spatial region.  Therefore, we have divided the molecules detected in the hot core of Sgr B2(N) into six ``families,'' and we assume a uniform source size within each family.  These families are defined based on having similar spatial distributions in the SMA observations, and by chemical relationships; e.g., having the same heavy-atom backbone or the same functional group, and thus possibly related chemistry.  The source size adopted for each family is calculated by averaging the values derived for each molecule in that family with detected emission lines in the SMA data set.  The molecular families, with the adopted source sizes, are given in Table 2.    For molecules for which no transitions are found in the SMA observations, the molecule is placed into the family to which it is most closely chemically related.  The only clear case of a molecule being placed outside its natural chemical family is ketene (H$_2$CCO), which is placed in the N- and O-bearing group rather than with the C- and O-containing organics.  This is because its morphology in the SMA data cube much more closely resembles that of formamide (NH$_2$CHO) than that of methanol and formaldehyde.

\begin{figure}
\figurenum{2}
\centering
\includegraphics[width=5.0in]{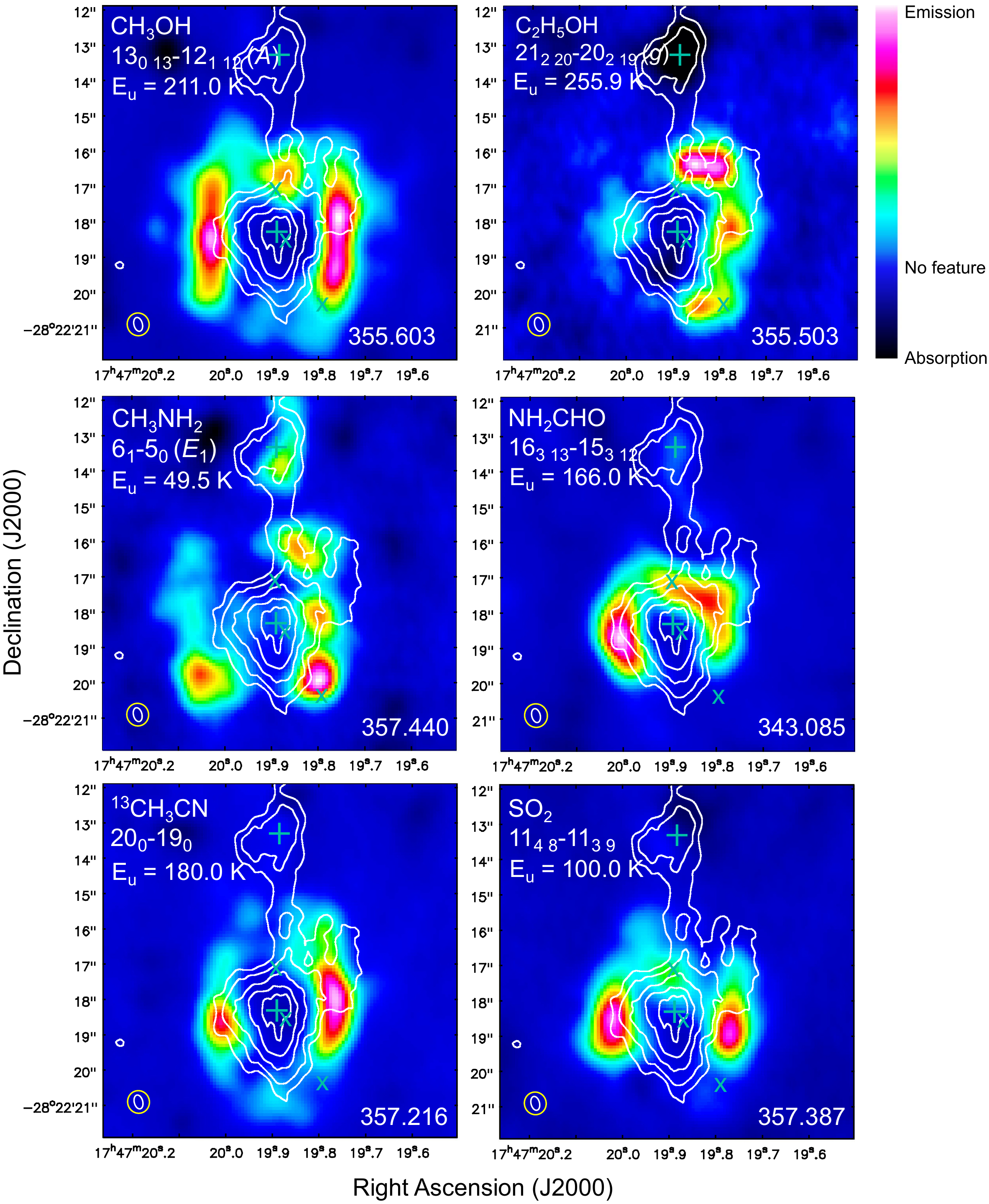}
\caption{Moment 0 images of transitions from the SMA observations.  In each panel, the white contours are the 350 GHz continuum from \cite{qin11}; the contour levels are (0.1, 0.2, 0.4, 0.6, 0.8) $\times$ 1.558 Jy beam$^{-1}$.  The color scale is the line identified in the panel, integrated from roughly 60--68 km s$^{-1}$.  The submillimeter dust cores SMA1 and SMA2, derived from \cite{qin11}, are labeled as teal plus signs, with the SMA1 core to the south and SMA2 to the north.  The three H II regions in the hot core region, K1, K2, and K3 (in order from south to north), from \cite{gaume95}, are labeled with teal `x' symbols.  The rest frequency of each transition (in GHz) is given in the lower-right corner of each panel.  A relative color scale is given in the top right corner; however, note that the absolute scale varies from panel to panel.}
\end{figure}

\begin{deluxetable}{l l c}
\tablenum{2}
\tablewidth{0 pt}
\tablecaption{Molecules detected in the Sgr B2(N) hot core by the HIFI survey, and the assumed source sizes based on SMA observations.}
\tablehead{Family & Molecules & Source Size $('')$}
\startdata
Simple O-bearing & CH$_3$OH, H$_2$CO, HCO$^+$\tablenotemark{a} & 3.1 \\
Complex O-bearing & CH$_3$OCH$_3$, C$_2$H$_5$OH, CH$_3$OCHO\tablenotemark{b}, HCOOH\tablenotemark{b} & 3.7 \\
NH-bearing &  NH$_3$\tablenotemark{a}, CH$_3$NH$_2$, CH$_2$NH\tablenotemark{a} & 2.5 \\
N- and O-bearing & NO\tablenotemark{a}, HNCO\tablenotemark{a}, NH$_2$CHO, H$_2$CCO & 2.3 \\
CN-bearing & CH$_3$CN, C$_2$H$_5$CN, C$_2$H$_3$CN, HC$_3$N\tablenotemark{b} & 2.0 \\
S-bearing & H$_2$CS, SO, SO$_2$, H$_2$S\tablenotemark{a}, OCS\tablenotemark{a}, CS, NS,  HCS$^+$\tablenotemark{b} & 2.8 \\
\enddata
\tablenotetext{a}{Indicates that this molecule was not imaged directly from the SMA data cube, but was placed into a family based on structurally similar molecules.}
\tablenotetext{b}{Indicates that this molecule was not detected in the HIFI survey, but this source size was used to derive an upper limit on its abundance.}
\end{deluxetable}

Extinction from dust is an important consideration toward the hot core in the submillimeter and THz, particularly for a massive source like Sgr B2(N).  In XCLASS, dust opacity is included using the equation

\begin{equation}
\tau_\textnormal{dust} (\nu) = 2\,m_\textnormal{H}\,N({\textnormal{H}_2})\,\chi_\textnormal{dust}\, \kappa (230\,\textnormal{GHz}) \left(\frac{\nu}{230\,\textnormal{GHz}}\right)^\beta
\end{equation}

\noindent where $N$(H$_2$) is the H$_2$ column density of the source, $m_\textnormal{H}$ is the mass of a hydrogen atom, and $\chi_\textnormal{dust}$ is the dust to gas mass ratio (assumed to be 0.01).  The dust opacity efficient $\kappa(\nu)$ at 230 GHz is taken from the first column of Table 1 in the calculations of \cite{ossenkopf94}, the case with a power law MRN distribution \citep{mathis77} of grain size and no coagulation, yielding 0.309 cm$^2$ g$^{-1}$.  \cite{etxaluze13} found an average $\beta$ of 1.8 across the HIFI spectral bandwidth in the Sgr B2 region, though slightly lower values ($\sim 1.5$) were seen at the positions of the N, M, and S hot cores.  However, those observations probe colder gas (fit temperatures of $\sim$25--30 K) than the hot core molecular species.  For the present analysis, a $\beta$ of 2 was assumed, the usual assumption for crystalline silicates and carbonaceous dust grains in star-forming regions \citep{ossenkopf94}.  We then use transitions of methanol, which has transitions spanning a wide range in energy (up to 1000 K) in each of bands 1--5, allowing the dust opacity to be fit independently from the excitation temperature in this survey.  From this, we derive $N(\textnormal{H}_2) = 8.2 \times 10^{24}$ cm$^{-2}$, corresponding to $\tau_\textnormal{dust} \sim 1$ at 800 GHz using Eq. (7).  A more detailed analysis of dust extinction in this spectrum will be presented in a separate study (Stephan et al., in preparation).

We have also estimated the H$_2$ column density in the hot core from the SMA continuum observations.  Sgr B2(N) is known from interferometric studies to have two peaks in the submillimeter continuum, referred to as SMA1 and SMA2 by \cite{qin11}.  These two hot cores are separated by $\sim 5''$, with the southern core SMA1 being more massive.  For most molecules, as Figure 2 shows, the emission at 64 km s$^{-1}$ arises primarily from SMA1.  The more northern SMA2 source, which is within the Herschel beam for bands 1--5, has its strongest emission at higher velocity, near 73 km s$^{-1}$ \citep{hollis03, belloche08}.  In the HIFI survey, the emission for most hot core molecules is centered at 64 km s$^{-1}$, so we focus on the SMA1 hot core. \cite{qin11}, using the SMA observations also used here in Figure 2, found that the submillimeter SMA1 dust emission is well modeled with a spherical distribution given by:

\begin{equation}
n(\textnormal{H}_2) = 1.7 \times 10^8\,\textnormal{cm}^{-3} \times \left( 1+\left(\frac{r}{11500\,\textnormal{AU}}\right)^2 \right)^{-2.5}
\end{equation}

\noindent This was calculated assuming $\kappa (\nu)$ = 0.6 cm$^{-2}$ g$^{-1}$ at 352 GHz, also derived from \cite{ossenkopf94}.

At the center of the hot core, integrating Eq. (8) yields an H$_2$ column density of $4.5 \times 10^{25}$ cm$^{-2}$.  However, Figure 2 shows that most molecules do not have significant submillimeter emission in the inner $\sim 1''$ of the SMA1 continuum peak.  The observed continuum level at the center is 1.55 Jy beam$^{-1}$ (corresponding to a brightness temperature of 170 K), so the dust optical depth at this position may already be high, and therefore may be at least partially responsible for the absence of molecular emission in this region.  Figure 2 shows that the molecular emission peaks lie along a ring with a radius of roughly $1.5''$.  At this radius, the H$_2$ column density, integrating Equation (8), is $8.0 \times 10^{24}$ cm$^{-2}$.  The observed continuum flux at the clumps where the molecular emission peaks in Figure 2 is 0.2--0.3 Jy beam$^{-1}$, or 23--35 K.  Assuming a dust temperature of 150 K, the median rotational temperature for molecules detected with a hot core component, yields $\tau_\textnormal{dust}$ (352 GHz) = 0.18--0.28.  This is in agreement with the value of 0.19 from Eq. (7).  This shows consistency between the dust opacity inferred from the HIFI survey and the continuum emission in the SMA observations.  We adopt $8.0 \times 10^{24}$ cm$^{-2}$ as the H$_2$ column density for the hot core HIFI emission in the abundance derivations presented in \S 5.1.  However, it should be noted that the $N$(H$_2$) derivations from both the dust absorption and emission are dependent on $\kappa (\nu)$.  Most other derivations of the H$_2$ column density toward Sgr B2(N) have been over a wider spatial extent than the hot core, and so derived lower values; see, e.g., \cite{nummelin00}.  \cite{belloche08} used 3 mm continuum observations with the Plateau de Bure Interferometer to derive an H$_2$ column density of $1.3 \times 10^{25}$ cm$^{-2}$.  \cite{etxaluze13} found an H$_2$ column density of $9.0 \times 10^{24}$ cm$^{-2}$ using 250 $\upmu$m continuum emission, with a FWHM $18.5''$ beam.

Since the hot core and the envelope both have similar kinematic properties (with the primary velocity component for most molecules located near $\sim$64 km s$^{-1}$), there are molecules for which the identification of an emission component to one or the other can be difficult.  Warm components ($T_\textnormal{rot} >$ 100 K) are in general assumed to originate from the hot core, while colder components are assumed to be extended.  For some optically thick molecules, i.e. H$_2$S, warm components are observed, but nevertheless a larger source size than that presented in Table 2 is required.  This is because the observed transitions are too strong for a small source size:  for example, an optically thick line at $T_\textnormal{rot} = 150$ K, and a source size of $3''$, and the dust opacity as given above, both typical for the hot core, has an intensity of $\sim 0.5$ K.  (The frequency dependences of the beam dilution factor and the dust opacity counteract each other when $\theta_\textnormal{source} << \theta_\textnormal{beam}$, so the product $\eta_\textnormal{bf}(\nu) e^{-\tau_\textnormal{d}(\nu)}$ varies by only about 25\% between 500--1000 GHz.)  Because these optically thick lines trace the surface layer of the hot core, they emit from a larger area than the true abundance distribution.  Therefore, for these molecules, rare isotopologues (e.g. $^{13}$C, $^{34}$S) are used to derive the abundance in the hot core, for which the source sizes from Table 2 are used, and the isotope ratios given in \S 3.3 are assumed.  Dust opacity is assumed to be negligible for components emitting from the envelope; \cite{lis90} find $N$(H$_2$) $\sim 10^{24}$ cm$^{-2}$, which indicates $\tau_\textnormal{dust} < 1$ in the HIFI spectral range.

In general the modeling was done by eye, rather than through an automated fitting program such as was done by, e.g. \cite{zernickel12}.  This is due to the extensive line blending present in the survey, particularly in the lower bands, and the baseline uncertainties in the spectrum.  We began with parameters from the previous surveys of Sgr B2(N) \citep{turner91, nummelin00, belloche09}.  The modeling of all species was done in parallel so as to account for blended lines in the model.  We do not present formal uncertainties on the abundances derived in the fitting procedure.  Typical uncertainties is these kinds of analyses, which are often dominated by degeneracies between the column density and excitation temperature, have been explored elsewhere \citep{nummelin00, comito05}.  Because this is the largest-bandwidth survey of this source at high spectral resolution, transitions spanning widely in excitation energy are detected for most molecules, so the column densities and excitation temperatures are fairly well constrained.

\subsection{Modeling absorption lines}

Because of the bright continuum of Sgr B2(N) in the HIFI spectral range, and the regions of colder and diffuse gas along the line of sight, many more absorption lines are detected in this survey than toward nearer sources such as Orion KL \citep{crockett13} or NGC6334(I) \citep{zernickel12}.  For modeling absorption lines, the data set with the continuum included is used.  The equations used to model the absorption lines in XCLASS are

\begin{equation}
T_A(\nu)_{m,c=1} = (T_\textnormal{cont} + \Delta T_\textnormal{em})\,e^{-\tau(\nu)_{m,c=1}} + \eta_{bf}(\theta_{m,c=1},\nu) J(T_{\textnormal{ex}_{m,c=1}}) (1-e^{-\tau(\nu)_{m,c=1}}) 
\end{equation}

\begin{equation}
T_A(\nu)_{m,c=i} = T_A(\nu)_{m,c=i-1}\,e^{-\tau(\nu)_{m,c=i}} + \eta_{bf}(\theta_{m,c=i},\nu)J(T_{\textnormal{ex}_{m,c=i}}) (1-e^{-\tau(\nu)_{m,c=i}})
\end{equation}

\noindent where $\Delta T_\textnormal{em}$ is the intensity of the emission components from Eq. (1) (including dust attenuation), if they exist for the molecule in question, and $T_\textnormal{cont}$ is the continuum observed by HIFI. In the case that only one absorption component is modeled, only Eq. (9) is used.  With multiple components, however, Eq. (10) is employed recursively, with the result of each absorption component being used as the background radiation field for the next.  The optical depth is calculated using Eqs. (2)--(5).  This effect is only significant when more than one temperature component is needed at a given velocity.  This is needed, for example, in cases where saturated absorption, suggestive of $J(T_\textnormal{ex}) \sim 0$, is seen for a molecule, but more excited transitions are also detected, requiring a component with a higher $T_\textnormal{ex}$ (e.g., for H$_2$S).  In general, the components are ordered from warmest to coldest, assuming an envelope decreasing in excitation temperature as the distance from the hot core increases.

For most molecules with cold absorption (defined here as $T_\textnormal{ex} < 40$ K), the absorption is assumed to be spatially extended across the HIFI beam.  Some molecules, e.g. HCN, also show ``hot'' absorption ($T_\textnormal{ex} \ge 40$ K) in high-energy transitions, which must be modeled as spatially compact.  This absorption is always either red-shifted or blue-shifted from the standard Sgr B2(N) source velocity of 64 km s$^{-1}$, and will be discussed more in \S 5.2.  Table 3 shows all species detected in the HIFI survey, indicating which are detected with lines in emission, cold absorption, and hot absorption, using the definitions given above.  More specific information on the fits for each molecule is given in \S 4.

For some molecules with complex lineshapes, particularly those displaying both absorption and emission, the source geometry is likely too complex, and optical depths too high for reliable column densities to be determined with the present modeling approach.  The fits for these cases presented in this manuscript should be regarded as purely effective lineshape fits to the spectrum.  There are also a few molecules with extremely complex lineshapes or complex excitation where the lines were not fit with XCLASS:  $^{12}$CO, H$_2$O and its rare isotopologues (H$_2$$^{17}$O, H$_2$$^{18}$O, and HDO), and H$_3$O$^+$.  For CO and water, most transitions show complex combinations of absorption and emission, and may also have some contribution from emission in the off-position.  For H$_3$O$^+$, transitions up to high energy are observed, but only involving metastable ($J=K$) levels.  For the present analysis, we have fit the detected lines for these species with Gaussian components, and included the lineshapes in the full model for line identification purposes.

\subsection{Isotope ratios}

Rare isotopologues are detected for a number of molecules.  The $^{12}$C/$^{13}$C ratio was taken to be 20 in Sgr B2 \citep{wilson94}, verified by modeling of the HNCO $K_a = 1-0$ $Q$-branches seen in absorption at 900 GHz (shown in Figure 3).  Minor isotopologues of oxygen are seen only in H$_2$O and CO (as well as for HCO$^+$ using previously published BIMA data, which will be discussed in \S 5.2.2), and the $^{18}$O/$^{16}$O and $^{18}$O/$^{17}$O ratios were taken to be 250 \citep{wilson94} and $3.2 \pm 0.4$, respectively.  The latter ratio is derived from the ground state transitions of H$_2$$^{18}$O and H$_2$$^{17}$O in absorption, as shown in Figure 4, and is also consistent with the study of \cite{penzias81}.  The ground state transitions of H$_2$$^{16}$O are saturated in absorption at the Sgr B2(N) envelope velocity.

Minor isotopologues of sulfur are seen in H$_2$S, SO, and CS.  \cite{frerking80} found an average $^{13}$CS/C$^{34}$S ratio of 0.67 toward three positions in the Galactic Center: Sgr A, Sgr B2, and a position offset from Sgr B2 by a few arcminutes, yielding $^{32}$S/$^{34}$S = 13 using the carbon isotope ratio given above.  For H$_2$S, ground state transitions for the \emph{ortho} species ($2_{12}-1_{01}$) are detected near 735 GHz for H$_2$$^{32}$S, H$_2$$^{34}$S, and H$_2$$^{33}$S in the HIFI survey.  H$_2$$^{32}$S is saturated in absorption in the Sgr B2 envelope, so the $^{32}$S/$^{34}$S ratio cannot be derived, but a comparison of H$_2$$^{34}$S and H$_2$$^{33}$S optical depths yields a $^{34}$S/$^{33}$S ratio of $6.9 \pm 0.6$, as seen in Figure 4.

\begin{figure}
\figurenum{3}
\includegraphics[width=6.5in]{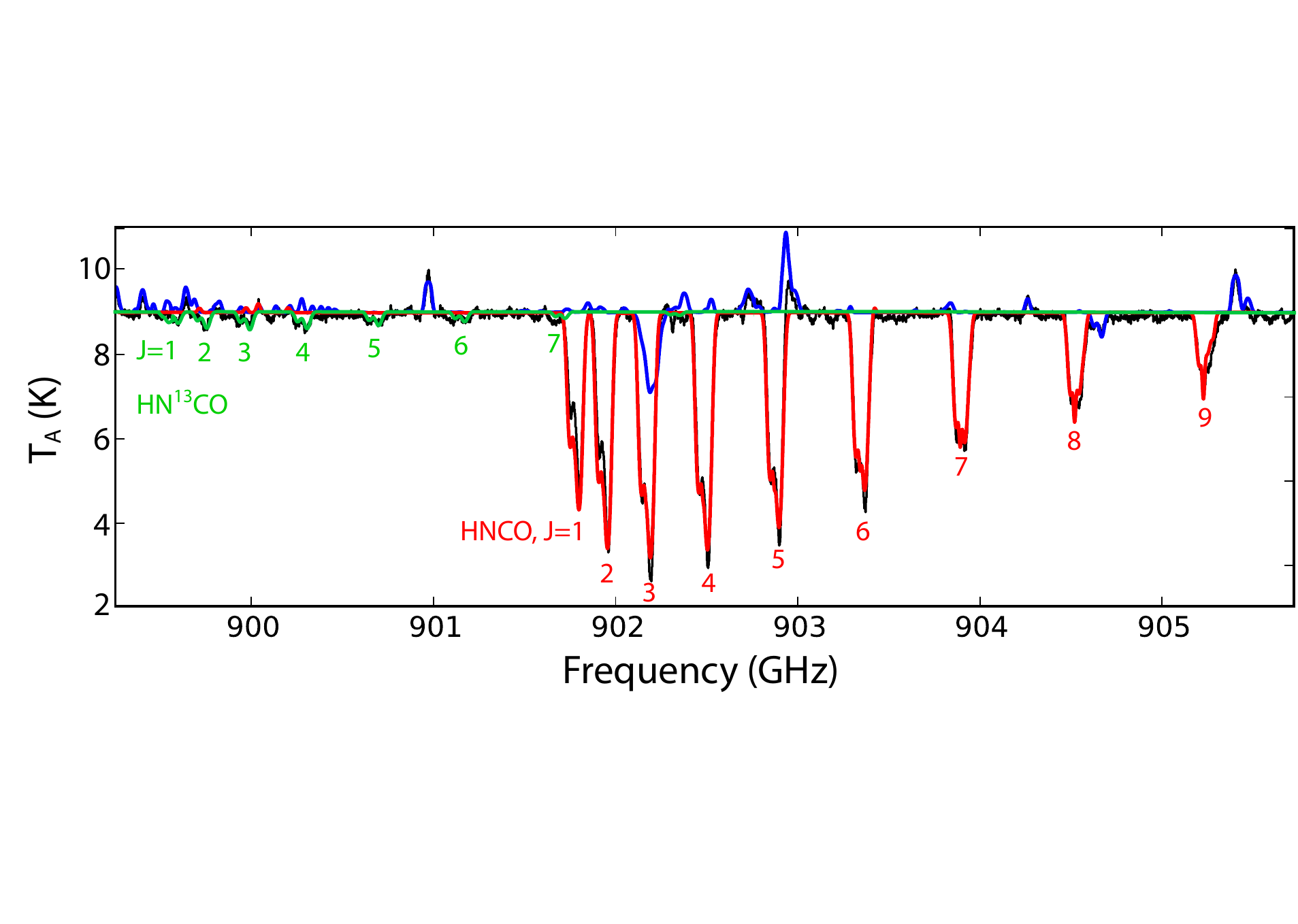}
\caption{$K_a = 1-0$ $Q$-branches of HNCO and HN$^{13}$CO.  The low-energy ($E_\textnormal{low}$ = 0--30 K) transitions in this figure are dominated by cold absorption from the envelope, which is discussed further in \S 5.2.  The two species are fit by the same model, assuming a $^{12}$C/$^{13}$C isotopic ratio of 20.  The blue trace shows the fullband model to all other molecules detected in the spectrum.}
\end{figure}

\begin{figure}
\figurenum{4}
\includegraphics[width=6.5in]{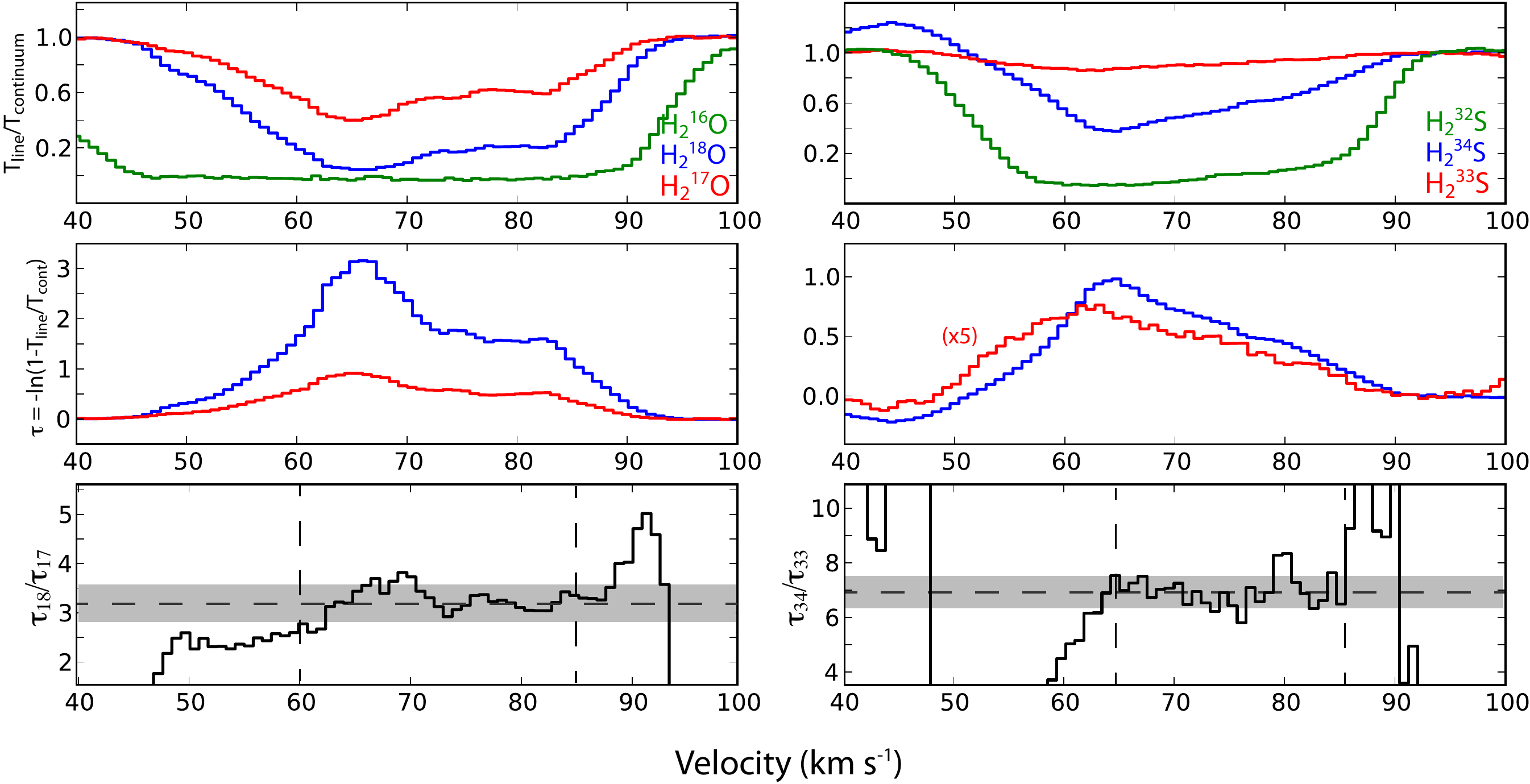}
\caption{Determination of the $^{18}$O/$^{17}$O and $^{34}$S/$^{33}$S ratios from ground-state transitions of isotopologues of H$_2$O and H$_2$S, respectively.  The top panel of each column shows the spectrum of each transition:  the $1_{11}-0_{00}$ ($para$ ground state) of H$_2$O, and the $2_{12}-1_{01}$ ($ortho$ ground state) of H$_2$S.  The middle panel shows the optical depths as a function of velocity for the minor isotopologues; because the transitions are in saturated absorption, optical depth cannot be determined for the main isotoplogue of each species.  The lowest panel shows the ratio of the opacities of the two minor isotopologues; the dashed line indicates the average value between velocities of 60 and 85 km s$^{-1}$ (65--85 km s$^{-1}$ for H$_2$S), while the gray box indicates the 1$\sigma$ error bar.  The transition of H$_2$$^{34}$S is contaminated by an emission line of CS ($J = 15-14$) on the blue wing (centered at 55 km s$^{-1}$ in the rest frame of the H$_2$$^{34}$S transition).}
\end{figure}

The $^{15}$N isotope is seen only in NH$_3$, and the $^{14}$N/$^{15}$N ratio is taken to be $182 \pm 22$ \citep{adande12}.  Both isotopes of chlorine, $^{35}$Cl and $^{37}$Cl, are seen in HCl and H$_2$Cl$^+$, and for both species the terrestrial $^{35}$Cl/$^{37}$Cl ratio of 3:1 is assumed.  The only detected deuterated species are HDO and DCN, but D/H ratios cannot be derived for either species from the present LTE analysis due to high optical depth in the normal isotopologues, and the complicated lineshapes.

\subsection{Statistics of the HIFI spectrum}

Figures 5 and 6 show several segments of the HIFI spectrum with the full model and those of individual molecules overlaid, spanning bands 1--5.  Bands 1 and 2, seen in Figure 5, show the rich spectrum associated with millimeter wave surveys of Sgr B2(N) and other molecule-rich star-forming regions, where complex organics occupy a large fraction of the spectral channels.  At higher frequencies (Figure 6), the spectral line density decreases, due to a falloff in the complex organic emission.  The Herschel beam width is decreasing at higher bands, increasing the beam coupling factor, but dust opacity, the Boltzmann distribution, and the increasing noise level of HIFI all suppress the contribution of these interstellar ``weeds'' \citep{herbst09, crockett10}.

\begin{figure}
\figurenum{5}
\centering
\includegraphics[width=6.0in]{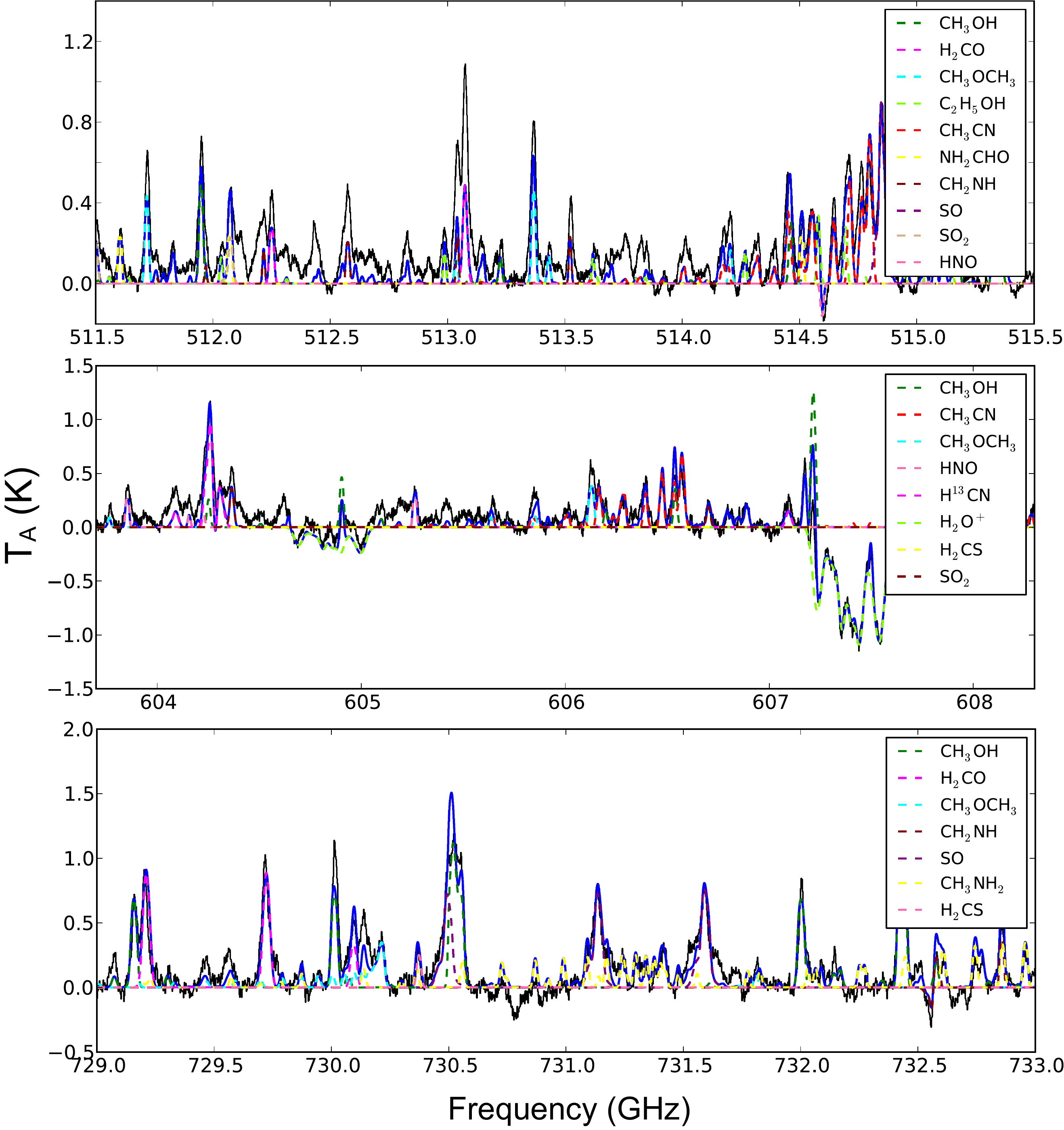}
\caption{Excerpts of the HIFI survey of Sgr B2(N) in bands 1--2, with the full model (solid, in blue) and those of the molecules contributing most of the flux to the spectral range of each panel.  For the purposes of this figure, a molecule's simulation includes that of all detected isotopologues of that molecule.}
\end{figure}

\begin{figure}
\figurenum{6}
\centering
\includegraphics[width=6.0in]{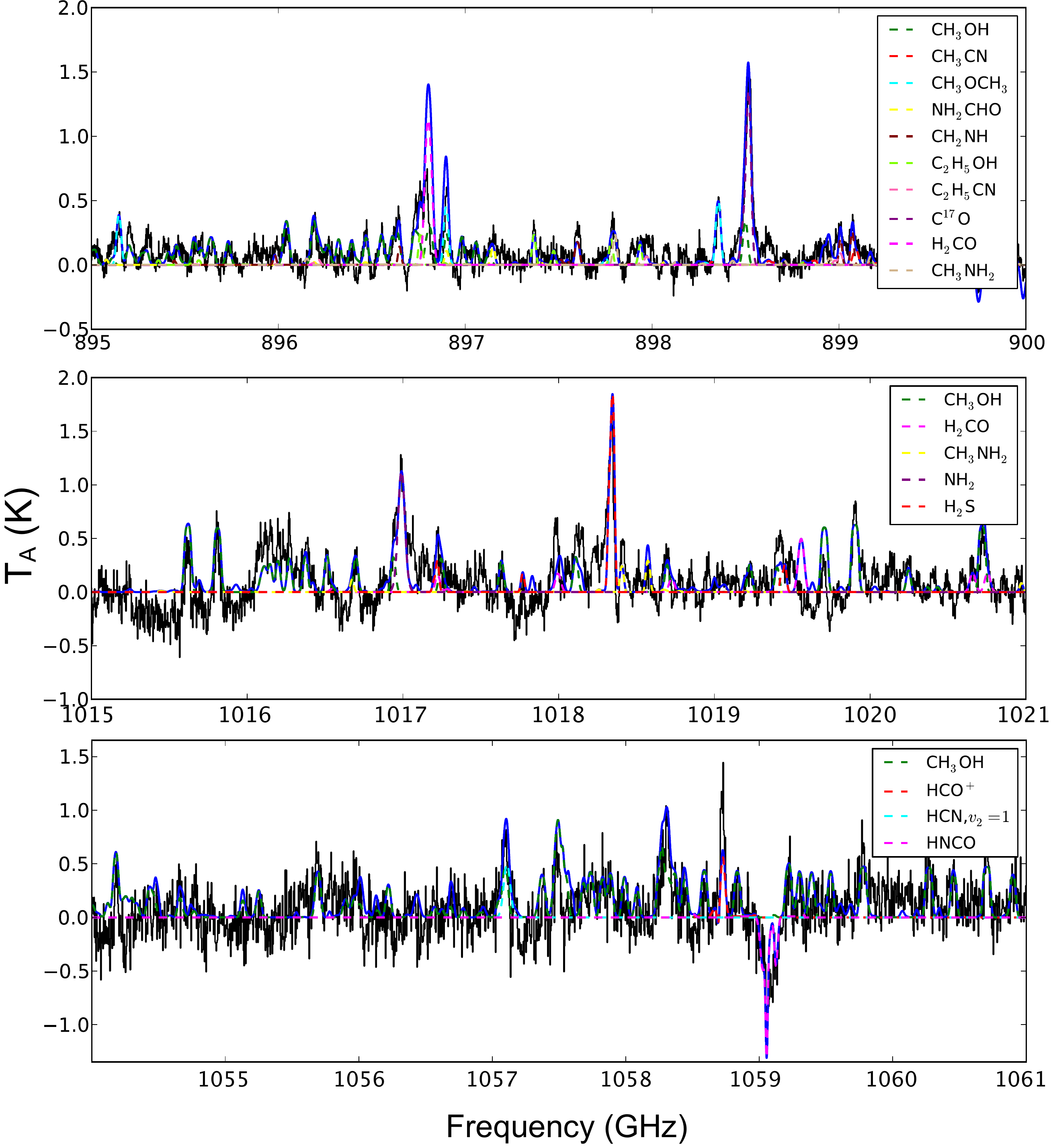}
\caption{Excerpts of the HIFI survey of Sgr B2(N) in bands 3--4, with the full model (solid, in blue) and those of the molecules contributing most of the flux to the spectral range of each panel.  For the purposes of this figure, a molecule's simulation includes that of all detected isotopologues of that molecule.}
\end{figure}

To assess the quality of the model to each molecule in describing the spectrum, we calculated a reduced-$\chi^2$ goodness-of-fit parameter for each model.  The method for this calculation is presented by \cite{crockett13}.  Channels judged to have unblended emission or absorption (defined as channels where the molecule in question contributes at least 90\% of the total flux in the model) are selected for each molecule for the $\chi^2$ calculation.  The results of this calculation are presented in Table 4.  The uncertainty contains error terms from the baseline uncertainty (defined to be 0.05 K), the fit rms noise, calibration uncertainty (10\%), pointing uncertainty ($2''$), and source size.  Molecules with only absorption features are not included in Table 4, but all are fit well by the model (though, since most have only a few transitions, this is not surprising that a good model could be developed).  Table 4 shows that for most molecules, the model describes the emission of the molecule within the uncertainty of the spectrum ($\chi^2_\textnormal{red}$ near 1).  The molecules that are poorly fit tend to be molecules with high opacities and are detected in both the hot core and the envelope: CH$_3$OH, HCN, HNC, H$_2$S, H$_2$CO, HCO$^+$, CS, and minor isotopologues of CO.  For each of these species, the models to the minor isotopologues are better fit.  Overall, the model describes the spectrum quite well.

In Figure 7 we show the fraction of emissive channels in the model in each band.  Absorption lines, which occupy a small fraction of the data channels, are neglected for this calculation.  The two panels differ in the threshold for the definition of an emissive channel.  In the top panel, the threshold is the 2$\sigma$ noise level for each band, taken from Table 1.  This shows that as the frequency increases, the fraction of channels with a detected transition drops from nearly 50\% at band 1a to the point where only one transition (of $^{12}$CO) is detected in each of the subbands in bands 6--7.  The bottom panel shows the same plot, except here the fraction of channels with emission calculated to be stronger than 30 mK (the 2 $\sigma$ noise level of band 1a) in each band is plotted; that is, this is the theoretical line density assuming a constant brightness sensitivity for each band.  This shows a roughly constant spectral density through bands 1--4, followed by a drop in bands 5--7.  Most of the drop in spectral density up to roughly 1 THz is therefore due to the increase in the HIFI noise level.  It should also be noted that a constant value in the fraction of channels that are emissive still represents a drop in the number of transitions per frequency interval, since the spectral linewidth (in frequency units), and therefore the number of channels a transition occupies, increases with frequency.

The top panel in Figure 7 also includes a correction for unidentified emissive channels.  The method for estimating the fraction of emissive channels that are unidentified is the same as is used in \cite{crockett13}.  First, the baseline and root-mean-square noise levels are determined over each successive \mbox{3 GHz} frequency range.  To do this, all channels with $>1\sigma$ emission in the LTE model (with $\sigma$ defined initially from Table 1) are first flagged and not included in the noise level calculation.  Other emissive channels are then flagged by a sigma-clipping algorithm:  the noise level is calculated, any channels at greater than $2\sigma$ are flagged, and the calculated baseline is subtracted.  This procedure is iterated until the noise level is unchanged within 1\% from one iteration to the next.  Once the final noise level is determined, emissive channels are defined as those greater than $3\sigma$.  An emissive channel is considered to be unidentified if the line intensity predicted by the model is at least a factor of 5 weaker than the observed line intensity.  We find that between 10--13\% of emissive channels are unidentified at bands 1a--5a, falling to 5\% for band 5b.  No unidentified lines are seen in bands 6 or 7.  There are no clear unidentified absorption lines.  A previously unidentified line located near 617 GHz has now been identified as $^{36}$ArH$^+$ in a number of sources, including the Crab Nebula \citep{barlow13}.  This line is seen toward Sgr B2 with extensive structure, but will not be analyzed in this manuscript, with a future report forthcoming (Schilke et al., in preparation).  From this analysis, we estimate a total of $\sim$8,000 features (including both assigned and unassigned) in the Sgr B2(N) survey.

\begin{figure}
\figurenum{7}
\centering
\includegraphics[width=4.0in]{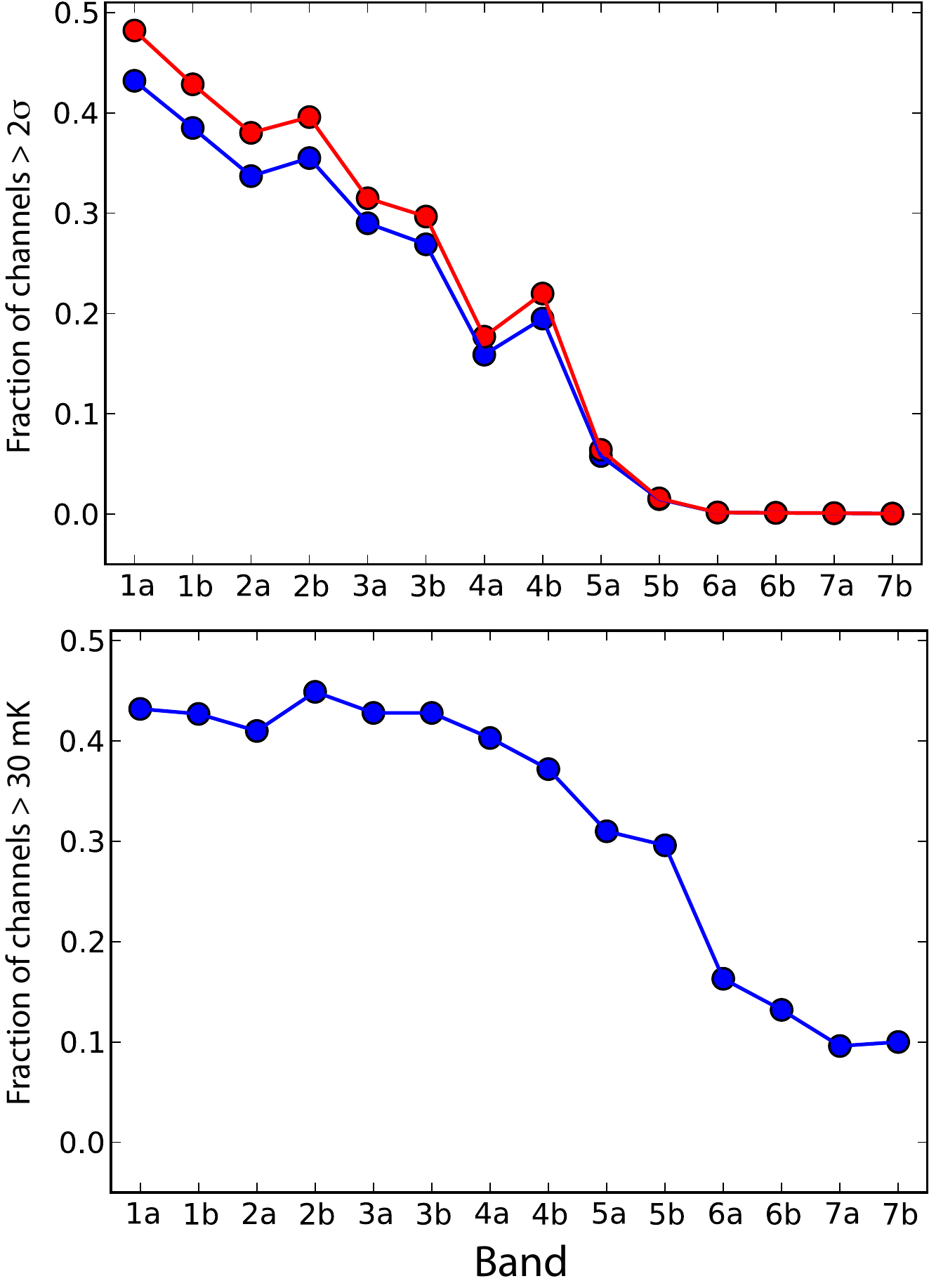}
\caption{Fraction of channels with emission above a given threshold in each HIFI band.  The blue points indicate the fraction of emissive channels derived from the LTE model; in the top panel, the red points indicate the fraction of emissive channels after correcting for unidentified channels using the method described in the text.  In the top panel, the RMS noise levels are taken from Table 1; in the bottom panel, the 2$\sigma$ noise level from band 1a is used.}
\end{figure}

The integrated emissive line intensity, luminosity (given in units of $L_\odot$), and line count of each molecule with detected emission features is presented in Table 5.  For $^{12}$CO and all isotopologues of H$_2$O, we sum the fit Gaussian components.  The integrated flux is therefore a lower limit because of absorption in the envelope.  These values are calculated from the best-fit LTE models, and exclude absorption components.  Because there are now two other high-mass sources for which full analyses of HIFI surveys have been completed, NGC6334(I) \citep{zernickel12} and Orion KL \citep{crockett13}, it is interesting to compare some of the basic properties of these two surveys.  As in the other two sources, methanol is the dominant emitter in the Sgr B2(N) HIFI spectrum.  We find about a factor of 2 more total integrated flux in molecular lines than in the NGC6334(I) survey, but less than Orion KL by about a factor of 2.   The conversion to luminosity is made assuming $d = 8$ kpc.

\section{Results}

In this section we describe the spectral fits to each of the molecules detected in the Sgr B2(N) HIFI survey.  In Tables 6--8 we provide the LTE parameters derived from this survey for the Sgr B2(N) source.  Table 6 presents emission components, Table 7 shows parameters for cold absorption components, and Table 8 describes the hot absorption.  In these tables we show only the parameters for the Sgr B2(N) envelope ($v_\textnormal{LSR}$ = 50--100 km s$^{-1}$).  Absorption in the spiral arm clouds is seen in a number of molecules toward Sgr B2 \citep{lis10b, qin10, schilke10}, and is fit as part of the fullband model, but we omit the fit parameters for the current analysis, and we do not discuss these components in this paper.

For each detected molecule with emission components, a figure showing the detected transitions (or, in the case with molecules with a large number of transitions, a representative sample of the detected transitions) can be found in Figure Set 8.

\begin{figure}
\figurenum{8}
\centering
\includegraphics[width=4.5in]{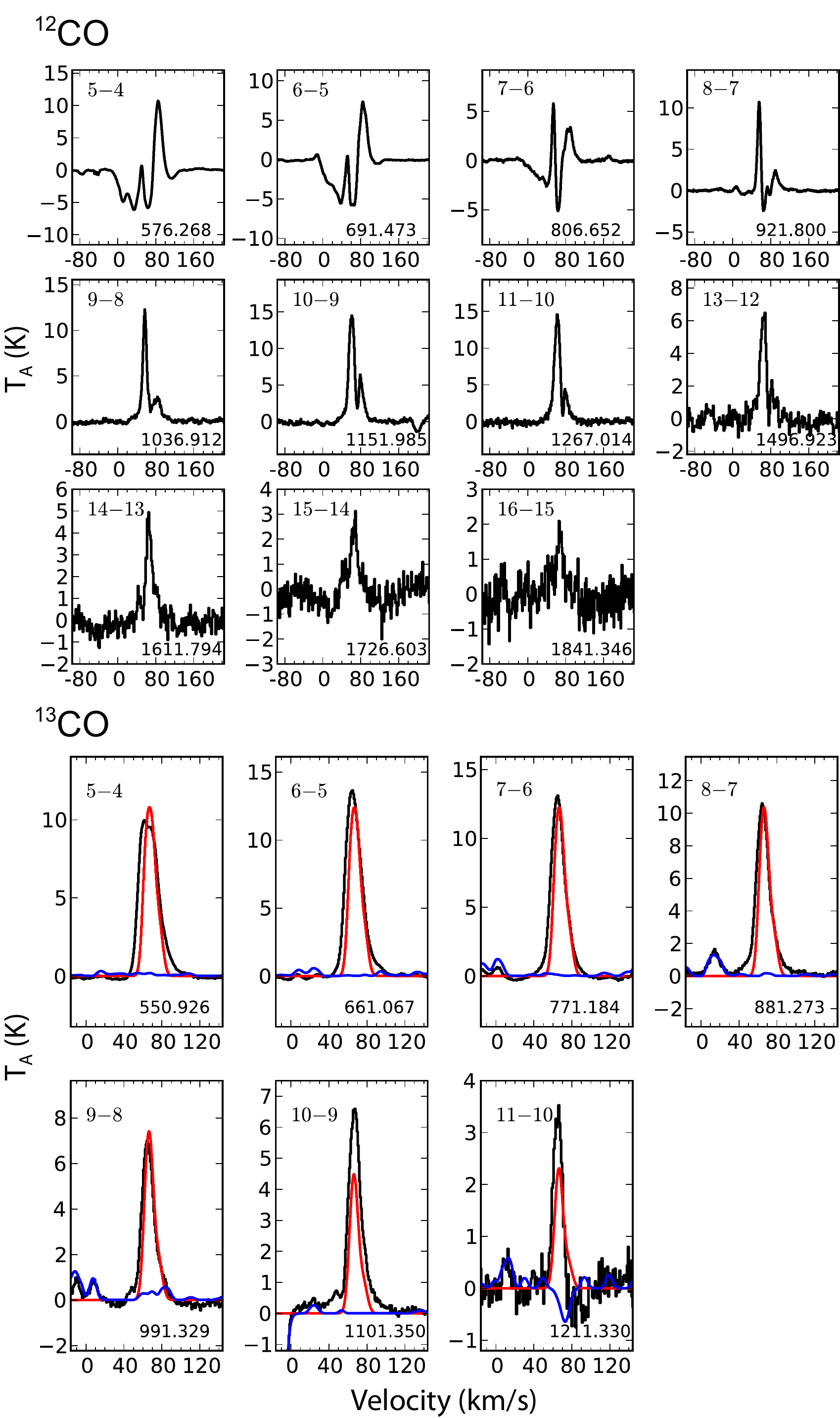}
\caption{Detected transitions of $^{12}$CO and $^{13}$CO.  In each panel of $^{13}$CO, the red curve is the model to the molecule in question, and the blue curve is the fullband model (with the subject molecule subtracted).  The quantum numbers of the transition are given in the upper-left corner, and the rest frequency (in GHz) is given in the lower-right corner.  The full Figure Set 8 (figures 8.1--8.30) can be found in the online edition of the journal.}
\end{figure}

\subsection{Simple O-bearing molecules}

\paragraph{CO} (Figs. 8.1, 8.2) Five isotoplogues of carbon monoxide are detected: $^{12}$CO, $^{13}$CO, C$^{18}$O, and C$^{17}$O, and $^{13}$C$^{18}$O; the last of these is tentative with only a few detected lines.  Transitions of the $^{12}$CO isotopologue have extremely complicated lineshapes, with mostly absorption at the Sgr B2(N) velocity up to $J' = 9$, though emission appears on the red and blue wings.  The lines continue to appear with a mix of absorption and emission, up to $J' = 13$, with the highest energy lines in bands 6 and 7 appearing as pure emission.

The minor isotopologues show only emission (except for perhaps some self-absorption in the $^{13}$CO $J= 5-4$), and are only detected in bands 1--5.  Each is fit with two velocity components centered at 66 and 76 km s$^{-1}$, and $T_\textnormal{rot} = 60$ K.  In the modeling of the minor isotopologues, the source size must be large to reproduce the emission, since even C$^{17}$O lines are observed with intensities of $\sim$1--2 K; therefore most of the emission must arise from the envelope, rather than from the $\theta=3''$ hot core.  A comparison of the fluxes of $^{13}$CO, C$^{18}$O, and C$^{17}$O transitions, given the isotope ratios in \S 3.3, indicates that both $^{13}$CO and C$^{18}$O are moderately optically thick.

\paragraph{CH$_3$OH} (Fig. 8.3) Methanol has more detected transitions than any other molecule in the HIFI spectrum, with transitions detected from the ground state to energies of about 1000 K.  An initial report on the HIFI spectrum by \cite{neill12}, focusing on the Q-branch at 830 GHz, showed that two temperature components were necessary to fit the spectrum succesfully.  One is warm ($T_\textnormal{rot} = 170$ K), which we take to be the hot core, and the second is cooler (60 K), which we take to be from the envelope.  Low-energy transitions of the $^{12}$CH$_3$OH isotopologue have high optical depths.  The $^{13}$CH$_3$OH isotopologue is also fit by the same model.

Methanol is also detected in the cold envelope of Sgr B2(N), as evidenced by the detection of many low-energy lines ($E_\textnormal{low} < 100$ K) in absorption.  The model is poor for some absorption lines, particularly the lines where both emission and absorption components are present.  For the absorption model, the $A$ and $E$ symmetry species are separated and have different column densities and temperatures in some components; non-LTE excitation is likely important.  This is a function of the different transitions detected in the two symmetry species.  In particular, a ground-state transition of CH$_3$OH($E$), the $2_{-2}-1_{-1}$ at 520179 MHz, is detected with absorption saturated to zero, so an optically thick component with $T_\textnormal{ex} \sim 2.7$ K is needed.

\paragraph{H$_2$CO} (Fig. 8.4) Most detected transitions of H$_2$$^{12}$CO have high optical depths, and achieving a satisfactory fit is very difficult (as seen by the relatively poor $\chi^2$ in Table 4).  Formaldehyde also may have emission from the envelope.  The model to H$_2$$^{12}$CO includes two cooler (80 K), extended components.  Some high-energy, high-frequency ($\nu > 1$ THz) transitions also have an absorption component near 53 km s$^{-1}$, resulting in P Cygni profiles.  The characteristics of this hot absorbing gas, which is detected at this velocity in a number of molecules, will be discussed in \S 5.2.  To derive the abundance of formaldehyde in the hot core, we use the fit to the H$_2$$^{13}$CO isotopologue.  H$_2$$^{13}$CO is detected through approximately 40 transitions ranging in energy from 70 to nearly 500 K, and fit well by a single component with $T_\textnormal{rot} = 100$ K.  Deuterated formaldehyde is not detected, with an upper limit to the [HDCO]/[H$_2$CO] ratio (assuming $T_\textnormal{ex} = 100$ K) of 0.005.

\paragraph{H$_2$CCO}  (Fig. 8.5) Ketene has weak ($\sim 0.1-0.2$ K) emission features detected throughout Bands 1 and 2.  The emission is well characterized with a single, warm ($T_\textnormal{rot} = 180$ K), optically thin component.

\paragraph{HCO$^+$}  (Fig 8.6) Nine transitions of H$^{12}$CO$^+$ are detected, ranging from $J_\textnormal{up} = 6-14$.  All transitions have complex lineshapes, and are fit with three emission components and a warm ($T_\textnormal{ex} = 55$ K) absorption component centered at 73 km s$^{-1}$.  Many transitions of H$^{13}$CO$^+$ are blended with transitions of other species.  The $J = 6-5$ and $7-6$ lines show an unusual profile suggesting absorption in the blue wing, but this is not seen in the normal isotopologue.  Neither HC$^{18}$O$^+$ nor DCO$^+$ are detected.  The $J = 8-7$ and $9-8$ lines of H$^{13}$CO$^+$ have clean profiles, which are assumed to arise from the hot core and are used to derive the abundance of HCO$^+$, assuming $T_\textnormal{rot} = 150$ K.

\paragraph{HOCO$^+$}  (Fig. 13) Protonated carbon dioxide is detected through absorption transitions in this survey, most notably through the $K_a = 1-0$ $Q$-branch at 780 GHz, as well as a few R-branch transitions.  The lines are fit with two velocity components at 64 and 78 km s$^{-1}$, both with rotation temperatures of 14 K, in agreement with millimeter measurements by \cite{minh91}.  These authors also found that HOCO$^+$ has emission extended over several arcminutes, peaking roughly $1'$ north of Sgr B2(N).  The HCO$^+$/HOCO$^+$ abundance ratio in the envelope will be discussed further in \S 5.3.

\paragraph{Others} OH$^+$ and H$_2$O$^+$ are detected in absorption through their ground state transitions in the Sgr B2 envelope and in the spiral arm clouds \citep{schilke10}.  Higher-excitation lines are not detected in either molecule.  OH is detected in absorption in the Sgr B2(N) envelope through two transitions at 1834 and 1837 GHz with components at 73 and 59 km s$^{-1}$.  Many more OH transitions have been detected at shorter wavelengths by ISO \citep{goicoechea02, polehampton07}. H$_3$O$^+$ is detected in absorption through metastable ($J=K$) transitions up to very high energy ($J=11$, $E_l = 1224$ K) levels, and its excitation is discussed in a separate study \citep{lis12}.  H$_2$O, H$_2$$^{18}$O, H$_2$$^{17}$O, and HDO are detected through a number of lines with complex lineshapes with both absorption and emission components.  The abundance of water in all components will require complex non-LTE modeling, which we have not pursued for this paper, though an estimate in the cold envelope based on the ground state transitions, which are detected in absorption, is presented in \S 5.2.  SiO is not detected in this survey.

\subsection{Complex O-bearing molecules}

\paragraph{CH$_3$OCH$_3$}  (Fig. 8.7) Dimethyl ether is detected through a large number of transitions in the survey, ranging from $E_l = 80$ K to over 500 K, all in emission.  The spectrum is well modeled by a single warm ($T_\textnormal{rot} = 130$ K) component arising from the hot core.

\paragraph{C$_2$H$_5$OH}  (Fig. 8.8) Ethanol is also detected through numerous transitions in emission arising from the hot core, most prominently through $b$-type transitions of the $trans$ isomer.  It is well modeled with a single $T_\textnormal{rot} = 100$ K component.

\paragraph{Others} Methyl formate (CH$_3$OCHO) is not conclusively detected in this survey, though some transitions appear to be weakly present in emission.  Assuming a $3.7''$ source size (Table 2) and $T_\textnormal{rot} = 130$ K, we derive an upper limit to the column density of $2.0 \times 10^{17}$ cm$^{-2}$.  Formic acid (HCOOH) is also not conclusively detected; assuming the same source size and temperature, we derive an upper limit of $1.6 \times 10^{16}$ cm$^{-2}$.  These upper limits are used in the discussion of molecular abundances to follow (\S 5.1).  These two upper limits are similar to the column densities of these molecules derived from their detection in a 3 mm survey with the IRAM 30 m telescope presented by \cite{belloche09}.

\subsection{NH-bearing molecules}

\paragraph{CH$_3$NH$_2$} (Fig. 8.9) Methylamine has a very rich spectrum due to two large amplitude internal motions (CH$_3$ internal rotation and NH$_2$ inversion), and is detected in this survey through numerous emission transitions.  Unlike several other molecules with transitions with $E_l < 100$ K lying in the HIFI bandwidth (e.g., CH$_3$OH, CH$_2$NH, HNCO), no absorption is seen in the low-energy lines.  The transitions are well modeled by a single component with a rotational temperature of 150 K.

\paragraph{CH$_2$NH} (Fig. 8.10) Low-energy transitions of methanimine are seen in absorption, most clearly the $2_{20}-1_{11}$ line at 623291 MHz and the $3_{21}-2_{12}$ at 693069 MHz.  The $2_{21}-1_{10}$ transition at 617873 MHz is blended with the strong 617 GHz unidentified absorption feature.  Higher energy lines are seen in emission, and are well modeled with a single component with $T_\textnormal{rot} = 150$ K.

\paragraph{NH$_2$} (Fig. 8.11) NH$_2$ is found in both emission and absorption toward Sgr B2(N).  The ground state transitions ($1_{11}-0_{00}$ for \emph{ortho} and $2_{12}-1_{01}$ for \emph{para}), as well as the $2_{02}-1_{11}$ and $3_{13}-2_{02}$ transitions of \emph{ortho}, are found in absorption, while higher energy transitions are seen in emission.  The $1_{11}-0_{00}$ transition is also seen in the line of sight clouds.  The emission is modeled with a single component with $T_\textnormal{rot} = 100$ K.  Due to the strength of these emission lines, however (1--2.5 K), most of the emission must be more extended than the hot core.  We assume a source size of $15''$.  The linewidth is somewhat broader than the typical hot core component (13 km s$^{-1}$), and is centered at 64 km s$^{-1}$.  In Fig. 8.11, the velocity scale of each transition is referenced to the hyperfine component in the multiplet with the strongest line strength.

\paragraph{NH$_3$} (Fig. 8.12) Each of the nine transitions of $^{14}$NH$_3$ and $^{15}$NH$_3$ in the bandwidth of the HIFI survey is seen in absorption in the envelope and in the line of sight clouds.  No emission is seen in any of these transitions, which all have $E_\textnormal{low} < 100$ K.  Therefore, to estimate the abundance of NH$_3$ in the hot core, centimeter inversion transitions must be used.  \cite{huttemeister93} measured a few dozen transitions of $^{14}$NH$_3$, both metastable ($J=K$) and nonmetastable ($J>K$), with the 100 m Effelsberg telescope, while \cite{vogel87} observed six transitions with the Very Large Array.  Ammonia has both emission and absorption components toward Sgr B2, with high optical depths, making the analysis difficult.  Ultimately both studies derived abundances of NH$_3$ using the ($J,K$) = (7,6) inversion transition, for which the optical depth ($\tau \sim 50$) could be well determined using the $^{14}$N quadrupole hyperfine structure.  This was used, assuming $T_\textnormal{rot} = 150-200$ K, to estimate $N$(NH$_3$) $\sim 10^{20}$ cm$^{-2}$.

The recent PRIMOS survey of Sgr B2(N) with the Green Bank Telescope (GBT)\footnote{http://www.cv.nrao.edu/~aremijan/PRIMOS/}  provides better constraint on the abundance of NH$_3$ in the hot core.  The beamwidth of the GBT ($34''$ at 22 GHz) is comparable to that of Herschel in the submillimeter, and the pointing center of the observations is the same within $2''$.  Metastable lines of $^{14}$NH$_3$ are dominated by absorption from the envelope; a number of $^{14}$NH$_3$ lines up to high energy were also observed in absorption with ISO by \cite{ceccarelli02}.  To reduce these complications, we use the $^{15}$NH$_3$ isotopologue to derive the ammonia abundance.  Approximately 25 transitions of $^{15}$NH$_3$ are detected, both in metastable and nonmetastable levels, which were modeled using XCLASS.  In addition to an emission component attributed to the hot core ($v_\textnormal{LSR} \sim 62$ km s$^{-1}$, $\Delta v \sim 5$ km s$^{-1}$), there are additional emissive components, and a warm absorption component with $v_\textnormal{LSR} = 68$ km s$^{-1}$.  We find that the hot core emission component has $T_\textnormal{rot} = 220$ K, and $N$($^{15}$NH$_3$) = $4.3 \times 10^{16}$ cm$^{-1}$, yielding $N$(NH$_3$) = $7.8 \times 10^{18}$ cm$^{-2}$.  This column density is about an order of magnitude lower than that derived by \cite{vogel87}.  However, that derivation also used $\Delta v \sim$ 20 km s$^{-1}$, about a factor of 4 broader than is seen in the (presumably) optically thin $^{15}$NH$_3$ lines. The $^{15}$N isotopologue is more likely to be a good indicator of the true ammonia abundance in the hot core.

\paragraph{N$_2$H$^+$} (Fig. 8.13) N$_2$H$^+$ is detected in emission through four transitions, from $J_\textnormal{up} = 6$ to 9, centered at a velocity of 62 km s$^{-1}$.  The emission is well modeled with a single cold ($T_\textnormal{rot} = 37$ K), extended component.

\paragraph{NH} NH is seen in absorption through its three ground state transitions, in both the Sgr B2(N) envelope and in the line of sight clouds.

\paragraph{N$^+$} The N II fine-structure transition at 1461.1 GHz is detected in absorption.  It is blended with the $J = 17-16$ transition of HC$^{15}$N, but our HCN model suggests that HC$^{15}$N contributes only a small fraction of the total opacity.  The [N II] absorption is detected at a velocity of 0 km s$^{-1}$. As 
\cite{etxaluze13} reported [N II] emission toward the Sgr B2(M) and (N) cores, this lineshape may be affected by emission in the off-position.

\subsection{N- and O-bearing molecules}

\paragraph{NO} (Fig. 8.14) Emission features of nitric oxide are detected in bands 1--4, ranging from $E_l$ = 84--420 K, in both the $\Omega = \frac{1}{2}$ and $\frac{3}{2}$ ladders.  The lines are fit with two components; one cooler ($T_\textnormal{rot} = 50$ K) and broader ($\Delta v = 16$ km s$^{-1}$), and one warmer (180 K) and narrower (8 km s$^{-1}$), both centered near 64 km s$^{-1}$.  The cooler component is attributed to the envelope, while the warmer component is attributed to the hot core.

\paragraph{HNCO} (Figs. 3, 8.15, 11) The abundance and excitation of isocyanic acid in Sgr B2 has been studied  based on its $a$-type transitions in the millimeter \citep{churchwell86}, who found emission from both cold (10 K) and warm (68 K) components.  The spacing between the $K_a = 0$ and $1$ ladders ($\sim A-B$) is 900 GHz, so all the $b$-type transitions between these ladders, in the $P$ ($\Delta J = -1$), $Q$ ($\Delta J = 0$), and $R$ ($\Delta J = +1$) branches, are observed in this survey up to $J = 17$.  Because of the strong continuum in the far-IR, these lines are all observed in absorption, with several velocity components, two cold ($T_\textnormal{rot} < 15$ K, at 64 and 82 km s$^{-1}$) and warmer (at 55 and 73 km s$^{-1}$).  These components will be discussed in more detail in \S 5.2.1.  In addition, high-energy ($J_\textnormal{up} \ge 22$) $a$-type transitions are detected in emission in the HIFI survey with lower-state energies ranging from 300--900 K, revealing the presence of hot HNCO likely arising from the hot core.  The high-energy lines are fit with a single LTE component with a rotational temperature of 280 K.  Because of the large $A$ rotational constant and large $b$-type dipole moment of HNCO, far-infrared pumping is likely very important in the excitation of HNCO in all components.

The $K_a = 1-0$ $Q$-branch of HN$^{13}$CO is also detected.  The change in the $B$ and $C$ rotational constants is very small from HNCO to HN$^{13}$CO ($\delta B/B$ and $\delta C/C \sim 3 \times 10^{-5}$), because the carbon nucleus is located very near the center of mass of the molecule.  Therefore, $a$-type transitions of HN$^{13}$CO fall in all cases within the linewidth of their HN$^{12}$CO counterparts.  The $A$ rotational constant, meanwhile, changes by $\sim 2$ GHz between isotopologues, which shifts the $b$-type transitions in frequency and enables their detection.

\paragraph{NH$_2$CHO}  (Fig. 8.16) Formamide has both allowed $a$-type ($\mu_a = 3.61$ debye) and $b$-type ($\mu_b = 0.85$ debye) transitions \citep{kurland57}.  Transitions of both types are detected in the HIFI survey in emission.  As in several previous millimeter surveys of the Sgr B2(N) region \citep{cummins86, turner91, nummelin00}, we find that simultaneous modeling of the $a$- and $b$-type transitions is challenging; namely, some models to the $a$-type transitions underpredict the intensity of the $b$-type transitions.  The $a$-type transitions found in the HIFI bandwidth have $J_\textnormal{up} \ge 23$ and $E_\textnormal{up} \ge 300$ K, and have line strengths ($S_{ij}\mu^2$) of greater than 200 debye$^2$.  Transitions are detected from levels up to $\sim$800 K.  The detected $b$-type transitions go to lower energy ($E_\textnormal{up} \ge 35$ K), and in general have line strengths a factor of 20 or more lower than the $a$-types because of the lower dipole moment component along the $b$ principal axis.  The $b$-type transitions are detected only up to $E_\textnormal{up} =$ 400 K.

Three possible explanations are posited for one type of transition being overpredicted by a rotation diagram analysis, as also outlined by \cite{turner91}.  First, the spectroscopic catalog intensities could be in error.  In this case, formamide is a simple asymmetric top molecule, and the orientation of the dipole moment has been experimentally measured \citep{kurland57}, so this is highly unlikely.  Next, an anomolous non-LTE excitation mechanism (e.g., radiative pumping) could be responsible.  We also dismiss this for this molecule; $a$- and $b$-type transitions are observed from each $K_a$ ladder up to $K_a = 7$.  The HIFI spectrum, due to its high frequency compared to millimeter surveys, does have some observational bias: as noted above, the $a$-type transitions in the HIFI bandwidth have moderate to high energy, while the most prominent $b$-type transitions have lower energies, so a model with multiple thermal components could be employed where in effect the $a$-type and $b$-type transitions emit from different spatial regions.  However, in the past millimeter surveys, the $a$- and $b$-type transitions covered the same energy space and the same effect was observed, so we do not consider this possibility further.

Therefore, we conclude that the different models for $a$- and $b$-type transitions, with a higher column density inferred from the $b$-type transitions, is due to significant optical depth in the $a$-type transitions.  Accurate catalog frequencies and intensities are also available for the NH$_2$$^{13}$CHO isotopologue, and some low-energy transitions in band 1a are tentatively detected and shown in Fig. 8.16.  These lines provide an upper limit on the opacity of NH$_2$$^{12}$CO $a$-type transitions.  We find that a model with a source size of $2.3''$ (from Table 2) and a rotational temperature of 130 K can reproduce the intensities of the NH$_2$$^{12}$CHO $b$-types and NH$_2$$^{13}$CHO $a$-types reasonably well with a $^{12}$C/$^{13}$C ratio of 20.  This model also predicts the NH$_2$$^{12}$CHO $a$-types reasonably well in most cases.  Because of the mix of optically thin and thick lines, the source size is well constrained by the HIFI data; with a significantly larger source size, any model that predicts the $b$-types to be detected significantly overpredicts the $a$-type transitions of the dominant isotopologue; this is also seen in the rotation diagrams of \cite{nummelin00} and \cite{turner91}, which assume all lines are optically thin.  With a significantly smaller source size (i.e. less than $2''$), not only is this value inconsistent with the SMA maps, the $a$-type transitions become optically thick and the beam dilution factor is too small to account for their intensity.  The $a$-type transitions with lowest upper-state energies detected in the HIFI survey ($E_u = 300-500$ K) have $\tau \sim$ 2--7 in the model.  We have chosen not to undertake a multi-component analysis for this model, as there is not strong enough evidence for multiple thermal components in this data set.  A recent study by \cite{halfen11} of millimeter-wave transitions of formamide in Sgr B2(N) found a cold envelope component with $T_\textnormal{rot} = 26$ K in addition to a hot core component, which could contribute emission to some of the lowest-energy $b$-type transitions detected in the HIFI survey.

\paragraph{HNO}  Three lines are definitively attributed to features of low-energy HNO transitions in absorption from the Sgr B2(N) envelope:  the $1_{11}-0_{00}$ at 592934 MHz, $2_{12}-1_{01}$ at 671259 MHz, and $1_{10}-1_{01}$ at 514595 MHz (though this line is in the middle of a CH$_3$CN band).  Interferometric observations of the $1_{01}-0_{00}$ transition show that HNO is distributed broadly across the Sgr B2 complex \citep{kuan94}.

\subsection{Cyanide molecules}

\paragraph{HCN}  (Fig. 8.17) Hydrogen cyanide and its minor isotopologues (H$^{13}$CN, HC$^{15}$N, and DCN) are detected with both emission and absorption components.  Similar lineshapes are seen in the HIFI survey toward Sgr B2(M), first presented by \cite{rolffs10}.  The emission is broad, with velocity components at 47, 67, and 78 km s$^{-1}$.  Absorption appears centered at 73 km s$^{-1}$ in lower-energy transitions of HCN (up to $J = 10-9$) with a narrow linewidth (6 km s$^{-1}$), and at 55 km s$^{-1}$ in higher-energy transitions, up to $J = 20-19$ in the normal species and up to $J = 14-13$ in H$^{13}$CN and HC$^{15}$N.  The highest energy transitions have only the absorption component.  Both absorption components, because they appear in the high-excitation transitions that are in the HIFI bandwidth, are warm; this is discussed in more detail in \S 5.2.1.  Both the emission and absorption have high optical depth as demonstrated by the fact that corresponding transitions of HCN, H$^{13}$CN, and HC$^{15}$N have nearly the same intensity.

DCN is also detected.  The same model as the other isotopes is used, but the absorption is weak and not very clearly detected (many transitions of DCN also suffer from blending with other molecules).  Emission from HCN and H$^{13}$CN in the $\nu_2 = 1$ vibrational state is also observed.  This emission is centered at 64 km s$^{-1}$, with a linewidth of 15 km s$^{-1}$ (broader than most hot core molecules).  Because of the complex lineshapes in the ground vibrational state, we cannot derive a vibrational temperature with the present analysis.

\paragraph{HNC}  (Fig. 8.18) Hydrogen isocyanide is detected with the same lineshapes as HCN, with emission and warm absorption components.  HN$^{13}$C is also detected, but H$^{15}$NC and DNC are not.

\paragraph{CH$_3$CN}  (Fig. 8.19) Methyl cyanide and its two $^{13}$C isotopologues are detected in emission through HIFI bands 1--3, up to $J_\textnormal{up} = 50$, corresponding to $E_\textnormal{up} \sim 1200$ K.  The normal isotopologue, as in previous studies \citep{nummelin00, belloche09}, is found to have high optical depth, so the $^{13}$C isotopologues are used to derive the CH$_3$CN abundance in the hot core.  The main isotopologue is fit with three velocity components, following \cite{belloche09}, but the $^{13}$C isotopologues can be fit well with a single component centered at 64 km s$^{-1}$, with a rotational temperature of 200 K.  Rotational transitions in the $\nu_8 = 1$ state are also detected.  Observations of lower-frequency transitions in this vibrational state have been presented elsewhere \citep{goldsmith83, nummelin00}, indicative of radiative excitation through the 27 $\upmu$m vibrational band.   A non-LTE analysis of these bands will be presented in a future paper (Vasyunina et al., in preparation).

\paragraph{C$_2$H$_5$CN}  (Fig. 8.20) Ethyl cyanide is detected through numerous emissive transitions.  This molecule has its strongest emission at 64 km s$^{-1}$, with $T_\textnormal{rot} = 150$ K, but also shows a weaker component at 73 km s$^{-1}$, with $T_\textnormal{rot} = 120$ K.  \cite{belloche09} also found a third component at 53 km s$^{-1}$, but this is the weakest of the three in their model and so its nondetection at HIFI frequencies is not surprising given the weak emission from the other two components.  The detected transitions are low-energy ($E_\textnormal{up} \ge 100$ K) $b$-type transitions ($S_\textnormal{ij} \mu^2$ $\sim$ 10--25 debye$^2$) and high-energy ($E_\textnormal{up}$ = 600--1000 K) $a$-type transitions ($S_\textnormal{ij} \mu^2 \sim$ 800--900 debye$^2$).  The strongest of each set of transitions have comparable optical depth in the LTE model ($\tau \sim$ 0.3--1 in the 64 km s$^{-1}$ component).  The three $^{13}$C isotopologues are not detected.

\paragraph{C$_2$H$_3$CN}  (Fig. 8.21) Vinyl cyanide is weak in the HIFI spectrum, but is detected and modeled with a single emission component centered at 64 km s$^{-1}$, with a rotational temperature of 150 K.  Its rotational temperature is assumed to be 150 K, the best-fit temperature for ethyl cyanide.  Vinyl cyanide has similar rotational constants and a similar dipole moment to ethyl cyanide, except the $b$-type transitions are weaker relative to the $a$-types (($\mu_a/\mu_b$)$^2$ = 18.2 for vinyl cyanide, compared to ($\mu_a/\mu_b$)$^2$ = 9.8 for ethyl cyanide).  Due to this and the lower abundance of C$_2$H$_3$CN as compared to C$_2$H$_5$CN, the $b$-type transitions are too weak to be detected for this species.

\paragraph{CN}  (Fig. 8.22) The cyano radical is detected in emission at a velocity of 64 km s$^{-1}$.  This species is best modeled with a single cold ($T_\textnormal{rot} = 40$ K), extended component, indicating that its primary emission is from the envelope, rather than the hot core.

\paragraph{Others} HC$_3$N, which was detected in the HIFI surveys of Orion KL \citep{crockett13} and NGC 6334(I) \citep{zernickel12}, is not conclusively detected in this survey.  An upper limit to its column density is set, assuming $T_\textnormal{rot} = 150$ K; interestingly, this limit ($N_\textnormal{tot} \le 1.0 \times 10^{16}$ cm$^{-2}$) is significantly lower than that derived in the millimeter wave study of \cite{nummelin00}.  That study found a HC$_3$N column density of $8.5 \times 10^{16}$ cm$^{-2}$, with a temperature of 121 K and a source size of $3.8''$.  Applying the model of \cite{nummelin00}, with the dust opacity included as described above, causes the HIFI HC$_3$N lines to be overpredicted.  The HIFI survey is biased towards the hottest gas for this molecule, since the lowest-energy transition in the HIFI bandwidth has $E_\textnormal{up} = 601$ K.  This suggests that a multi-component model is needed to explain the emission of HC$_3$N across all energy levels.

\subsection{S-bearing molecules}

\paragraph{H$_2$S}  (Fig. 8.23) Three isotopologues of hydrogen sulfide, H$_2$$^{32}$S, H$_2$$^{34}$S, and H$_2$$^{33}$S, are detected with both emission and absorption components toward Sgr B2(N).  A ground state transition of \emph{ortho}-H$_2$S near 736 GHz ($2_{12}-1_{01}$, Fig. 4) is seen in absorption in all three isotopologues at the Sgr B2(N) velocity, along with a second peak near 82 km s$^{-1}$.  Higher energy transitions are a mix of absorption and emission features; in general, higher frequency transitions are seen in absorption, while lower frequency transitions are seen in emission.  For example, in the supplemental figure available online, the $2_{21}-2_{12}$ transition near 505 GHz ($E_\textnormal{low}$ = 55.1 K) can be seen in emission in all three isotopologues, while the $3_{30}-2_{21}$ at 1865 GHz ($E_\textnormal{low}$ = 79.4 K) is seen in absorption.  The emission features are centered at 64 km s$^{-1}$, while the absorption features at higher energies are centered near 53 km s$^{-1}$.  

The source of the emission likely has a significant contribution from the hot core, from which emission features in bands 6 and 7 are heavily obscured by dust; the stronger dust continuum at these frequencies can also make absorption features from foreground gas easier to detect.  A full analysis of the abundance of H$_2$S in the various spatial/velocity components will require non-LTE modeling.  To derive an estimate for the H$_2$S abundance in the hot core, we use the H$_2$$^{33}$S model (for which transitions may still not be optically thin) and use a source size of $2.8''$ from Table 2.  The high rotational temperature (130 K) for this component indicates that the emission likely has a predominantly hot core origin.  For H$_2$$^{32}$S and H$_2$$^{34}$S, a larger source size is required, as noted in \S 3.1.

\paragraph{SO}  (Fig. 8.24) SO and $^{34}$SO are detected in emission.  The lineshapes of the weaker $^{34}$SO lines are fit well with a single component centered at 64 km s$^{-1}$, with a rotational temperature of 150 K.  This component is likely optically thick in $^{32}$SO.  The model to the $^{32}$SO lines also contains three additional optically thin components, one narrow and centered at 67 km s$^{-1}$, and two broader components.  The SO abundance in the hot core is calculated from the $^{34}$SO model.  The $^{33}$SO isotopologue is not detected.

\paragraph{SO$_2$} (Fig. 8.25) Sulfur dioxide is detected in emission and fit with the same model as SO, except the broadest component ($\Delta v = 30$ km s$^{-1}$) is not seen.  $^{34}$SO$_2$ is only tenatively detected and is not included in the model.

\paragraph{CS} (Fig. 8.26) The normal isotopologue of CS is observed with both emission and absorption components, while the minor isotopologues are seen only in emission.  The emission is fit with two velocity components, one with the velocity width of most hot core species ($\Delta v = 10$ km s$^{-1}$), and the second broader with $\Delta v = 35$ km s$^{-1}$.  The broader component is less clear in the minor isotopologues.  Absorption is only seen in the main isotopologue of CS, which is centered at 55 km s$^{-1}$.  CS likely has high optical depth in the hot core, and more extended emission based on the high line intensities, so  the $^{13}$CS species, which is more likely to be optically thin, is used to derive the abundance in the hot core.  The hot core component, centered at 64 km s$^{-1}$, has a rotational temperature of 120 K in the model.

\paragraph{H$_2$CS}  (Fig. 8.27) Thioformaldehyde is detected through emission features in bands 1 and 2, and is well modeled by a single warm ($T_\textnormal{rot} = 120$ K) velocity component centered at 64 km s$^{-1}$.

\paragraph{OCS} (Fig. 8.28) Transitions of OCS ranging from $J_\textnormal{up} = 41$ to 57 ($E_\textnormal{up}$ = 500--1000 K) are seen in this survey in bands 1a, 1b, and 2a, with emission well modeled by a single component at 64 km s$^{-1}$ and with a rotational temperature of 200 K.  OC$^{34}$S is very tenatively detected, but not included in the model, because most transitions of OC$^{34}$S are blended.

\paragraph{NS} (Fig. 8.29) Emission transtions of NS in the lower-energy ($\Omega = \frac{1}{2}$) ladder are detected.  The lines are modeled with $T_\textnormal{rot} = 90$ K, and two velocity components centered at 64 and 73 km s$^{-1}$.

\paragraph{Others} SH$^+$ is detected in absorption through ground state transitions at 526 and 683 GHz; higher-excitation lines are not detected.  The model is based on that of \cite{godard12}, though in that study only the 526 GHz multiplet was modeled.  It was necessary to change the model somewhat to adequately fit the 683 GHz line as well.  SiS is not detected.

\subsection{Carbon and hydrocarbons}

\paragraph{CCH} (Fig. 8.30) Ethynyl radical is detected through four doublets from $N_\textnormal{up}$ = 6 to 9.  The lines are observed in emission, and are well modeled with a single cold ($T_\textnormal{rot} = 50$ K), extended component centered at 64 km s$^{-1}$.

\paragraph{Others} Neutral and singly ionized atomic carbon have fine-structure transitions in the HIFI bandwidth, at 492 and 809 GHz for C I and at 1901 GHz for C II.  The neutral carbon transitions have emission at the Sgr B2(N) envelope velocity and absorption in the spiral arms, while the [C II] line has little absorption at 64 km s$^{-1}$, and its strongest absorption near 0 km s$^{-1}$.  For both, emission in the off-position may be significant, so the line profiles may be unreliable.

CH has two ground state transitions at 532.7 and 536.6 GHz in its $\Omega = \frac{3}{2}$ state ($N = 1, J = \frac{3}{2}-\frac{1}{2}$), detected in absorption both in the Sgr B2(N) envelope and in the line of sight clouds.  An analysis of these features toward Sgr B2(M) was presented by \cite{qin10}.  The ground state transitions of CH$^+$ and $^{13}$CH$^+$ are detected, also in the envelope and line of sight.  Higher-excitation lines are not detected for both CH and CH$^+$.  The model of \cite{godard12} is used for this species.  The linear carbon trimer, C$_3$, which is nonpolar and therefore lacks a pure rotational spectrum, has its $\nu_2$ vibrational band origin at 1.89 THz.  Transitions in the $Q$ and $P$ branch fall in the HIFI bandwidth and are detected in absorption at the velocity of the Sgr B2(N) envelope.  This band has previously been detected with ISO by \cite{cernicharo00}.

\subsection{Halogen molecules}

Three halogen-containing molecules are detected in absorption in the HIFI survey: HF, HCl, and H$_2$Cl$^+$.  The ground state transition of HF at 1232 GHz is detected in absorption in the Sgr B2(N) envelope (where it is saturated) and in the spiral arms.  H$^{35}$Cl and H$^{37}$Cl are detected in absorption in the envelope in the $J = 1-0$ and $2-1$ transitions.  H$_2$Cl$^+$ and H$_2$$^{37}$Cl$^+$ are detected through their ground state transitions only; absorption is seen at 2 and 16 km s$^{-1}$, but not at the Sgr B2(N) velocity.

\section{Discussion}

\subsection{Molecular abundances in the hot core}

In Figure 9 we show the abundances of all species detected in the hot core with respect to H$_2$, using an H$_2$ column density of $8.0 \times 10^{24}$ cm$^{-2}$ as described above in \S 3.1.  These abundances are compared in this figure to those in the two regions of compact molecular emission in Orion KL, the Hot Core and Compact Ridge, as derived by the HIFI survey of that source by \cite{crockett13}.  These two sources are the richest in the sky for investigations of the chemistry associated with high-mass star formation, so an examination of the abundance similarities and differences is useful.  The complex spectra observed towards both sources are indicative of warm gas where it is believed that ices have evaporated, releasing their chemical content into the gas phase.  Additionally, the surveys presented in \cite{crockett13} and here were obtained, reduced, and analyzed with similar methods, so observational and modeling effects are minimized.  Most species found in the Hot Core region of Orion KL have rotational temperatures similar to those seen of hot-core species in Sgr B2(N) (150--300 K), while the Compact Ridge is associated with cooler temperatures (80--125 K). 

The molecules associated with each family are presented in separate panels.  As outlined in \S 3.1, the molecules within a single panel were modeled with a uniform source size (Table 2); however, the source size differs between panels.  Figure 10 shows the same abundances, but given relative to a molecule in each family detected in the HIFI survey.  The results in this figure are therefore independent of the H$_2$ column density,\footnote{The abundances of the complex oxygen-bearing organics are given with respect to methanol, which was fit with a different source size.  For the purposes of this figure, the column densities are scaled to a uniform source size, using $N_{\textnormal{tot},2} = N_{\textnormal{tot},1} \left(\frac{\theta_1}{\theta_2}\right)^2$.  Here, ($N_{\textnormal{tot},1}, \theta_1$) and ($N_{\textnormal{tot},2}, \theta_2$) are two independent models to the HIFI data.  Because the source size for the hot core is much smaller than the HIFI beam at all frequencies, these two models are equivalent on optically thin lines.} and so may be more reliable for comparing Sgr B2(N) to Orion KL.  In the following paragraphs, we discuss the similarities and differences between the molecular abundances in Sgr B2(N) and Orion KL surveys for each molecule group.

\begin{figure}
\figurenum{9}
\centering
\includegraphics[width=6.0in]{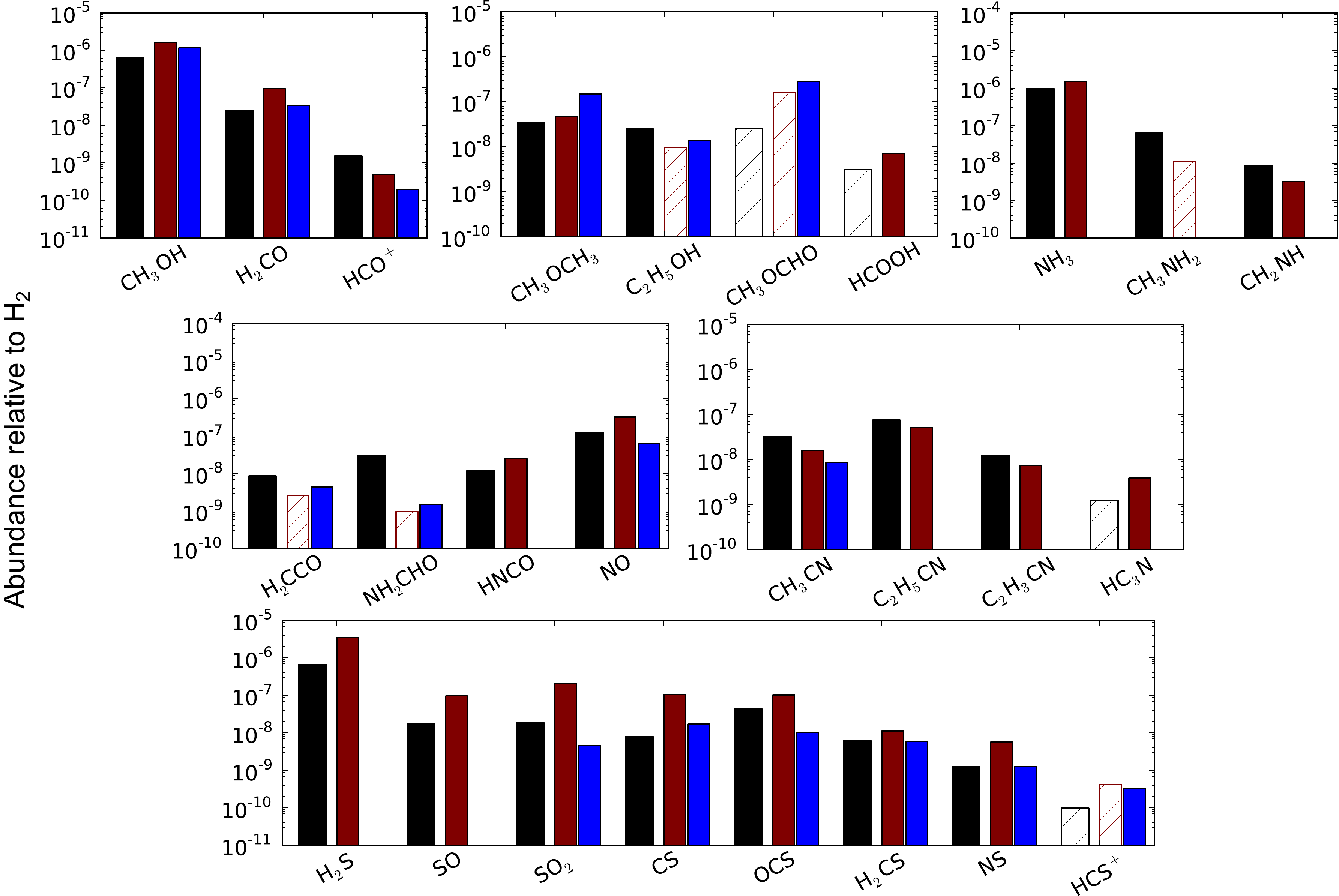}
\caption{Abundances of molecules in the Sgr B2(N) hot core with respect to H$_2$ (black bars), compared to abundances in the Orion KL Hot Core (maroon) and Compact Ridge (blue), from \cite{crockett13}.  Bars in solid colors are detected species, while those shown with a hatch pattern are upper limits.}
\end{figure}

\begin{figure}
\figurenum{10}
\centering
\includegraphics[width=5.0in]{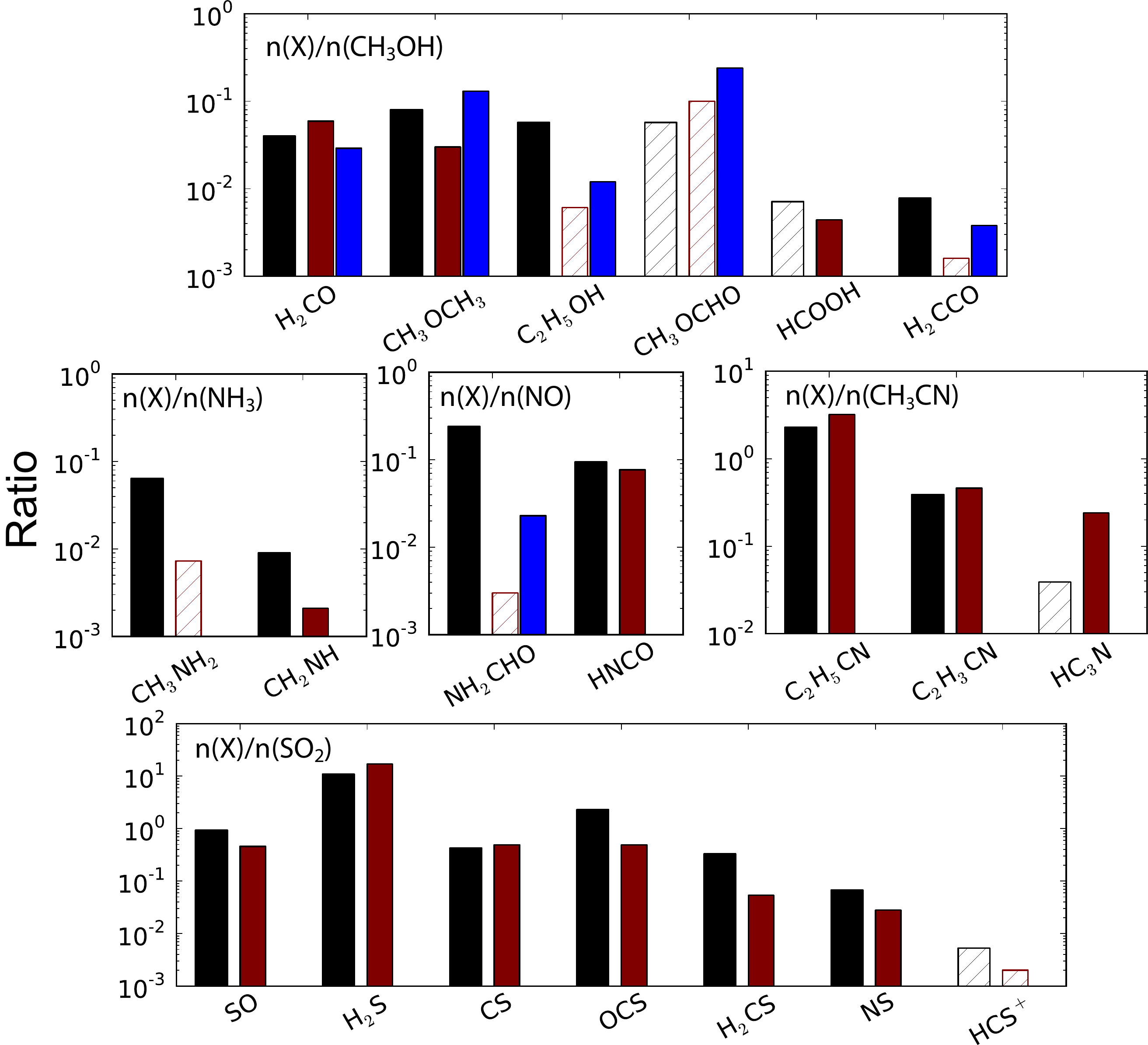}
\caption{The same values as Figure 9, except given as ratios between molecules detected in the HIFI survey, rather than with respect to H$_2$.}
\end{figure}

\paragraph{O-bearing species}  The abundances of formaldehyde and methanol are similar within the two sources.  These molecules are generally believed to have a common grain surface origin from the hydrogenation of CO on grain surfaces (see, e.g., \cite{fuchs09}).   Complex oxygen-bearing organics are known to have their highest abundances within the Orion KL region toward the Compact Ridge \citep{blake87, crockett13}.  Dimethyl ether (CH$_3$OCH$_3$) and methyl formate (CH$_3$OCHO) have lower abundances in the Sgr B2(N) hot core than in the Orion Compact Ridge. Methyl formate is in fact not detected in the HIFI survey, though it has been detected in lower-frequency millimeter surveys of this region \citep{turner91, nummelin00, belloche09}.  These species likely derive from methanol, and plausible formation processes have been studied that occur both on grain surfaces \citep{garrod06, garrod08, oberg09} and in the gas phase \citep{bouchoux97, laas11, neill11}.

Ethanol (C$_2$H$_5$OH) and ketene (H$_2$CCO) are both more abundant in Sgr B2(N) than either compact region in Orion KL.  These two molecules have the same (C-C-O) heavy atom backbone, and in the hot core model of \cite{garrod08} are formed through grain surface chemistry:  ethanol through the reaction of CH$_3$ and CH$_2$OH, and ketene from both C$_2$ and C$_2$O.  In principle, ethanol could also be formed by the further hydrogenation of ketene.  Vinyl alcohol (CH$_2$CHOH), which would be an intermediate between the two species if this were the case, has been detected in Sgr B2(N) \citep{turner01}, but not in the HIFI survey.

\paragraph{NH-bearing species}  Ammonia has a similar abundance in the Orion KL and Sgr B2(N) hot cores, but CH$_3$NH$_2$ (methylamine) and CH$_2$NH (methanamine) both have substantially higher abundances in Sgr B2(N).  These two molecules were discussed in more detail in conjunction with millimeter wave observations by \cite{halfen13}.   CH$_2$NH in the envelope of Sgr B2(N), seen most clearly through low-energy transitions in absorption, may have a different chemical origin than the warm (150 K) component in the hot core.  CH$_2$NH and CH$_3$NH$_2$ could both form by sequential hydrogenation of HCN in ices, a process that has also been investigated in the laboratory by \cite{theule11}.  They found that both HCN and CH$_2$NH were both readily hydrogenated to form CH$_3$NH$_2$; they inferred that CH$_2$NH was a short-lived intermediate in the synthesis of CH$_3$NH$_2$.  However, it should be noted that the H-atoms in this experiment were warm ($E_\textnormal{kin} \sim 300$ K) in order to give enough energy for reaction barriers to be surpassed.  At lower temperatures, these reactions may have lower efficiency, which could allow CH$_2$NH to have a longer lifetime than in these laboratory experiments.

These molecules were proposed to instead have a gas-phase origin by \cite{halfen13}, involving the reaction of NH$_3$ with CH (to form CH$_2$NH) and CH$_3$ (to form CH$_3$NH$_2$).  The reaction of CH and NH$_3$ has been found in the laboratory to have a reasonable rate coefficient at 100 K \citep{bocherel96}; while the product species was not measured in that study, CH$_2$NH was postulated as the most likely result of this reaction.  If the radical species CH and CH$_3$ have higher abundances toward Sgr B2(N), then the NH-bearing species could have higher abundances by this chemistry.

\paragraph{N- and O-bearing species} NO and HNCO have similar abundances in Orion KL and Sgr B2(N).  However, by far the most notable difference in chemical abundances between the two sources is in formamide (NH$_2$CHO), which has a higher abundance in Sgr B2(N) by more than an order of magnitude, implied as discussed above by the detection of the low-$S_{ij}\mu^2$ $b$-type transitions (which are not detected in the Orion KL fullband survey).  To date there are two other star-forming regions toward which formamide has been detected.  \cite{zernickel12} report only 8 very weak lines in the HIFI survey of NGC6334I, and a fractional abundance of $1.6 \times 10^{-9}$, suggesting greater similarity with Orion KL.  In the solar-mass protostar IRAS 16293-2422, formamide was recently detected with an abundance $\sim 10^{-10}$, also more consistent with Orion KL \citep{kahane13}.  The latter source, being a low-mass protostar, may have some significant chemical differences, however.  The suggestion based on the available data is therefore that formamide is unusually abundant in Sgr B2(N).

Formamide, like most complex molecules, has multiple possible formation routes.  In the model of \cite{garrod08}, it forms from both the hydrogenation of OCN on grain surfaces at low temperature, and then in the hot core phase by the following gas-phase reaction:

\begin{equation}
\textnormal{NH}_2 + \textnormal{H}_2\textnormal{CO} \rightarrow \textnormal{NH}_2\textnormal{CHO} + \textnormal{H}
\end{equation}

\noindent This reaction is exothermic and was assumed to proceed without a barrier.  A greater abundance is formed by this gas-phase mechanism--ultimately reaching a maximum abundance of $\sim 10^{-7}$, a factor of a few greater than detected toward Sgr B2(N).  NH$_2$ and H$_2$CO do both have high abundances in the Sgr B2(N) hot core; however, both molecules also have similar abundances in the Orion KL hot core.  Additionally, \cite{kahane13} point out that the most likely destruction route for gas-phase formamide is through protonation by HCO$^+$, which also has similar abundances in the two sources (Figure 9).  So it is not clear that this chemistry would lead to such a dramatic difference in the formamide abundance between the two sources.  However, if formamide is produced in the gas phase, the time elapsed since grain mantle evaporation, as well as the physical conditions in the hot core phase, would strongly affect the abundance.

\paragraph{CN-bearing species} Methyl cyanide (CH$_3$CN), ethyl cyanide (C$_2$H$_5$CN), and vinyl cyanide (C$_2$H$_3$CN) have similar abundances in Sgr B2(N) and the Orion KL Hot Core.  Cyanoacetylene (HC$_3$N) is not detected in Sgr B2(N), with an upper limit a factor of 3 below the detected abundance in Orion (assuming $T_\textnormal{rot} = 150$ K); however, as noted in \S 4.5, the $J$ levels probed by HIFI are much higher in energy than those seen in millimeter wave observations.  Since HC$_3$N has been detected in a wide range of interstellar environments, including with an extended distribution over the Sgr B2 envelope \citep{morris76}, the favored pathways involve gas-phase chemistry, which could be either ion-molecule or neutral-neutral \citep{knight86, woon97, tagaki99}.  Hydrogen cyanide (HCN) transitions have very complex lineshapes in this survey, and multiple components with high excitation temperatures as will be discussed in \S 5.2.1, so deriving an abundance in the hot core was not possible.

\paragraph{S-bearing species}  Sulfur chemistry is generally less well understood than for the other molecule classes discussed above, and in dense clouds a large fraction of the sulfur may be depleted onto grains or in another unknown form, but the degree to which this is the case is disputed \citep{ruffle99, tieftrunk94, wakelam11, goicoechea06}.  A comparison of the abundances relative to SO$_2$ (Figure 10) shows close agreement in the SO/SO$_2$, H$_2$S/SO$_2$, and CS/SO$_2$ ratios between the two sources, while OCS, H$_2$CS, and NS are more abundant toward Sgr B2(N) relative to SO$_2$.  The abundance ratios of S-bearing molecules have been proposed as ``chemical clocks'' \citep{vandertak03, wakelam11}.   Recent chemical modeling was presented by \cite{wakelam11} to interpret the abundances of S-bearing species toward massive YSOs.  In their model, the sulfur is initially divided in the ice mantles evenly between OCS and H$_2$S.  After evaporation, the early-time chemistry (up to a few times $10^4$ years) in the layer with $T = 100$ K is characterized by OCS/SO$_2$, H$_2$S/SO$_2$, and SO/SO$_2$ ratios exceeding unity.  (They note that CS is underpredicted in their modeling compared to their observations by a few orders of magnitude.)   With time, SO$_2$ becomes the dominant sulfur carrier.

In both Orion KL and Sgr B2(N), most of the detected sulfur is found in H$_2$S; this value is better constrained for Orion KL, because of the envelope contribution in Sgr B2(N), even in warm lines; on the other hand, the H$_2$S abundance for Orion KL comes from a detailed analysis to all three sulfur isotopes \citep{crockett13a}.  SO, SO$_2$, CS, OCS, and H$_2$CS are all significant carriers of the measured sulfur not in H$_2$S, suggesting a relatively young chemistry.  H$_2$CS was not included in the modeling of \cite{wakelam11}, but can form from either OCS or H$_2$S through reactions with CH$_3$$^+$ to make the H$_3$CS$^+$ ion \citep{smith78}, followed by dissociative recombination.

\subsection{Kinematics and chemistry of the Sgr B2 envelope}

\subsubsection{Kinematic components}

In this section, we discuss the velocity components of the absorption lines detected in the HIFI survey, focusing on the Sgr B2(N) envelope itself ($v_\textnormal{LSR}$ = 50--90 km s$^{-1}$).  We leave the analysis of features in the line of sight clouds detected toward Sgr B2(N) to a future paper.  As noted in \S 3, only LTE modeling was pursued for this paper.  Due to the lower densities in the envelope \citep[$\sim 10^4-10^5$ cm$^{-3}$,][]{lis91, cernicharo06} than the hot core, LTE is likely a poor assumption and so the temperature and column densities derived for these components in this study are likely inaccurate.  A more comprehensive understanding of the physical conditions and chemical abundances in the envelope will require more sophisticated modeling.  In this section, the rotational temperatures derived for different velocity components in the LTE model are discussed.  These should be interpreted as a proxy for the energy levels over which a particular component is detected, and can indicate the relative temperature or density of various spectral components.

Figure 11 shows five transitions of the HNCO $K_a = 1-0$ $Q$-branch near 900 GHz, ranging from $J$ = 2 to 17 ($E_\textnormal{low}$ = 3--161 K).  The fullband model fit to the HNCO spectrum is overlaid in red.  HNCO provides a unique perspective on the envelope kinematics; because of the availability of transitions varying widely in excitation energy within a narrow frequency space, the Herschel beamwidth and the continuum level is uniform.  There are also $P$-branch ($\Delta J = -1$) and $R$-branch ($\Delta J = +1$) transitions in the HIFI survey, which are also used to constrain the fullband LTE model.

\begin{figure}
\figurenum{11}
\centering
\includegraphics[width=2.5in]{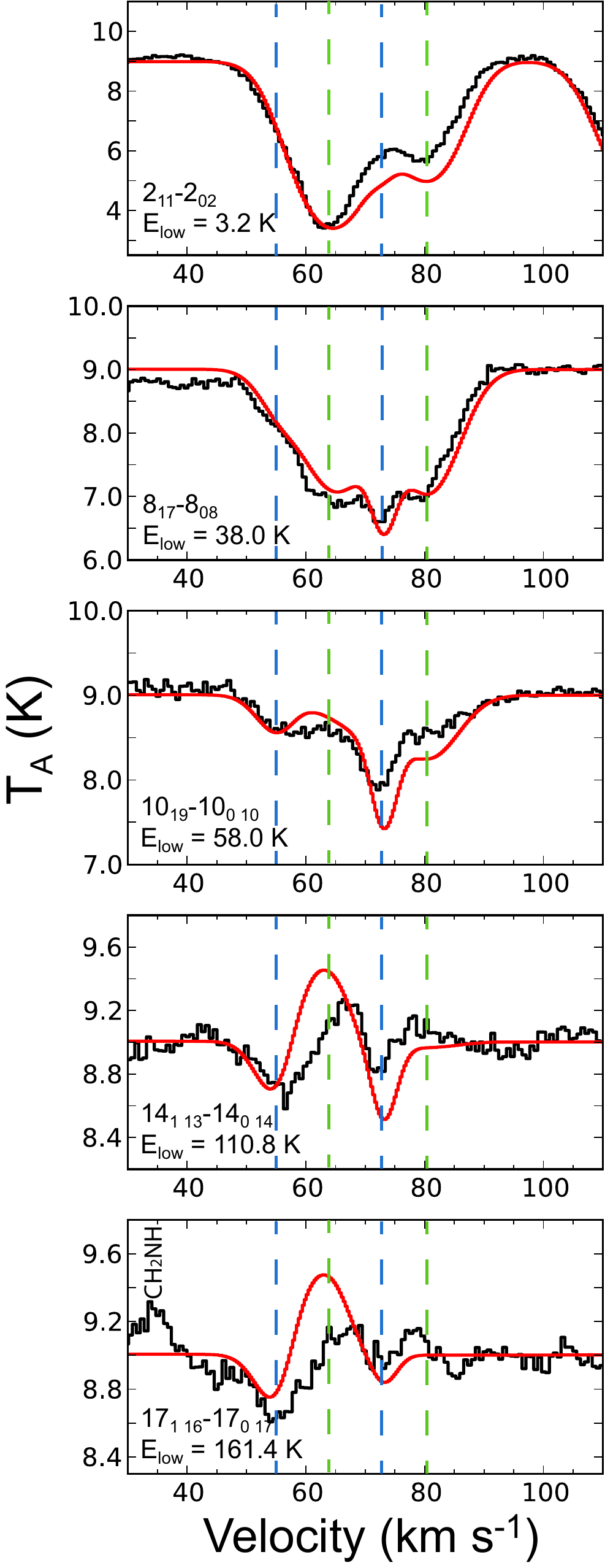}
\caption{Selected transitions of the HNCO $Q$-branch showing velocity components in the Sgr B2(N) envelope.  Each panel is centered on a single transition, increasing in energy from top to bottom.  The red curve is the LTE model to the HNCO line profiles.  The vertical lines indicate the velocities of the absorption components:  the cold components (at 64 and 80 km s$^{-1}$) are in green, and the warm components (at 55 and 73 km s$^{-1}$) are in blue.}
\end{figure}

There are five velocity components in the full model to the HNCO spectrum.  One is an emission component with $T_\textnormal{rot} = 280$ K, as seen most clearly through the $a$-type transitions in Bands 1 and 2 (see Figure 8.15); this component is attributed to the hot core.  This component contributes little flux to the lowest-$J$ transitions because of their low statistical weights, but is calculated to be optically thick (with $T_\textnormal{A} \sim 0.5$ K) in the higher energy transitions in Figure 11.  The other four components are absorption arising from the Sgr B2 envelope.  The lowest energy transitions, represented by the $J = 2$ line in Figure 11, are dominated by two velocity components, centered at 64 and 81 km s$^{-1}$.  These components are fit with rotational temperatures of 13 and 12 K, respectively.  This is consistent with the finding of \cite{churchwell86}, who found a rotational temperature of 10 K for transitions with $E_\textnormal{low} \le 40$ K in the millimeter ($K_a = 0$ $a$-type transitions).  The $J = 8$ and $J = 10$ transitions of the HNCO $Q$-branch prominently show a narrower ($\Delta v = 5$ km s$^{-1}$) component centered at a velocity of 73 km s$^{-1}$.  This component has a rotational temperature of 40 K.  Finally, at higher energies still, absorption becomes visible at velocities blueshifted relative to the predominant Sgr B2(N) envelope velocity, centered at $v_\textnormal{LSR} = 55$ km s$^{-1}$, with a width of 8 km s$^{-1}$.  This component has a fit rotational temperature of 150 K.  The $J=14$ and $J=17$ transitions in Figure 11 show these two components most clearly, with emission from the hot core at 64 km s$^{-1}$ and absorption on each side.  At this energy, the envelope absorption seen in the low-$J$ transitions no longer contributes to the lineshape.

This general pattern is seen in other molecules with absorption in the Sgr B2(N) envelope:  absorption at the primary envelope velocity (64 km s$^{-1}$) tends to have low excitation temperatures, while absorption in higher-energy lines is either red-shifted or blue-shifted from this velocity.  In Figure 12, we show transitions from 12 molecules with absorption components in this survey.  In the left column, lines with very low excitation energy are shown, where absorption is generally seen with two velocity components matching those for low-energy HNCO.  The middle column shows the three molecules that show absorption in a component red-shifted from the envelope velocity, while the right column shows the five molecules with absorption in a blue-shifted component.  All of the lines that show this absorption are  higher in energy, and so require elevated rotational temperatures (see Table 8).

Absorption components with high rotational temperatures (many 100 K or greater) imply highly excited gas that must be in front of a very bright continuum source, presumably the dust continuum of the Sgr B2(N) hot cores.  This suggests that these lines are tracing the gas directly in front of the hot core, rather than the extended Sgr B2 envelope; over wider spatial scales the continuum is weaker.  Notably, there are two molecules with high-excitation lines at the Sgr B2(N) envelope velocity that have unique molecular excitation due to their symmetry.  As noted in \S 4, H$_3$O$^+$ is detected in energy levels up to 1200 K in bands 6 and 7 of this survey.  All of these transitions originate from metastable levels, which cannot radiate to lower energy levels except through weakly allowed $\Delta K = 3$ transitions.  Therefore, the individual excitation temperature $T_\textnormal{ex}$ of each transition is likely very low ($< 10$ K).  Likewise, the nonpolar carbon trimer, C$_3$, is detected in this spectrum through the $\nu_2$ vibrational band up to fairly high energies ($J = 14$, or $E_\textnormal{low} = 130$ K), and rotational temperatures of 32 and 36 K are fit to the two velocity components at 65 and 76 km s$^{-1}$, respectively.  Because this molecule is nonpolar, pure rotational radiative transitions within the ground vibrational ladder are very weak, so population can also be maintained in high $J$ levels at lower densities.  For each of the high-excitation transitions shown in the center and right columns in Figure 12, on the other hand, the lower levels are able to radiatively relax to even lower energy levels, so these transitions have high critical densities.

The redshifted and blueshifted absorption components in the gas in front of the Sgr B2(N) hot core suggest infalling and expanding gas along the line of sight, respectively.  The blueshifted ($v_\textnormal{LSR} \sim 55$ km s$^{-1}$) components are characterized by very high excitation temperatures, observed in transitions with lower-state energies of 400 K or greater, while the 73 km s$^{-1}$ component is observed only in energy levels up to $\sim 200$ K.  In the two molecules that show absorption in both the redshifted and blueshifted components, HNCO and HCN, the blueshifted component is detected at higher energies.  This suggests that the expanding gas, associated with the blueshifted components, is warmer than the infalling gas associated with the redshifted velocities, and therefore likely located closer to the protostar(s) in the Sgr B2(N) hot cores. \cite{rolffs10}, analyzing HCN transitions from $J_\textnormal{up}$ = 6--13 in the nearby Sgr B2(M) source, found the best agreement with the lineshapes with a physical model with infall at large radii that reverses to outflow at smaller radii.  Figure 12 suggests a similar scenario for Sgr B2(N), though, as noted above, a more sophisticated model will be needed to fully interpret these observations.

\begin{figure}
\figurenum{12}
\centering
\includegraphics[width=6.0in]{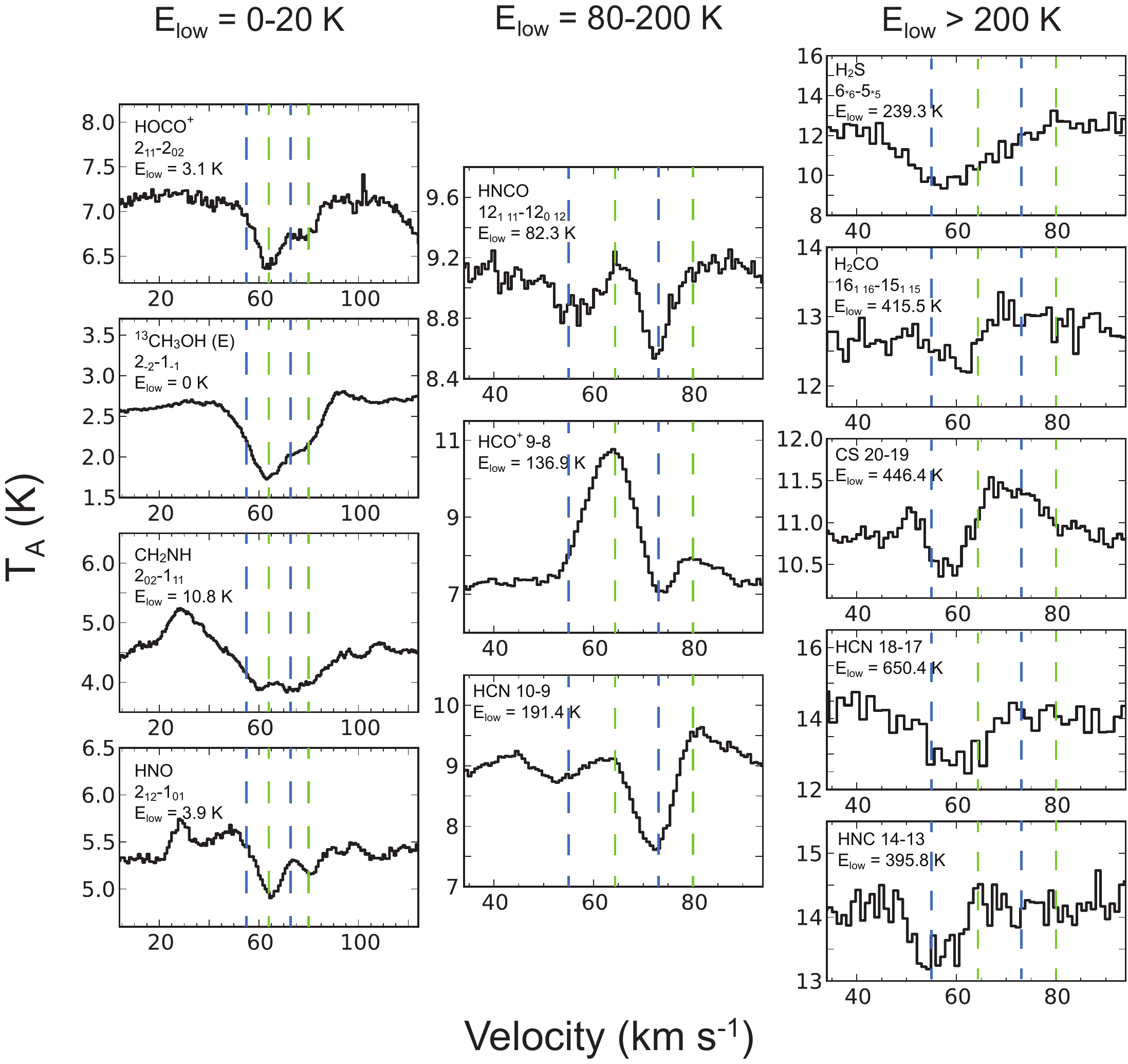}
\caption{Single transitions of molecules with absorption detected in the Sgr B2(N) envelope.  The left column shows molecules with absorption in low-energy transitions ($T_\textnormal{rot} \sim 10$ K).  The middle panel indicates molecules with absorption detected at higher redshift than the most prominent Sgr B2(N) envelope velocity (64 km s$^{-1}$) in higher energy levels (suggesting infall).  The right panel shows molecules with absorption detected at lower redshift than the most prominent Sgr B2(N) envelope velocity in high-energy transitions (suggesting outflow).  The green and blue lines are as in Figure 11.}
\end{figure}

\subsubsection{HOCO$^+$ and the gas-phase CO$_2$ abundance}

Carbon dioxide (CO$_2$) is known from surveys with the Infrared Space Observatory (ISO) and the Spitzer Space Telescope to be an important constituent of interstellar ices, and so likely participates heavily in the chemistry in ices and potentially in the gas phase as well at warmer temperatures.  Observations of a number of infrared-bright regions investigated CO$_2$ in both the solid and gas phases through its two infrared-active vibrational modes at 4.3 and 15.0 $\upmu$m. The different wavelengths and band profiles allow for gas and solid CO$_2$ to be distinguished, particularly in the 15 $\upmu$m band \citep{degraauw96, vandishoeck96, boonman03, sonnentrucker06}.  As CO$_2$ lacks a permanent dipole moment, it is not emissive at mm/submm wavelengths.  In the solid phase, the typical abundance is 10-20\% that of solid H$_2$O, and a fractional abundance of $\sim 10^{-6}$ relative to total H$_2$ \citep{gerakines99, gibb04}.  Modeling of the band profile has suggested that CO$_2$ is found in two phases, one polar (H$_2$O-dominated) and the other nonpolar (CO/CO$_2$ dominated).  In the gas, on the other hand, the typical detected abundance is $(1-3) \times 10^{-7}$ \citep{vandishoeck96, boonman03, sonnentrucker06}.  Whereas most of the CO is typically found in the gas phase due to its low evaporation temperature, and H$_2$O has similar abundances in the gas and solid phases, CO$_2$ is depleted in the gas phase relative to its ice abundance  \citep{vandishoeck96}.  However, CO$_2$ can only be sought toward sources with a bright mid-IR continuum, so toward the Sgr B2 envelope, where the envelope is too optically thick at these wavelengths, CO$_2$ has not been detected directly.

Protonated carbon dioxide, HOCO$^+$, was first proposed by \cite{herbst77} as an indirect tracer of gas-phase CO$_2$.  To date HOCO$^+$ has been sought toward a number of sources, but has only been detected toward the Galactic Center, several starless translucent clouds \citep{turner99}, and a single low-mass Class 0 protostar \citep{sakai08}.  Toward the Galactic Center, HOCO$^+$ is found with an extended distribution, including toward Sgr B2 and Sgr A \citep{thaddeus81, minh88, minh91, deguchi06}.  A map of the $4_{04}-3_{03}$ transition revealed maximum emission at a position about an arcminute north of Sgr B2(N) \citep{minh88}.  Through a comparison of the abundances of HOCO$^+$ with that of protonated carbon monoxide, HCO$^+$, the abundance of gas-phase CO$_2$ relative to that of CO can be constrained \citep{herbst77, minh88}.  The likely formation and destruction processes of HOCO$^+$ and HCO$^+$ are summarized in Table 9, along with rates taken from the UMIST database\footnote{http://www.udfa.net/}.  It is possible that HOCO$^+$ could also form through the endothermic reaction of HCO$^+$ with O$_2$, and this reaction could be significant in shocks \citep{pineaudesforets89}, in which case HOCO$^+$ does not trace the abundance of CO$_2$.  However, since HOCO$^+$ is rotationally cold and extended over the Galactic Center region, we consider it more likely that HOCO$^+$ is produced from CO$_2$.  By invoking the steady state approximation, the following expressions can be derived for the abundances of HOCO$^+$ and HCO$^+$, as well as an expression for the CO$_2$/CO abundance ratio:

\begin{equation}
[\textnormal{HCO}^+] = \frac{k_1 [\textnormal{H}_3^+][\textnormal{CO}]}{k_4 [e^-] + k_5 [\textnormal{H}_2\textnormal{O}]}
\end{equation}

\begin{equation}
[\textnormal{HOCO}^+] = \frac{k_2 [\textnormal{H}_3^+][\textnormal{CO}_2]}{k_3 [\textnormal{CO}] + k_6 [e^-]+ k_7 [\textnormal{H}_2\textnormal{O}]}
\end{equation}

\begin{equation}
\frac{[\textnormal{CO}_2]}{[\textnormal{CO}]} = \frac{k_1}{k_2} \frac{(k_3[\textnormal{CO}] + k_6[e^-] + k_7 [\textnormal{H}_2\textnormal{O}])}{(k_4[e^-] + k_5 [\textnormal{H}_2\textnormal{O}])} \frac{[\textnormal{HOCO}^+]}{[\textnormal{HCO}^+]}
\end{equation}

\noindent where the rate constants are numbered according to Table 9.  Quantities in brackets indicate volume densities.  

Figure 13 shows observations of HOCO$^+$ and HCO$^+$ in the cold envelope toward Sgr B2(N).  Two velocity components are observed in each line, centered at $v_\textnormal{LSR} = 64$ and 78 km s$^{-1}$.  For HOCO$^+$, a rotational temperature of 14 K is found for both components.  The column densities derived are $3.0 \times 10^{13}$ cm$^{-2}$ for the 64 km s$^{-1}$ component, and $1.5 \times 10^{13}$ cm$^{-2}$ for the 78 km s$^{-1}$ component, assuming LTE.  While 14 K is very unlikely to be the gas kinetic temperature, the column densities we derive are reasonably accurate; at these excitation temperatures, most of the population is in the $K_a = 0$ ladder, which we probe directly through these transitions.

\begin{figure}
\figurenum{13}
\centering
\includegraphics[width=6.0in]{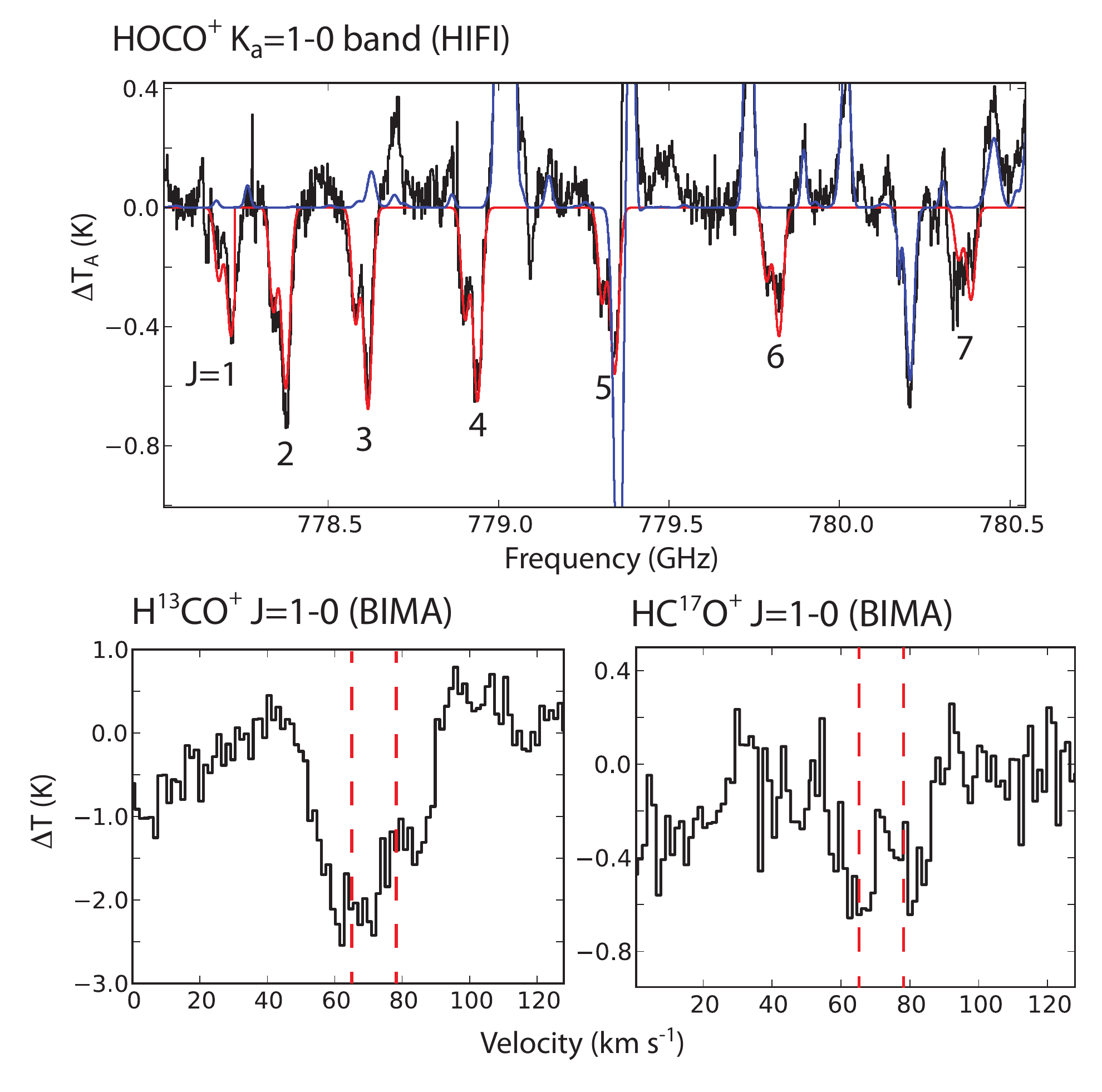}
\caption{Observations of HOCO$^+$ and HCO$^+$ in the cold envelope around Sgr B2(N).  The top panel is the $K_a = 1-0$ $Q$-branch of HOCO$^+$ at 779 GHz detected in the HIFI survey; the red curve is the LTE model to HOCO$^+$, while the blue curve is the model to all other molecules identified in the spectrum.  The strong absorption line in the model at 779.4 GHz is a low-excitation line of CH$_3$OH ($6_{24}-5_{15}$ of the $A$ species) that is not well reproduced by the model.  The lower panels are transitions of HCO$^+$ isotopologues observed in the BIMA observations of \cite{friedel04}.  The two dashed red lines indicate velocities of 64 and 78 km s$^{-1}$, the two velocity components modeled in the HOCO$^+$ transitions.}
\end{figure}

For HCO$^+$, HIFI only measured transitions with $J_\textnormal{up} \ge$ 6, and so does not probe the colder gas in which HOCO$^+$ is detected.  Observations of the ground state line of two HCO$^+$ isotopologues, H$^{13}$CO$^+$ and HC$^{17}$O$^+$, in the 3 mm spectral window were made in a joint survey with the NRAO 12 m telescope and the BIMA interferometric array \citep{friedel04}.  In the 12 m telescope data, with a beamwidth of $71''$, a combination of broad emission and narrower absorption at the Sgr B2(N) velocity is observed.  In the BIMA observations, with a synthesized beamwidth of $32.2''$ (in declination) $\times$ 6.5$''$ (in right ascension), only absorption is observed in both lines, as seen in Figure 13.  The emission seen in the 12 m telescope presumably arises from warmer gas, possibly associated with the fairly broad HCO$^+$ emission seen by HIFI (Fig. 8.6).  We use the BIMA observations here to derive the column density of HCO$^+$.  The HC$^{17}$O$^+$ is seen at approximately the 2$\sigma$ noise level, but peaks at the velocities of the two components also detected in HOCO$^+$ (indicated by the dashed red lines) makes its identification secure.

Using the isotopic ratios $^{12}$C/$^{13}$C = 20 and $^{16}$O/$^{17}$O = 800 (\S 3.3), the detection of HC$^{17}$O$^+$ requires that the H$^{13}$CO$^+$ transition has high optical depth.  Under that condition, the line intensities are given by 

\begin{equation}
\Delta T(\textnormal{H}^{13}\textnormal{CO}^+) = J(T_\textnormal{ex,13}) - J(T_\textnormal{CMB}) - T_\textnormal{cont}
\end{equation}

\begin{equation}
\Delta T(\textnormal{HC}^{17}\textnormal{O}^+) = (1 - e^{-\tau_{17}}) (J(T_\textnormal{ex,17}) - J(T_\textnormal{CMB}) - T_\textnormal{cont})
\end{equation}

\noindent Assuming $T_\textnormal{ex,13} = T_\textnormal{ex,17}$, we can derive

\begin{equation}
\tau_{17} = -\textnormal{ln}\left(1-\frac{\Delta T_{17}}{\Delta T_{13}}\right)
\end{equation}

\noindent For the 64 km s$^{-1}$ component, we find $\tau_{17} = 0.26 \pm 0.11$, while for the 78 km s$^{-1}$ component, $\tau_{17} = 0.32 \pm 0.16$.  Deriving a HC$^{17}$O$^+$ column density from this value requires us to assume an excitation temperature.  Because of the weak continuum at 87 GHz (as compared to in the submillimeter and far-infrared), $T_\textnormal{ex}$ must be low ($\le 5$ K) in order for the line to be seen in absorption.  The column density in all levels of HC$^{17}$O$^+$ derived for the 64 km s$^{-1}$ component, assuming $T_\textnormal{ex}$ is between 2.7 and 5 K and LTE conditions, is $(5.6 \pm 3.1) \times 10^{12}$ cm$^{-2}$.  This implies $N$(HCO$^+$) = $(4.5 \pm 2.5) \times 10^{15}$ cm$^{-2}$.  This yields an HOCO$^+$/HCO$^+$ ratio of $(6.7 \pm 3.7) \times 10^{-3}$, with the uncertainty dominated by the HCO$^+$ column density.

As Equation (14) shows, the CO$_2$/CO ratio also depends on the formation and destruction mechanisms for HCO$^+$ and HOCO$^+$.  We assume that the formation of both molecules is from H$_3$$^+$.  For the destruction, the dominant mechanisms depend on the relative abundances of CO, water, and electrons.  From the ground-state transitions of H$_2$$^{18}$O and H$_2$$^{17}$O in the HIFI survey (e.g. the $para$ ground state transitions in Figure 4), assuming $T_\textnormal{ex} = 2.7$ K, we derive an H$_2$$^{18}$O column density of $(2-4) \times 10^{14}$ cm$^{-2}$, or a H$_2$O column density of $(5-10) \times 10^{16}$ cm$^{-2}$.  Assuming an H$_2$ column density of $\sim 10^{24}$ cm$^{-2}$ \citep{lis89, lis90}, this implies a water abundance of $(3-10) \times 10^{-8}$.  \cite{neufeld97} derived an abundance of $3 \times 10^{-8}$ in water with temperatures less than 90 K, while \cite{wirstrom10} found a water abundance of $10^{-7}$ from the ground state \emph{ortho} transition, in general agreement with the HIFI ground state lines.  \cite{cernicharo06} found a much higher water abundance of $10^{-5}$, but this was a hot water component likely spatially distinct from the gas considered here.

From this, in Figure 14 the CO$_2$/CO ratio is plotted as a function of the electron abundance relative to H$_2$, $\chi(e^-)$.  At high $\chi(e^-)$ ($\ge 10^{-7}$), electrons are the primary destroyer of both HCO$^+$ and HOCO$^+$.  At low electron abundance, meanwhile, CO destroys HOCO$^+$ while H$_2$O destroys HCO$^+$.  A key parameter is therefore the H$_2$O/CO ratio.  A gas-phase CO fractional abundance of $10^{-4}$ is assumed in Figure 14.  Curves are plotted for water abundances of $3 \times 10^{-7}$ and $3 \times 10^{-8}$, and it can be seen that it is only in the low-$e^-$ regime that this value is important.  In between these two limits, HCO$^+$ is destroyed by electrons while HOCO$^+$ is destroyed by CO.

\begin{figure}
\figurenum{14}
\centering
\includegraphics[width=4.0in]{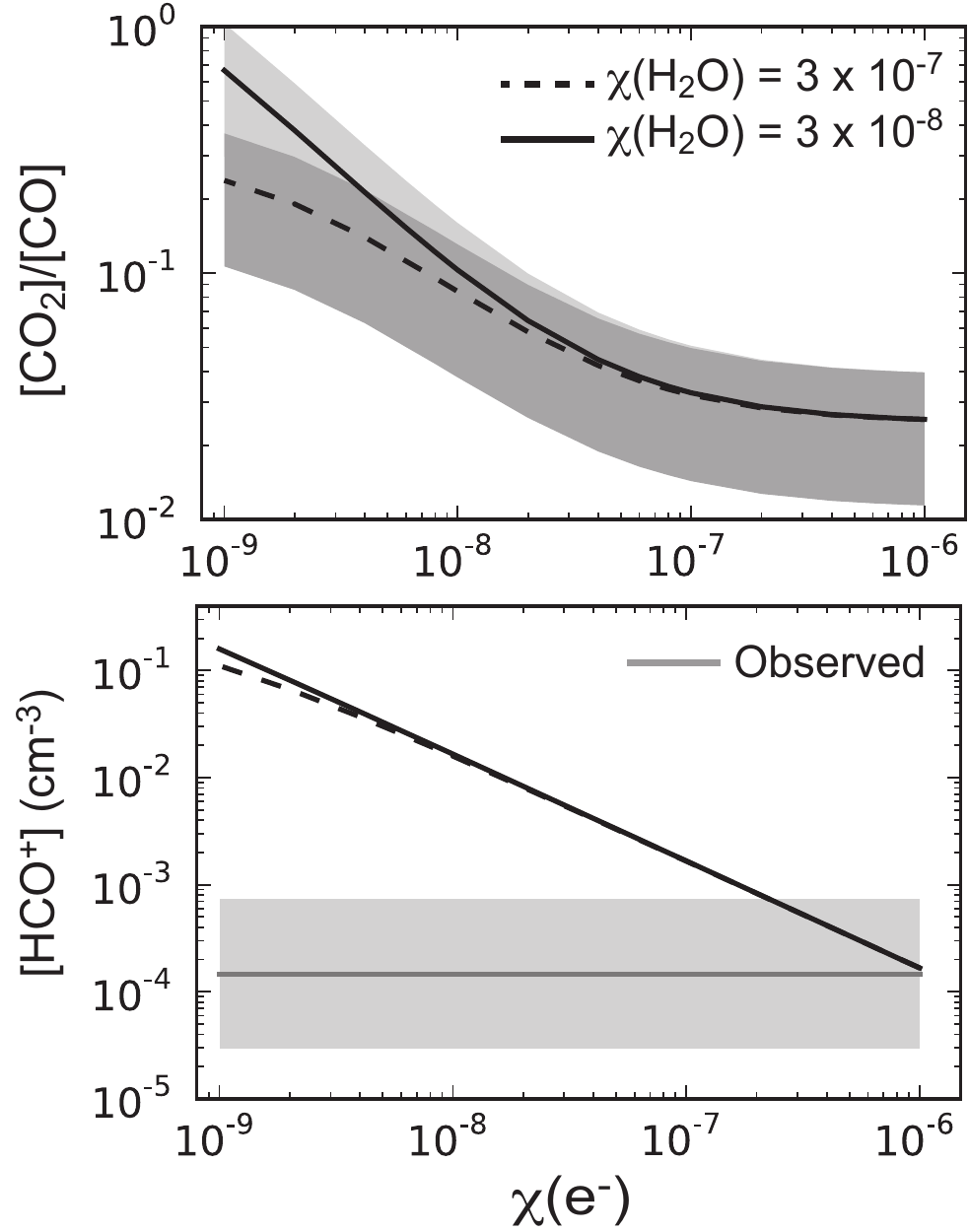}
\caption{The top panel shows the CO$_2$/CO abundance ratio derived from Eq. (14) as a function of the electron abundance.  Calculations are shown with two assumed H$_2$O fractional abundances, which only diverge at very low electron abundance.  The gray shaded areas show the roughly factor of 2 uncertainty in the HCO$^+$ column density.  The bottom panel shows the HCO$^+$ volume density derived from equation (12), assuming a CO fractional abundance of $10^{-4}$ and a H$_3$$^+$ volumn density of $2.5 \times 10^{-3}$ cm$^{-3}$.  The gray line indicates the observed HCO$^+$ volumn density, assuming a cloud thickness of 10 pc, and the gray region indicates a factor of 5 total uncertainty.}
\end{figure}

The bottom panel in Figure 14 shows the volume density of HCO$^+$, from Eq. (12), for the same models.  For this calculation, the volume density of H$_3$$^+$ must be assumed.  We use a value of $2.5 \times 10^{-3}$, which is calculated from the usual steady-state equation for dense clouds \citep{mccall99}:

\begin{equation}
[\textnormal{H}_3^+] = \frac{\zeta}{k_\textnormal{CO} \chi(\textnormal{CO})}
\end{equation}

\noindent where $k_\textnormal{CO}$ is from reaction 1 in Table 9 (the primary destruction mechanism for H$_3$$^+$ in dense regions) and $\zeta$ is the cosmic ray ionization rate, which we take to be $4 \times 10^{-16}$ s$^{-1}$ from \cite{vandertak06}, based on observations of H$_3$O$^+$.  This shows that at very low electron abundances, the HCO$^+$ abundance gets very high, due to a very slow destruction rate.  We estimate the observed HCO$^+$ volume density by dividing the observed column density by a pathlength of 10 pc, a rough estimate but probably correct within an order of magnitude.  This shows that the HCO$^+$ abundance is most consistent with a CO$_2$/CO ratio of between 0.01--0.1.  

Given a CO fractional abundance of $10^{-4}$, this CO$_2$/CO ratio implies a CO$_2$ gas-phase abundance of $10^{-6}-10^{-5}$.  This is consistent with the finding of \cite{minh88, minh91}.  A lower limit to the CO$_2$ gas-phase abundance of $10^{-6}$ relative to H$_2$ is an order of magnitude higher than toward other massive star-forming regions where gas-phase CO$_2$ has been detected directly \citep{vandishoeck96, boonman03}.  In the ices, CO$_2$ has been detected toward Sgr A* \citep{degraauw96, gibb04} with an abundance of 14\% relative to H$_2$O, and a CO$_2$/CO ratio of at least 1.  While the ice abundance of CO$_2$ toward Sgr A* is typical for star-forming regions, the abundance in the gas is clearly enhanced.  Also, while we find a H$_2$O/CO ratio of about $3 \times 10^{-4}$ in the gas phase, the ratio in the ices is at least 8, indicating that most of the water is still frozen out in ices in the Sgr B2 envelope.

Several chemical explanations have been proposed to explain the relatively high gas-phase abundance of CO$_2$ toward the Galactic Center as compared to other sources.  \cite{minh98}, based on the high abundances of HOCO$^+$ and HNCO at the position $1'$ north of Sgr B2(N), propose that the ices have been released into the gas phase through sputtering in shocks, possibly associated with a cloud collapse.  \cite{charnley00} suggest that while shocks could be responsible for destroying CO$_2$ after grain mantle evaporation through endothermic reactions with either H or H$_2$, at the relatively low density of the Sgr B2 envelope ($\sim 10^4$ cm$^{-3}$) CO$_2$ can survive in relatively strong ($\le 45$ km s$^{-1}$) C-shocks.

The possibility also exists that CO can be convered into CO$_2$ in the gas phase through the reaction

\begin{equation}
\textnormal{CO} + \textnormal{OH} \rightarrow \textnormal{CO}_2 + \textnormal{H}
\end{equation}

\noindent This reaction has a reasonable rate at moderate temperatures in the laboratory ($10^{-13}$ cm$^3$ s$^{-1}$ at 100 K, \cite{frost93}).  We only detect two transitions of OH in the HIFI survey (the $J = 3/2-1/2$ doublet in the $\Omega = 1/2$ ladder), from which it is difficult to derive a column density, but \cite{goicoechea02} found a column density of $(1.5-2.5) \times 10^{16}$ cm$^{-2}$, or an abundance of $10^{-8}-10^{-7}$ assuming $N$(H$_2$) is in the range $\sim 10^{23}-10^{24}$ cm$^{-2}$, with ISO.  At higher temperatures ($\ge 300$ K), OH can react quickly with H$_2$ to form H$_2$O, but in the 50--150 K temperature range, Eq. (19) might be important.  \cite{charnley00} suggested that the gas-phase formation may be largely responsible for the gas-phase CO$_2$ toward YSOs detected with ISO.

\section{Conclusion}

We have presented a full HIFI chemical survey of the Sagittarius B2(N) star-forming region.  A total of 44 molecules are detected in this survey through transitions in emission and absorption.  We have performed LTE modeling to the observed features of nearly all molecules in the spectrum, along with Gaussian fits to lines to some molecules with very high optical depth and complex lineshapes ($^{12}$CO and all isotopologues of H$_2$O) or unique excitation  (H$_3$O$^+$).  About 12\% of the data channels observed to have emission over the 2$\sigma$ rms noise level are currently unidentified.

For each molecule found to arise from the $\sim 3''$ Sgr B2(N) hot core, assumed primarily on the basis of a component with rotation temperature $> 100$ K, we have derived an abundance relative to H$_2$.  The source sizes used for the LTE modeling derive from high spatial resolution Submillimeter Array observations at 350 GHz.  We have compared these abundances to those derived for the HIFI survey of Orion KL \citep{crockett13}, and discussed the similarities and differences between various molecules in the two massive star forming regions.  Some molecules (namely, C$_2$H$_5$OH, CH$_3$NH$_2$, CH$_2$NH, and H$_2$CCO, NH$_2$CHO) are far more abundant toward Sgr B2(N), while others (CH$_3$OCHO, HC$_3$N, SO, and SO$_2$) are less abundant.  These differences could hold a clue into the evolutionary differences between these two regions.

We have also presented a kinematic analysis of the envelope of Sgr B2(N) as detected through numerous absorption lines against the strong far-infrared dust continuum.  At least two cold ($T_\textnormal{rot} \sim$ 10--20 K) absorption compnents, centered at approximately 64 and 80 km s$^{-1}$, are detected in a number of molecules, along with two warm absorption components (with energy levels of 100--600 K excited) presumably associated with gas directly in front of the bright hot core, rather than the more extended envelope.  One of these components is red-shifted from the Sgr B2(N) envelope velocity, signifying infall, while the other is blue-shifted, suggesting outflow.  We show that while both components are detected in high-energy transitions requiring warm and dense gas for excitation, the 
blue-shifted ($v_\textnormal{LSR} \sim 55$ km s$^{-1}$) gas is more highly excited, suggesting an origin  closer to the protostars in the central hot core region.  This is consistent with the scenario posited by \cite{rolffs10} for the nearby Sgr B2(M) star-forming region, in which infall is occuring at large radii, reversing to outflow closer to the star, evidence for feedback from the deeply embeddded protostar(s).  We also use the detection of protonated carbon dioxide, HOCO$^+$, in conjunction with previously published observations of HCO$^+$ isotopologues, to infer the gas-phase CO$_2$ abundance in the Sgr B2(N) region.

Along with the other HEXOS sources, this spectral scan and the model we present here may aid future investigations of molecular spectra of star-forming regions in the far-infrared, both with future space-based missions and ground-based observatories such as the Atacama Large Millimeter Array (ALMA).  The reduced, deconvolved spectrum and the best-fit LTE models will be made available to the public through the Herschel Science Center website.

\begin{acknowledgements}
HIFI has been designed and built by a consortium of institutes and university departments from across Europe, Canada, and the United States under the leadership of SRON Netherlands Institute for Space Research, Groningen, The Netherlands and with major contributions from Germany, France, and the US.  Consortium members are: Canada: CSA, U.Waterloo; France: CESR, LAB, LERMA, IRAM; Germany: KOSMA, MPIfR, MPS; Ireland: NUI Maynooth; Italy: ASI, IFSI-INAF, Osservatorio Astrofisico di Arcetri-INAF; Netherlands: SRON, TUD; Poland: CAMK, CBK; Spain: Observatorio Astron\'{o}mico Nacional (IGN), Centro de Astrobiolog\'{i}a (CSIC-INTA); Sweden: Chalmers University of Technology--MC2, RSS \& GARD, Onsala Space Observatory, Swedish National Space Board, Stockholm Observatory; Switzerland: ETH Zurich, FHNW; USA: Caltech, JPL, NHSC.  Support for this work was provided by NASA through an award issued by JPL/Caltech.

We thank Anthony Remijan and Joanna Corby for providing complementary data from other telescopes (GBT and BIMA) used in this manuscript.
\end{acknowledgements}

\bibliography{SgrB2}

\begin{thebibliography}{132}
\expandafter\ifx\csname natexlab\endcsname\relax\def\natexlab#1{#1}\fi

\bibitem[{Adams {et~al.}(1978)Adams, Smith, \& Grief}]{adams78}
Adams, N., Smith, D., \& Grief, D. 1978, IJMSI, 26, 405

\bibitem[{Adande \& Ziurys(2012)}]{adande12}
Adande, G., \& Ziurys, L. 2012, ApJ, 744, 194

\bibitem[{Barlow {et~al.}(2013)Barlow, Swinyard, Owen, Cernicharo, Gomez,
  Ivison, Krause, Lim, Matsuura, Miller, Olofsson, \& Polehampton}]{barlow13}
Barlow, M., Swinyard, B., Owen, P., {et~al.} 2013, Sci, 342, 1343

\bibitem[{Belloche {et~al.}(2009)Belloche, Garrod, M\"{u}ller, Menten, Comito,
  \& Schilke}]{belloche09}
Belloche, A., Garrod, R., M\"{u}ller, H., {et~al.} 2009, A\&A, 499, 215

\bibitem[{Belloche {et~al.}(2008)Belloche, Menten, Comito, a, a, \&
  a}]{belloche08}
Belloche, A., Menten, K., Comito, C., {et~al.} 2008, A\&A, 482, 179

\bibitem[{Belloche {et~al.}(2013)Belloche, M\"{u}ller, Menten, Schilke, \&
  Comito}]{belloche13}
Belloche, A., M\"{u}ller, H., Menten, K., Schilke, P., \& Comito, C. 2013,
  A\&A, 559, A47

\bibitem[{Bergin {et~al.}(2010)Bergin, Phillips, Comito, blah, blah, \&
  blah}]{bergin10}
Bergin, E., Phillips, T., Comito, C., {et~al.} 2010, A\&A, 521, L20

\bibitem[{Blake {et~al.}(1987)Blake, Sutton, Masson, \& Phillips}]{blake87}
Blake, G., Sutton, E., Masson, C., \& Phillips, T. 1987, ApJ, 315, 621

\bibitem[{Bocherel {et~al.}(1996)Bocherel, Herbert, Rowe, Sims, Smith, \&
  Travers}]{bocherel96}
Bocherel, P., Herbert, L., Rowe, B., {et~al.} 1996, JPhCh, 100, 3063

\bibitem[{Boonman {et~al.}(2003)Boonman, van Dishoeck, Lahuis, \&
  Doty}]{boonman03}
Boonman, A., van Dishoeck, E., Lahuis, F., \& Doty, S. 2003, A\&A, 399, 1063

\bibitem[{Bouchoux \& Choret(1997)}]{bouchoux97}
Bouchoux, G., \& Choret, N. 1997, RCMS, 11, 1799

\bibitem[{Brian \& Mitchell(1990)}]{brian90}
Brian, J., \& Mitchell, A. 1990, PhR, 186, 215

\bibitem[{Ceccarelli {et~al.}(2002)Ceccarelli, Baluteau, Walmsley, Swinyard,
  Caux, Sidher, Cox, Gry, Kessler, \& Prusti}]{ceccarelli02}
Ceccarelli, C., Baluteau, J.-P., Walmsley, M., {et~al.} 2002, A\&A, 383, 603

\bibitem[{Cernicharo {et~al.}(2000)Cernicharo, Goicoechea, \&
  Caux}]{cernicharo00}
Cernicharo, J., Goicoechea, J., \& Caux, E. 2000, ApJ, 534, L199

\bibitem[{Cernicharo {et~al.}(2006)Cernicharo, Goicoechea, Pardo, \&
  Asensio-Ramos}]{cernicharo06}
Cernicharo, J., Goicoechea, J., Pardo, J., \& Asensio-Ramos, A. 2006, ApJ, 642,
  940

\bibitem[{Charnley \& Kaufman(2000)}]{charnley00}
Charnley, S., \& Kaufman, M. 2000, ApJ, 529, L111

\bibitem[{Churchwell {et~al.}(1986)Churchwell, Wood, Myers, \&
  Myers}]{churchwell86}
Churchwell, E., Wood, D., Myers, P., \& Myers, R. 1986, ApJ, 305, 405

\bibitem[{Comito {et~al.}(2005)Comito, Schilke, Phillips, Lis, Motte, \&
  Mehringer}]{comito05}
Comito, C., Schilke, P., Phillips, T., {et~al.} 2005, ApJS, 156, 127

\bibitem[{Crockett {et~al.}(2014{\natexlab{a}})Crockett, Bergin, Neill, a, a,
  \& a}]{crockett13}
Crockett, N., Bergin, E., Neill, J., {et~al.} 2014{\natexlab{a}}, ApJ, in press

\bibitem[{Crockett {et~al.}(2014{\natexlab{b}})Crockett, Bergin, Neill, Black,
  Blake, \& Kleshcheva}]{crockett13a}
---. 2014{\natexlab{b}}, ApJ, 781, 114

\bibitem[{Crockett {et~al.}(2010)Crockett, Bergin, Wang, a, a, \&
  a}]{crockett10}
Crockett, N., Bergin, E., Wang, S., {et~al.} 2010, A\&A, 521, L21

\bibitem[{Cummins {et~al.}(1986)Cummins, Linke, \& Thaddeus}]{cummins86}
Cummins, S., Linke, R., \& Thaddeus, P. 1986, ApJS, 60, 819

\bibitem[{de~Graauw {et~al.}(2010)de~Graauw, Helmich, Phillips, a, a, \&
  a}]{degraauw10}
de~Graauw, T., Helmich, F., Phillips, T., {et~al.} 2010, A\&A, 518, L6

\bibitem[{de~Graauw {et~al.}(1996)de~Graauw, Whittet, Gerakines, Bauer,
  Beintema, Boogert, Boxhoorn, {et~al.}}]{degraauw96}
de~Graauw, T., Whittet, D., Gerakines, P., {et~al.} 1996, A\&A, 315, L345

\bibitem[{Deguchi {et~al.}(2006)Deguchi, Miyazaki, \& Minh}]{deguchi06}
Deguchi, S., Miyazaki, A., \& Minh, Y. 2006, PASJ, 58, 979

\bibitem[{Eisenhauer {et~al.}(2003)Eisenhauer, Sch\"{o}del, Genzel, Ott, Tecza,
  Abuter, Eckert, \& Alexander}]{eisenhauer03}
Eisenhauer, F., Sch\"{o}del, R., Genzel, R., {et~al.} 2003, ApJ, 597, L121

\bibitem[{Etxaluze {et~al.}(2013)Etxaluze, Goicoechea, Cernicharo, Polehampton,
  Noriega-Crespo, Molinari, Swinyard, Wu, \& Bally}]{etxaluze13}
Etxaluze, M., Goicoechea, J., Cernicharo, J., {et~al.} 2013, A\&A, 556, A137

\bibitem[{Frerking {et~al.}(1980)Frerking, Wilson, Linke, \&
  Wannier}]{frerking80}
Frerking, M., Wilson, R., Linke, R., \& Wannier, P. 1980, ApJ, 240, 65

\bibitem[{Friedel {et~al.}(2004)Friedel, Snyder, Turner, \&
  Remijan}]{friedel04}
Friedel, D., Snyder, L., Turner, B., \& Remijan, A. 2004, ApJ, 600, 234

\bibitem[{Frost {et~al.}(1993)Frost, Sharkey, \& Smith}]{frost93}
Frost, M., Sharkey, P., \& Smith, I. 1993, JPhCh, 97, 12254

\bibitem[{Fuchs {et~al.}(2009)Fuchs, Cuppen, Ioppolo, Romanzin, Bisschop, a, a,
  \& a}]{fuchs09}
Fuchs, G., Cuppen, H., Ioppolo, S., {et~al.} 2009, A\&A, 505, 629

\bibitem[{Garrod \& Herbst(2006)}]{garrod06}
Garrod, R., \& Herbst, E. 2006, A\&A, 457, 927

\bibitem[{Garrod {et~al.}(2008)Garrod, Weaver, \& Herbst}]{garrod08}
Garrod, R., Weaver, S.~W., \& Herbst, E. 2008, ApJ, 682, 283

\bibitem[{Gaume \& Claussen(1990)}]{gaume90}
Gaume, R., \& Claussen, M. 1990, ApJ, 351, 538

\bibitem[{Gaume {et~al.}(1995)Gaume, Claussen, Pree, Goss, \&
  Mehringer}]{gaume95}
Gaume, R., Claussen, M., Pree, C.~D., Goss, W., \& Mehringer, D. 1995, ApJ,
  449, 663

\bibitem[{Geppert {et~al.}(2004)Geppert, Thomas, Ehlerding, a, a, \&
  a}]{geppert04}
Geppert, W., Thomas, R., Ehlerding, A., {et~al.} 2004, FaDi, 127, 425

\bibitem[{Gerakines {et~al.}(1999)Gerakines, Whittet, Ehrenfreund, Boogert,
  Tielens, Schutte, Chiar, {et~al.}}]{gerakines99}
Gerakines, P., Whittet, D., Ehrenfreund, P., {et~al.} 1999, ApJ, 357

\bibitem[{Gibb {et~al.}(2004)Gibb, Whittet, Boogert, \& Tielens}]{gibb04}
Gibb, E., Whittet, D., Boogert, A., \& Tielens, A. 2004, ApJS, 151, 35

\bibitem[{Godard {et~al.}(2012)Godard, Falgarone, Gerin, a, a, \& a}]{godard12}
Godard, B., Falgarone, E., Gerin, M., {et~al.} 2012, A\&A, 540, A87

\bibitem[{Goicoechea \& Cernicharo(2002)}]{goicoechea02}
Goicoechea, J., \& Cernicharo, J. 2002, ApJ, 576, L77

\bibitem[{Goicoechea {et~al.}(2006)Goicoechea, Pety, Gerin, Teyssier, Roueff,
  Hily-Blant, \& Baek}]{goicoechea06}
Goicoechea, J., Pety, J., Gerin, M., {et~al.} 2006, A\&A, 456, 565

\bibitem[{Goicoechea {et~al.}(2004)Goicoechea, Rodr\'{i}guez-Fern\'{a}ndez, \&
  Cernicharo}]{goicoechea04}
Goicoechea, J., Rodr\'{i}guez-Fern\'{a}ndez, N., \& Cernicharo, J. 2004, ApJ,
  600, 214

\bibitem[{Goldsmith {et~al.}(1983)Goldsmith, Krotkov, Snell, Brown, \&
  Godfrey}]{goldsmith83}
Goldsmith, P., Krotkov, R., Snell, R., Brown, R., \& Godfrey, P. 1983, ApJ,
  274, 184

\bibitem[{Goldsmith {et~al.}(1987)Goldsmith, Snell, \& Lis}]{goldsmith87}
Goldsmith, P., Snell, R., \& Lis, D. 1987, ApJ, 313, L5

\bibitem[{Halfen {et~al.}(2011)Halfen, Ilyushin, \& Ziurys}]{halfen11}
Halfen, D., Ilyushin, V., \& Ziurys, L. 2011, ApJ, 743, 60

\bibitem[{Halfen {et~al.}(2013)Halfen, Ilyushin, \& Ziurys}]{halfen13}
---. 2013, A\&A, 767, 66

\bibitem[{Herbst {et~al.}(1977)Herbst, Green, Thaddeus, \&
  Klemperer}]{herbst77}
Herbst, E., Green, S., Thaddeus, P., \& Klemperer, W. 1977, ApJ, 215, 503

\bibitem[{Herbst \& van Dishoeck(2009)}]{herbst09}
Herbst, E., \& van Dishoeck, E. 2009, ARA\&A, 47, 427

\bibitem[{Hollis {et~al.}(2004{\natexlab{a}})Hollis, Jewell, Lovas, \&
  Remijan}]{hollis04a}
Hollis, J., Jewell, P., Lovas, F., \& Remijan, A. 2004{\natexlab{a}}, ApJ, 613,
  L45

\bibitem[{Hollis {et~al.}(2004{\natexlab{b}})Hollis, Jewell, Lovas, Remijan, \&
  Mollendal}]{hollis04b}
Hollis, J., Jewell, P., Lovas, F., Remijan, A., \& Mollendal, H.
  2004{\natexlab{b}}, ApJ, 610, L21

\bibitem[{Hollis {et~al.}(2007)Hollis, Jewell, Remijan, \& Lovas}]{hollis07}
Hollis, J., Jewell, P., Remijan, A., \& Lovas, F. 2007, ApJ, 660, L125

\bibitem[{Hollis {et~al.}(2006)Hollis, Lovas, Remijan, Jewell, Ilyushin, \&
  Kleiner}]{hollis06}
Hollis, J., Lovas, F., Remijan, A., {et~al.} 2006, ApJ, 643, L25

\bibitem[{Hollis {et~al.}(2003)Hollis, Pedelty, Boboltz, a, a, \& a}]{hollis03}
Hollis, J., Pedelty, J., Boboltz, D., {et~al.} 2003, ApJ, 596, L235

\bibitem[{H\"{u}ttemeister {et~al.}(1993)H\"{u}ttemeister, Wilson, Henkel, \&
  Mauersberger}]{huttemeister93}
H\"{u}ttemeister, S., Wilson, T., Henkel, C., \& Mauersberger, R. 1993, A\&A,
  276, 445

\bibitem[{Jones {et~al.}(2012)Jones, Burton, Cunningham, a, a, \& a}]{jones12}
Jones, P., Burton, M., Cunningham, M., {et~al.} 2012, MNRAS, 419, 2961

\bibitem[{Kahane {et~al.}(2013)Kahane, Ceccarelli, Faure, \& Caux}]{kahane13}
Kahane, C., Ceccarelli, C., Faure, A., \& Caux, E. 2013, ApJL, 763, L38

\bibitem[{Klippenstein {et~al.}(2010)Klippenstein, Georgievskii, \&
  McCall}]{klippenstein10}
Klippenstein, S., Georgievskii, Y., \& McCall, B. 2010, JPCA, 114, 278

\bibitem[{Knight {et~al.}(1986)Knight, Freeman, McEwan, Smith, Adams, \&
  Smith}]{knight86}
Knight, J., Freeman, C., McEwan, M., {et~al.} 1986, MNRAS, 219, 89

\bibitem[{Kuan \& Snyder(1994)}]{kuan94}
Kuan, Y.-J., \& Snyder, L. 1994, ApJS, 94, 651

\bibitem[{Kuan \& Snyder(1996)}]{kuan96}
---. 1996, ApJ, 470, 981

\bibitem[{Kurland \& Wilson(1957)}]{kurland57}
Kurland, R., \& Wilson, E. 1957, JChPh, 27, 585

\bibitem[{Laas {et~al.}(2011)Laas, Garrod, Herbst, \& Weaver}]{laas11}
Laas, J., Garrod, R., Herbst, E., \& Weaver, S.~W. 2011, ApJ, 728, 71

\bibitem[{Lis \& Goldsmith(1989)}]{lis89}
Lis, D., \& Goldsmith, P. 1989, ApJ, 337, 704

\bibitem[{Lis \& Goldsmith(1990)}]{lis90}
---. 1990, ApJ, 356, 195

\bibitem[{Lis \& Goldsmith(1991)}]{lis91}
---. 1991, ApJ, 369, 157

\bibitem[{Lis {et~al.}(2012)Lis, Schilke, Bergin, Emprechtinger, a, a, \&
  a}]{lis12}
Lis, D., Schilke, P., Bergin, E., {et~al.} 2012, RSPTA, 370, 5162

\bibitem[{Lis {et~al.}(2010)Lis, Phillips, Goldsmith, Neufeld, Herbst, Comito,
  Schilke, M\"{u}ller, {et~al.}}]{lis10b}
Lis, D., Phillips, T., Goldsmith, P., {et~al.} 2010, A\&A, 521, L26

\bibitem[{Loomis {et~al.}(2013)Loomis, Zaleski, Steber, a, a, \& a}]{loomis13}
Loomis, R., Zaleski, D., Steber, A., {et~al.} 2013, ApJL, 765, L9

\bibitem[{Mathis {et~al.}(1977)Mathis, Rumpl, \& Nordsieck}]{mathis77}
Mathis, J., Rumpl, W., \& Nordsieck, K. 1977, ApJ, 217, 425

\bibitem[{McCall {et~al.}(1999)McCall, Geballe, Hinkle, \& Oka}]{mccall99}
McCall, B., Geballe, T., Hinkle, K., \& Oka, T. 1999, ApJ, 522, 338

\bibitem[{Menten \& Wyrowski(2011)}]{menten11b}
Menten, K., \& Wyrowski, F. 2011, Springer Tracts in Modern Physics, Vol. 241,
  Interstellar Molecules (Springer-Verlang Berlin Heidelberg), 27

\bibitem[{Miao {et~al.}(1995)Miao, Mehringer, Kuan, \& Snyder}]{miao95}
Miao, Y., Mehringer, D., Kuan, Y.-J., \& Snyder, L. 1995, ApJ, 445, L59

\bibitem[{Minh {et~al.}(1991)Minh, Brewer, Irvine, Friberg, \&
  Johansson}]{minh91}
Minh, Y., Brewer, M., Irvine, W., Friberg, P., \& Johansson, L. 1991, A\&A,
  244, 470

\bibitem[{Minh {et~al.}(1998)Minh, Haikala, Hjalmarson, \& Irvine}]{minh98}
Minh, Y., Haikala, L., Hjalmarson, A., \& Irvine, W. 1998, ApJ, 498, 261

\bibitem[{Minh {et~al.}(1988)Minh, Irvine, \& Ziurys}]{minh88}
Minh, Y., Irvine, W., \& Ziurys, L. 1988, ApJ, 334, 175

\bibitem[{Molinari {et~al.}(2011)Molinari, Bally, Noriega-Crespo, a, a, \&
  a}]{molinari11}
Molinari, S., Bally, J., Noriega-Crespo, A., {et~al.} 2011, ApJL, 735, L33

\bibitem[{Morris \& Serabyn(1996)}]{morris96}
Morris, M., \& Serabyn, E. 1996, ARA\&A, 34, 645

\bibitem[{Morris {et~al.}(1976)Morris, Turner, Palmer, \& Zuckerman}]{morris76}
Morris, M., Turner, B., Palmer, P., \& Zuckerman, B. 1976, ApJ, 205, 82

\bibitem[{M\"{u}ller {et~al.}(2005)M\"{u}ller, Scl\"{o}der, Stutzki, \&
  Winnewisser}]{muller05}
M\"{u}ller, H., Scl\"{o}der, F., Stutzki, J., \& Winnewisser, G. 2005, JMoSt,
  742, 215

\bibitem[{M\"{u}ller {et~al.}(2001)M\"{u}ller, Thorwirth, Roth, \&
  Winnewisser}]{muller01}
M\"{u}ller, H., Thorwirth, S., Roth, D., \& Winnewisser, G. 2001, A\&A, 370,
  L49

\bibitem[{Neill {et~al.}(2012)Neill, Bergin, Lis, Phillips, Emprechtinger, \&
  Schilke}]{neill12}
Neill, J., Bergin, E., Lis, D., {et~al.} 2012, JMoSp, 280, 150

\bibitem[{Neill {et~al.}(2011)Neill, Steber, Muckle, a, a, \& a}]{neill11}
Neill, J., Steber, A., Muckle, M., {et~al.} 2011, JPCA, 115, 6472

\bibitem[{Neufeld {et~al.}(1997)Neufeld, Zmuidzinas, Schilke, \&
  Phillips}]{neufeld97}
Neufeld, D., Zmuidzinas, J., Schilke, P., \& Phillips, T. 1997, ApJ, 488, L141

\bibitem[{Nobukawa {et~al.}(2011)Nobukawa, Ryu, Tsuru, \& Koyama}]{nobukawa11}
Nobukawa, M., Ryu, S., Tsuru, G., \& Koyama, K. 2011, ApJL, 739, L52

\bibitem[{Nummelin {et~al.}(2000)Nummelin, Bergman, \& \r{A}.
  Hjalmarson}]{nummelin00}
Nummelin, A., Bergman, P., \& \r{A}. Hjalmarson. 2000, ApJS, 128, 213

\bibitem[{\"{O}berg {et~al.}(2009)\"{O}berg, Garrod, van Dishoeck, \&
  Linnartz}]{oberg09}
\"{O}berg, K., Garrod, R., van Dishoeck, E., \& Linnartz, H. 2009, A\&A, 504,
  891

\bibitem[{Oka {et~al.}(2005)Oka, Geballe, Goto, Usuda, \& McCall}]{oka05}
Oka, T., Geballe, T., Goto, M., Usuda, T., \& McCall, B. 2005, ApJ, 632, 882

\bibitem[{Ossenkopf \& Henning(1994)}]{ossenkopf94}
Ossenkopf, V., \& Henning, T. 1994, A\&A, 291, 943

\bibitem[{Ott(2010)}]{ott10}
Ott, S. 2010, ASP Conference Series, 434, 139

\bibitem[{Penzias(1981)}]{penzias81}
Penzias, A. 1981, ApJ, 249, 518

\bibitem[{Pickett {et~al.}(1998)Pickett, Poynter, Cohen, a, a, \&
  a}]{pickett98}
Pickett, H., Poynter, I., Cohen, E., {et~al.} 1998, JQSRT, 60, 883

\bibitem[{Pilbratt {et~al.}(2010)Pilbratt, Riedinger, Passvogel, a, a, \&
  a}]{pilbratt10}
Pilbratt, G., Riedinger, J., Passvogel, T., {et~al.} 2010, A\&A, 518, L1

\bibitem[{Pineau~des For\^{e}ts {et~al.}(1989)Pineau~des For\^{e}ts, Roueff, \&
  Flower}]{pineaudesforets89}
Pineau~des For\^{e}ts, G., Roueff, E., \& Flower, D. 1989, JCSFT, 85, 1665

\bibitem[{Polehampton {et~al.}(2007)Polehampton, Baluteau, Swinyard,
  Goicoechea, Brown, White, Cernicharo, \& Grundy}]{polehampton07}
Polehampton, E., Baluteau, J.-P., Swinyard, B., {et~al.} 2007, MNRAS, 377, 1122

\bibitem[{Ponti {et~al.}(2010)Ponti, Terrier, Goldwurm, Belanger, \&
  Trap}]{ponti10}
Ponti, G., Terrier, R., Goldwurm, A., Belanger, G., \& Trap, G. 2010, ApJ, 714,
  732

\bibitem[{Prasad \& Huntress(1980)}]{prasad80}
Prasad, S., \& Huntress, W. 1980, ApJS, 43, 1

\bibitem[{Qin {et~al.}(2010)Qin, Schilke, Comito, M\"{o}ller, Rolffs,
  M\"{u}ller, Belloche, {et~al.}}]{qin10}
Qin, S.-L., Schilke, P., Comito, C., {et~al.} 2010, A\&A, 521, L14

\bibitem[{Qin {et~al.}(2011)Qin, Schilke, Rolffs, Comito, Lis, \&
  Zhang}]{qin11}
Qin, S.-L., Schilke, P., Rolffs, R., {et~al.} 2011, A\&A, 530, L9

\bibitem[{Rakshit(1982)}]{rakshit82}
Rakshit, A. 1982, IJSMI, 41, 185

\bibitem[{Reid(1993)}]{reid93}
Reid, M. 1993, ARA\&A, 31, 345

\bibitem[{Reid {et~al.}(2009)Reid, Menten, Zheng, Brunthaler, \& Xu}]{reid09}
Reid, M., Menten, K., Zheng, X., Brunthaler, A., \& Xu, Y. 2009, ApJ, 705, 1548

\bibitem[{Remijan {et~al.}(2008)Remijan, Hollis, Lovas, Stork, Jewell, \&
  Meier}]{remijan05}
Remijan, A., Hollis, J., Lovas, F., {et~al.} 2008, ApJ, 675, L85

\bibitem[{Requena-Torres {et~al.}(2008)Requena-Torres, Mart\'{i}n-Pintado,
  Mart\'{i}n, \& Morris}]{requenatorres08}
Requena-Torres, M., Mart\'{i}n-Pintado, J., Mart\'{i}n, S., \& Morris, M. 2008,
  ApJ, 672, 352

\bibitem[{Requena-Torres {et~al.}(2006)Requena-Torres, Mart\'{i}n-Pintado,
  Rodr\'{i}guez-Franco, Mart\'{i}n, Rodr\'{i}guez-Fern\'{a}ndez, \&
  de~Vicente}]{requenatorres06}
Requena-Torres, M., Mart\'{i}n-Pintado, J., Rodr\'{i}guez-Franco, A., {et~al.}
  2006, A\&A, 455, 971

\bibitem[{Roelfsema {et~al.}(2012)Roelfsema, Helmich, Teyssier, a, a, \&
  a}]{roelfsema12}
Roelfsema, P., Helmich, F., Teyssier, D., {et~al.} 2012, A\&A, 537, A17

\bibitem[{Rolffs {et~al.}(2010)Rolffs, Schilke, Comito, Bergin, a, a, \&
  a}]{rolffs10}
Rolffs, R., Schilke, P., Comito, C., {et~al.} 2010, A\&A, 521, L46

\bibitem[{Ruffle {et~al.}(1999)Ruffle, Hartquist, Caselli, \&
  Williams}]{ruffle99}
Ruffle, D., Hartquist, T., Caselli, P., \& Williams, D. 1999, MNRAS, 306, 691

\bibitem[{Ryu {et~al.}(2013)Ryu, Nobukawa, Nakashima, Tsuru, Koyama, \&
  Uchiyama}]{ryu13}
Ryu, S., Nobukawa, M., Nakashima, S., {et~al.} 2013, PASJ, 65, 33

\bibitem[{Sakai {et~al.}(2008)Sakai, Sakai, Aikawa, \& Yamamoto}]{sakai08}
Sakai, N., Sakai, T., Aikawa, Y., \& Yamamoto, S. 2008, ApJ, 675, L89

\bibitem[{Schilke {et~al.}(2010)Schilke, Comito, M\"{u}ller, Bergin, Herbst,
  Lis, Neufeld, Phillips, Bell, {et~al.}}]{schilke10}
Schilke, P., Comito, C., M\"{u}ller, H., {et~al.} 2010, A\&A, 521, L11

\bibitem[{Smith \& Adams(1978)}]{smith78}
Smith, D., \& Adams, N. 1978, ApJ, 220, L87

\bibitem[{Sonnentrucker {et~al.}(2006)Sonnentrucker, Gonz\'{a}lex-Alfonso,
  Neufeld, Bergin, Melnick, Forrest, Pipher, \& Watson}]{sonnentrucker06}
Sonnentrucker, P., Gonz\'{a}lex-Alfonso, E., Neufeld, D., {et~al.} 2006, ApJ,
  650, L71

\bibitem[{Sunyaev {et~al.}(1993)Sunyaev, Markevitch, \& Pavlinsky}]{sunyaev93}
Sunyaev, R., Markevitch, M., \& Pavlinsky, M. 1993, ApJ, 407, 606

\bibitem[{Takagi {et~al.}(1999)Takagi, Fukuzawa, Osamura, \&
  Schaefer}]{tagaki99}
Takagi, N., Fukuzawa, K., Osamura, Y., \& Schaefer, H. 1999, ApJ, 525, 791

\bibitem[{Thaddeus {et~al.}(1981)Thaddeus, Guelin, \& Linke}]{thaddeus81}
Thaddeus, P., Guelin, M., \& Linke, R. 1981, ApJ, 246, L41

\bibitem[{Theule {et~al.}(2011)Theule, Borget, Mispelaer, a, a, \&
  a}]{theule11}
Theule, P., Borget, F., Mispelaer, F., {et~al.} 2011, A\&A, 534, A64

\bibitem[{Tieftrunk {et~al.}(1994)Tieftrunk, des Forets, Schilke, \&
  Walmsley}]{tieftrunk94}
Tieftrunk, A., des Forets, G.~P., Schilke, P., \& Walmsley, C. 1994, A\&A, 289,
  579

\bibitem[{Turner(1991)}]{turner91}
Turner, B. 1991, ApJS, 76, 617

\bibitem[{Turner \& Apponi(2001)}]{turner01}
Turner, B., \& Apponi, A. 2001, ApJ, 561, L207

\bibitem[{Turner {et~al.}(1999)Turner, Terzieva, \& Herbst}]{turner99}
Turner, B., Terzieva, R., \& Herbst, E. 1999, ApJ, 518, 699

\bibitem[{van~der Tak {et~al.}(2006)van~der Tak, Belloche, Schilke, a, a, \&
  a}]{vandertak06}
van~der Tak, F., Belloche, A., Schilke, P., {et~al.} 2006, A\&A, 454, L99

\bibitem[{van~der Tak {et~al.}(2003)van~der Tak, Boonman, Braakman, \& van
  Dishoeck}]{vandertak03}
van~der Tak, F., Boonman, A., Braakman, R., \& van Dishoeck, E. 2003, A\&A,
  412, 133

\bibitem[{van Dishoeck {et~al.}(1996)van Dishoeck, Helmich, de~Graauw, Black,
  Boogert, Ehrenfreund, Gerakines, {et~al.}}]{vandishoeck96}
van Dishoeck, E., Helmich, F., de~Graauw, T., {et~al.} 1996, A\&A, 315, L349

\bibitem[{Vogel {et~al.}(1987)Vogel, Genzel, \& Palmer}]{vogel87}
Vogel, S., Genzel, R., \& Palmer, P. 1987, ApJ, 316, 243

\bibitem[{Wakelam {et~al.}(2011)Wakelam, Hersant, \& Herpin}]{wakelam11}
Wakelam, V., Hersant, F., \& Herpin, F. 2011, A\&A, 529, A112

\bibitem[{Wilson \& Rood(1994)}]{wilson94}
Wilson, T., \& Rood, R. 1994, ARA\&A, 32, 191

\bibitem[{Wirstr\"{o}m {et~al.}(2010)Wirstr\"{o}m, Bergman, Black, Hjalmarson,
  Larsson, Olofsson, {et~al.}}]{wirstrom10}
Wirstr\"{o}m, E., Bergman, P., Black, J., {et~al.} 2010, A\&A, 522, A19

\bibitem[{Woon \& Herbst(1997)}]{woon97}
Woon, D., \& Herbst, E. 1997, ApJ, 477, 204

\bibitem[{Yusef-Zadeh {et~al.}(2013{\natexlab{a}})Yusef-Zadeh, Cotton, Viti,
  Wardle, \& Royster}]{yusefzadeh13}
Yusef-Zadeh, F., Cotton, W., Viti, S., Wardle, M., \& Royster, M.
  2013{\natexlab{a}}, ApJL, 764, L19

\bibitem[{Yusef-Zadeh {et~al.}(2013{\natexlab{b}})Yusef-Zadeh, Hewitt, Wardle,
  Tatischeff, Roberts, Cotton, Uchiyama, Nobukawa, Tsuru, Heinke, \&
  Royster}]{yusefzadeh13b}
Yusef-Zadeh, F., Hewitt, J., Wardle, M., {et~al.} 2013{\natexlab{b}}, ApJ, 762,
  33

\bibitem[{Zaleski {et~al.}(2013)Zaleski, Seifert, Steber, a, a, \&
  a}]{zaleski13}
Zaleski, D., Seifert, N., Steber, A., {et~al.} 2013, ApJL, 765, L10

\bibitem[{Zernickel {et~al.}(2012)Zernickel, Schilke, Schmiedeke, a, a, \&
  a}]{zernickel12}
Zernickel, A., Schilke, P., Schmiedeke, A., {et~al.} 2012, A\&A, 546, A87

\end{thebibliography}

\appendix
\section{Appendix 1: Derivation of radiative transfer equation}

Here we derive Eq. (1), the case with gas intermixed with dust, in the case that $T_\textnormal{gas} = T_\textnormal{dust}$, and LTE conditions so that $T_\textnormal{ex} = T_\textnormal{gas}$.  Here we assume a single thermal component for simplicity.  First, consider the antenna temperature measured on-source at a frequency at which only thermal dust continuum emission is observed:

\begin{equation}
{T_\textnormal{cont}}^\textnormal{ON} = \eta_\textnormal{bf} (J(T_\textnormal{bg}) e^{-\tau_\textnormal{dust}} + J(T_\textnormal{dust}) (1-e^{-\tau_\textnormal{dust}}))
\end{equation}

\noindent In this equation, $T_\textnormal{bg}$ is the cosmic microwave background temperature; the antenna temperature measured at the off position will be this temperature, so in an on-off measurement the measured antenna temperature will be 

\begin{equation}
{T_\textnormal{cont}}^\textnormal{ON-OFF} = \eta_\textnormal{bf} (J(T_\textnormal{dust}) - J(T_\textnormal{bg}))  (1-e^{-\tau_\textnormal{dust}}) 
\end{equation}

\noindent On the line center, where both continuum and line emission contribute to the observed signal, the observed antenna temperature will be (for an on-source/off-source measurement):

\begin{equation}
{T_\textnormal{cont+line}}^\textnormal{ON-OFF} = \eta_\textnormal{bf} (J_\textnormal{dust+line} - J(T_\textnormal{bg})) (1-e^{-(\tau_\textnormal{dust} + \tau_\textnormal{line})})
\end{equation}

\noindent where $J_\textnormal{cont+line}$ is defined as

\begin{equation}
J_\textnormal{dust+line} = \frac{\epsilon_\textnormal{dust} + \epsilon_\textnormal{line}}{\kappa_\textnormal{dust} + \kappa_\textnormal{line}}
\end{equation}

\noindent Here, $\epsilon$ and $\kappa$ refer to the emission and absorption coefficients, respectively, of the continuum and line emission.  However, in the instance that the dust and gas are at the same temperature,

\begin{equation}
J_\textnormal{dust+line} = J(T_\textnormal{dust}) = J(T_\textnormal{ex,line})
\end{equation}

\noindent Therefore, in order to evaluate $\Delta T_\textnormal{line}$, the continuum-subtracted line intensity fit in the modeling procedure, we can subtract Eq. (A2) from (A3), which yields

\begin{equation}
\Delta T_\textnormal{line} = \eta_\textnormal{bf} (J(T_\textnormal{ex,line}) - J(T_\textnormal{bg})) (1-e^{-\tau_\textnormal{line}}) e^{-\tau_\textnormal{dust}}
\end{equation}

\noindent When $J(T_\textnormal{bg})$ is neglected, as described in the main text, this is Eq. (1), used for the modeling of emission lines in this analysis.

\begin{deluxetable}{l c c c}
\tablenum{3}
\tablewidth{0pt}
\tablecaption{Molecules detected in each environment in the Sgr B2(N) HIFI survey.}
\tablehead{Molecule & Emission & Cold absorption & Hot absorption \\
                                 &                 & ($T_\textnormal{rot} < 40$ K) & ($T_\textnormal{rot} > 40$ K)}
\startdata
CO & X & & X  \\
$^{13}$CO & X \\
C$^{18}$O & X \\
C$^{17}$O & X \\
$^{13}$C$^{18}$O & X \\
CH$_3$OH & X & X \\
$^{13}$CH$_3$OH & X & X \\
H$_2$CO & X & & X \\
H$_2$$^{13}$CO & X \\
H$_2$CCO & X \\
HCO$^+$ & X & & X \\
H$^{13}$CO$^+$ & X \\
HOCO$^+$ & & X \\
OH$^+$ & & X \\
H$_2$O$^+$ & & X \\
H$_3$O$^+$ & & X & X \\
H$_2$O & X & X & X \\
H$_2$$^{18}$O & X & X \\
H$_2$$^{17}$O & X & X \\
HDO & X & X \\
CH$_3$OCH$_3$ & X \\
C$_2$H$_5$OH & X \\
CH$_3$NH$_2$ & X \\
CH$_2$NH & X & X \\
NH$_2$ & X & X  \\
NH$_3$ & & X & \\
$^{15}$NH$_3$ & & X & \\
N$_2$H$^+$ & X \\
NH & & X & \\
NO & X \\
HNCO & X & X & X \\
HN$^{13}$CO & & X \\
NH$_2$CHO & X \\
NH$_2$$^{13}$CHO & X \\
HNO & & X & \\
HCN & X & & X \\
H$^{13}$CN & X & & X \\
HC$^{15}$N & X & & X \\
DCN & X \\
HNC & X & & X \\
HN$^{13}$C & X & & X \\
CH$_3$CN & X \\
$^{13}$CH$_3$CN & X \\
CH$_3$$^{13}$CN & X \\
C$_2$H$_5$CN & X \\
C$_2$H$_3$CN & X \\
CN & X \\
H$_2$S & X & X & X \\
H$_2$$^{34}$S & X & X & X \\
H$_2$$^{33}$S & X & X & X \\
SO & X \\
$^{34}$SO & X \\
SO$_2$ & X \\
CS & X & & X \\
C$^{34}$S & X \\
$^{13}$CS & X \\
C$^{33}$S & X \\
H$_2$CS & X \\
OCS & X \\
NS & X \\
SH$^+$ & & X \\
CCH & X \\
CH & & X \\
CH$^+$ & & X \\
$^{13}$CH$^+$ & & X \\
C$_3$ & & X \\
HF & & X \\
H$^{35}$Cl & & X \\
H$^{37}$Cl & & X \\
H$_2$$^{35}$Cl$^+$ & & X \\
H$_2$$^{37}$Cl$^+$ & & X \\

\enddata
\end{deluxetable}

\begin{deluxetable}{c c c c}
\tablenum{4}
\tablewidth{0pt}
\tablecaption{Goodness-of-fit calculations for the Sgr B2(N) fullband model, sorted in order from best-fit to worst.}
\tablehead{Molecule & $N_\textnormal{chan}$\tablenotemark{a} & $\chi^2$$_\textnormal{red}$ & Catalog}

\startdata
DCN & 48 & 0.5 & JPL \\
C$_2$H$_3$CN & 200 & 0.5 & CDMS \\
H$^{13}$CO$^+$ & 45 & 0.7 & CDMS \\
HC$^{15}$N & 147 & 0.8 & CDMS \\
C$_2$H$_5$CN & 3993 & 0.8 & JPL \\
C$_2$H$_5$OH & 2185 & 0.9 & JPL \\
CH$_3$$^{13}$CN & 247 & 0.9 & CDMS \\
C$^{33}$S & 129 & 0.9 & CDMS \\
CH$_3$NH$_2$ & 3487 & 1.0 & JPL \\
$^{13}$CH$_3$CN & 395 & 1.0 & CDMS \\
CN & 104 & 1.1 & CDMS \\
CCH & 235 & 1.1 & CDMS \\
NH$_2$CHO & 2875 & 1.1 & CDMS \\
H$_2$CCO & 334 & 1.1 & CDMS \\
NH$_2$$^{13}$CHO & 114 & 1.1 & CDMS \\
$^{34}$SO & 213 & 1.2 & CDMS \\
NH$_2$ & 1174 & 1.2 & CDMS \\
CH$_3$CN,$v_8 = 1$ & 1673 & 1.2 & JPL \\
NS & 273 & 1.3 & JPL \\
CH$_3$CN & 1586 & 1.4 & CDMS \\
$^{13}$CH$_3$OH & 5923 & 1.4 & CDMS \\
N$_2$H$^+$ & 100 & 1.4 & CDMS \\
H$_2$$^{13}$CO & 315 & 1.5 & CDMS \\
NO & 272 & 1.5 & JPL \\
CH$_3$OCH$_3$ & 3088 & 1.6 & JPL \\
HN$^{13}$C & 276 & 1.7 & CDMS \\
H$_2$CS & 447 & 1.8 & CDMS \\
SO & 1101 & 1.9  & JPL \\
CH$_2$NH & 1218 & 1.9 & JPL \\
$^{13}$C$^{18}$O & 56 & 1.9 & CDMS \\
H$_2$$^{33}$S & 102 & 1.9 & CDMS \\
H$^{13}$CN & 330 & 2.0 & CDMS \\
HNCO & 2485 & 2.0 & CDMS \\
HCN,$v_2=1$ & 353 & 2.0 & CDMS \\
SO$_2$ & 2601 & 2.0 & CDMS \\
C$^{34}$S & 111 & 2.0 & CDMS \\
$^{13}$CS & 224 & 2.1 & CDMS \\
OCS & 258 & 2.2 & CDMS \\
H$_2$$^{34}$S & 382 & 2.3 & CDMS \\
H$^{13}$CN,$v_2=1$ & 168 & 2.4 & CDMS \\
C$^{18}$O & 185 & 2.8 & CDMS \\
CS & 706 & 3.4 & CDMS \\
HCN & 464 & 4.2 & CDMS \\
HCO$^+$ & 281 & 4.9 & JPL \\
H$_2$CO & 1706 & 6.1 & CDMS \\
C$^{17}$O & 80 & 6.3 & CDMS \\
$^{13}$CO & 250 & 7.8 & CDMS \\
CH$_3$OH & 30174 & 8.5 & JPL \\
H$_2$S & 1071 & 13.0 & CDMS \\
HNC & 438 & 18.9 & CDMS \\

\enddata
\tablenotetext{a}{Number of channels used for the goodness-of-fit calculation, following the method described in the text from \cite{crockett13}.}
\end{deluxetable}

\begin{deluxetable}{l c c c c c c}
\tablenum{5}
\tablewidth{0pt}
\tablecaption{Integrated line intensities, luminosities, and line count for each molecule in the model to the Sgr B2(N) spectral survey.}
\tablehead{Molecule & \multicolumn{3}{c}{$\int T_\textnormal{A}dV$ (K km s$^{-1}$)} &  \multicolumn{2}{c}{Luminosity ($L_\odot$)} & $N_\textnormal{lines}$  \\
                                 & Hot Core & Envelope & Total  & Hot Core & Envelope}
\startdata
CH$_3$OH & 11500.3 & 2970.9 & 14471.2 & 330 & 69 & 1759 \\
CO & 0 & $\ge$1742.6 & $\ge$1742.6 & 0 & $\ge 37$ & 11 \\
$^{13}$CH$_3$OH & 1356.8 & 52.6 & 1409.4 & 34 & 2.6 & 655 \\
NH$_2$ & 0 & 1068.3 & 1068.3 & 0 & 31 & 34 \\
H$_2$S & 924.1 & 0 & 924.1 & 25 & 0 & 46 \\
$^{13}$CO & 0 & 833.7 & 833.7 & 0 & 20 & 7 \\
CH$_3$NH$_2$ & 864.6 & 0 & 864.6 & 38 & 0 & 715 \\
CH$_3$OCH$_3$ & 852.3 & 0 & 852.3 & 24 & 0 & 391 \\
H$_2$CO & 552.6 & 252.6 & 805.2 & 18 & 5.5 & 98 \\
H$_2$O & 0 & $\ge$752.8 & $\ge$752.8 & 0 & $\ge 18$ & 10 \\
C$_2$H$_5$CN & 707.5 & 0 & 707.5 & 23 & 0 & 667 \\
SO$_2$ & 616.8 & 20.7 & 637.5 & 23 & 1.6 & 248 \\
CS & 210.9 & 332.5 & 543.4 & 4.5 & 6.9 & 13 \\
CH$_3$CN & 522.2 & 0 & 522.2 & 12 & 0 & 177 \\
C$_2$H$_5$OH & 489.5 & 0 & 489.5 & 17 & 0 & 578 \\
HCO$^+$ & 329.0 & 118.9 & 447.9 & 7.1 & 2.1 & 9 \\
HNCO & 441.1 & 0 & 441.1 & 13 & 0 & 199 \\
SO & 291.6 & 138.0 & 429.6 & 7.4 & 3.2 & 42 \\
CH$_3$CN,$v_8=1$ & 390.1 & 0 & 390.1 & 11 & 0 & 226 \\
CH$_2$NH & 352.4 & 0 & 352.4 & 13 & 0 & 215 \\
C$^{18}$O & 0 & 330.1 & 330.1 & 0 & 7.2 & 7 \\
HCN & 0 & 286.2 & 286.2 & 0 & 5.4 & 9 \\
H$_2$$^{18}$O & 0 & 217.7 & 217.7 & 0 & 6.2 & 9 \\
NH$_2$CHO & 205.8 & 0 & 205.8 & 18 & 0 & 122 \\
H$_2$$^{34}$S & 192.2 & 0 & 192.2 & 6.2 & 0 & 32 \\
HNC & 0 & 182.8 & 182.8 & 0 & 4.3 & 8 \\
HCN,$v_2=1$ & 169.8 & 0 & 169.8 & 5.0 & 0 & 16 \\
H$^{13}$CN & 0 & 126.6 & 126.6 & 0 & 2.3 & 8 \\
NO & 45.8 & 74.9 & 120.7 & 1.7 & 1.9 & 21 \\
H$_2$CS & 116.1 & 0 & 116.1 & 2.8 & 0 & 84 \\
NS & 116.1 & 0 & 116.1 & 2.6 & 0 & 23 \\
C$^{17}$O & 0 & 104.0 & 104.0 & 0 & 2.5 & 6 \\
H$_2$$^{13}$CO & 99.6 & 0 & 99.6 & 2.9 & 0 & 55 \\
HDO & 0 & 90.5 & 90.5 & 0 & 2.2 & 13 \\
$^{13}$CH$_3$CN & 88.0 & 0 & 88.0 & 2.0 & 0 & 96 \\
CH$_3$$^{13}$CH & 87.3 & 0 & 87.3 & 2.0 & 0 & 89 \\
H$_2$$^{33}$S & 85.5 & 0 & 85.5 & 2.8 & 0 & 20 \\
H$_2$$^{17}$O & 0 & 84.0 & 84.0 & 0 & 2.4 & 6 \\
HC$^{15}$N & 0 & 79.6 & 79.6 & 0 & 1.7 & 7 \\
H$_2$CCO & 64.6 & 0 & 64.6 & 2.0 & 0 & 97 \\
OCS & 63.4 & 0 & 63.4 & 1.2 & 0 & 26 \\
HN$^{13}$C & 0 & 60.6 & 60.6 & 0 & 1.4 & 7 \\
CN & 0 & 52.1 & 52.1 & 0 & 1.1 & 8 \\
N$_2$H$^+$ & 0 & 50.1 & 50.1 & 0 & 1.0 & 5 \\
C$^{34}$S & 24.7 & 23.9 & 48.6 & 0.61 & 0.71 & 10 \\
$^{13}$CS & 22.4 & 19.7 & 42.1 & 0.56 & 0.62 & 10 \\
C$_2$H$_3$CN & 39.8 & 0 & 39.8 & 3.7 & 0 & 135 \\
$^{34}$SO & 36.0 & 0 & 36.0 & 1.1 & 0 & 27 \\
CCH & 0 & 35.8 & 35.8 & 0 & 0.82 & 10 \\
H$^{13}$CN,$v_2=1$ & 25.7 & 0 & 25.7 & 1.0 & 0 & 12 \\
H$^{13}$CO$^+$ & 22.3 & 0 & 22.3 & 0.52 & 0 & 6 \\
C$^{33}$S & 12.6 & 7.2 & 19.8 & 0.35 & 0.36 & 8 \\
$^{13}$C$^{18}$O & 0 & 14.1 & 14.1 & 0 & 0.38 & 5 \\
DCN & 0 & 10.0 & 10.0 & 0 & 0.65 & 5 \\

\enddata
\end{deluxetable}

\begin{deluxetable}{c c c c c c}
\tablenum{6}
\tablewidth{0pt}
\tablecaption{LTE parameters for the Sgr B2(N) fullband model (emission components).}
\tablehead{Molecule & Source size & $T_\textnormal{rot}$\tablenotemark{a} & $N_\textnormal{tot}$\tablenotemark{a} & $\Delta v$ & $v_\textnormal{LSR}$ \\
		                  &     $('')$     &           (K)                    & (cm$^{-2}$) &  (km s$^{-1}$)  & (km s$^{-1}$)}
\startdata
$^{13}$CO & 25 & 60 & ... & 10 & 66 \\
		& 25 & 60 & ... & 10 & 76 \\
C$^{18}$O & 25 & 60 & $1.0 \times 10^{17}$ & 10 & 66 \\
		& 25 & 60 & $2.0 \times 10^{16}$ & 10 & 76 \\
C$^{17}$O & 25 & 60 & $3.1 \times 10^{16}$ & 10 & 66 \\
		& 25 & 60 & $6.3 \times 10^{15}$ & 10 & 76 \\
$^{13}$C$^{18}$O & 25 & 60 & $5.0 \times 10^{15}$ & 10 & 66 \\
		& 25 & 60 & $1.0 \times 10^{15}$ & 10 & 76 \\
CH$_3$OH & 3.1 & 170 & $5.0 \times 10^{18}$ & 9 & 64 \\
		& 30 & 80 & $1.0 \times 10^{16}$ & 9 & 64 \\
$^{13}$CH$_3$OH & 3.1 & 170 & $2.5 \times 10^{17}$ & 9 & 64 \\
		& 30 & 80 & $5.0 \times 10^{14}$ & 9 & 64 \\
H$_2$CO & 3.1 & ... & ... & 12 & 64 \\
		& 30 & 80 & $1.5 \times 10^{14}$ & 10 & 66 \\
		& 30 & 80 & $2.5 \times 10^{14}$ & 15 & 66 \\
H$_2$$^{13}$CO & 3.1 & 100 & $1.0 \times 10^{16}$ & 10 & 64 \\
H$_2$CCO & 2.3 & 180 & $7.0 \times 10^{16}$ & 8 & 62 \\
HCO$^+$ & 25 & ... & ... & 10 & 64 \\
		& 25 & 60 & $5.0 \times 10^{13}$ & 15 & 76 \\
		& 25 & 20 & $3.5 \times 10^{15}$ & 10 & 58 \\
H$^{13}$CO$^+$ & 3.1 & 150 & $6.0 \times 10^{14}$ & 10 & 64 \\
CH$_3$OCH$_3$ & 3.7 & 130 & $2.8 \times 10^{17}$ & 8.5 & 64 \\
C$_2$H$_5$OH & 3.7 & 100 & $2.0 \times 10^{17}$ & 7.0 & 64 \\
CH$_3$NH$_2$ & 2.5 & 150 & $5.0 \times 10^{17}$ & 7.0 & 64 \\
CH$_2$NH & 2.5 & 150 & $7.0 \times 10^{16}$ & 6.0 & 64 \\
NH$_2$ & 15 & 100 & $8.0 \times 10^{15}$ & 13 & 64 \\
N$_2$H$^+$ & 25 & 37 & $6.5 \times 10^{13}$ & 13 & 62 \\
NO & 2.3 & 180 & $1.0 \times 10^{18}$ & 8 & 65 \\
	& 8.0 & 50 & $1.5 \times 10^{17}$ & 16 & 67 \\
HNCO & 2.3 & 280 & $9.5 \times 10^{16}$ & 8 & 62 \\
NH$_2$CHO & 2.3 & 130 & $2.4 \times 10^{17}$ & 7 & 64 \\
NH$_2$$^{13}$CHO & 2.3 & 130 & $1.2 \times 10^{16}$ & 7 & 64 \\
HCN & 12 & 63 & ... & 22 & 47 \\
	& 12 & 40 & ... & 11 & 58 \\
	& 12 & ... & ... & 14 & 78 \\
H$^{13}$CN & 12 & 63 & ... & 22 & 47 \\
	& 12 & 63 & ... & 11 & 67 \\
	& 12 & 63 & ... & 14 & 78 \\
HC$^{15}$N & 12 & 63 & $4.0 \times 10^{13}$ & 22 & 47 \\
	& 12 & 63 & $6.0 \times 10^{13}$ & 11 & 67 \\
	& 12 & 63 & $5.0 \times 10^{13}$ & 14 & 78 \\
DCN & 12 & 63 & $2.5 \times 10^{13}$ & 22 & 47 \\
	& 12 & 63 & $3.0 \times 10^{13}$ & 11 & 67 \\
	& 12 & 63 & $2.0 \times 10^{13}$ & 14 & 78 \\
HNC & 12 & 50 & ... & 14 & 47 \\
	& 12 & 63 & ... & 8 & 65 \\
	& 12 & 80 & ... & 14 & 78 \\
HN$^{13}$C & 12 & 63 & $4.0 \times 10^{13}$ & 22 & 47 \\
	& 12 & 63 & $4.5 \times 10^{13}$ & 11 & 67 \\
	& 12 & 63 & $3.0 \times 10^{13}$ & 14 & 78 \\
CH$_3$CN & 4 & 90 & $8.5 \times 10^{15}$ & 6.5 & 57 \\
	& 3 & 250 & ... & 6.5 & 63 \\
	& 2 & 300 & $1.4 \times 10^{16}$ & 6.5 & 69 \\
C$^{13}$H$_3$CN & 2 & 200 & $1.3 \times 10^{16}$ & 6.5 & 63 \\
CH$_3$$^{13}$CN & 2 & 200 & $1.3 \times 10^{16}$ & 6.5 & 63 \\
C$_2$H$_5$CN & 2 & 150 & $6.0 \times 10^{17}$ & 8.0 & 62 \\
	& 2 & 120 & $2.5 \times 10^{17}$ & 6.5 & 72 \\
C$_2$H$_3$CN & 2 & 150 & $1.0 \times 10^{17}$ & 8.0 & 62 \\
CN & 20 & 40 & $7.0 \times 10^{14}$ & 10 & 64 \\
H$_2$S & 7 & 160 & $1.5 \times 10^{17}$ & 7 & 64 \\
	& 7 & 190 & $4.0 \times 10^{16}$ & 6 & 64 \\
H$_2$$^{34}$S & 7 & 160 & $1.5 \times 10^{16}$ & 7 & 64 \\
	& 7 & 190 & $4.0 \times 10^{15}$ & 6 & 64 \\
H$_2$$^{33}$S & 2.8 & 120 & $5.9 \times 10^{16}$ & 6 & 64 \\
SO & 2.8 & 150 & $1.0 \times 10^{17}$ & 10 & 64 \\
	& 2.8 & 180 & $1.0 \times 10^{16}$ & 14 & 78 \\
	& 50 & 120 & $1.0 \times 10^{14}$ & 8 & 67 \\
	& 50 & 120 & $3.0 \times 10^{14}$ & 30 & 65 \\
$^{34}$SO & 2.8 & 150 & $1.0 \times 10^{16}$ & 10 & 64 \\
SO$_2$ & 2.8 & 150 & $1.5 \times 10^{17}$ & 10 & 64 \\
	& 2.8 & 180 & $1.5 \times 10^{16}$ & 14 & 80 \\
	& 50 & 120 & $1.5 \times 10^{14}$ & 8 & 67 \\
CS & 6 & 120 & ... & 10 & 64 \\
	& 10 & 85 & $2.8 \times 10^{15}$ & 35 & 67 \\
	& 10 & 20 & $2.8 \times 10^{15}$ & 35 & 67 \\
C$^{34}$S & 2.8 & 120 & $4.6 \times 10^{15}$ & 10 & 64 \\
	& 10 & 85 & $2.8 \times 10^{14}$ & 35 & 67 \\
$^{13}$CS & 2.8 & 120 & $4.6 \times 10^{15}$ & 10 & 64 \\
	& 10 & 85 & $2.8 \times 10^{14}$ & 35 & 67 \\
C$^{33}$S & 2.8 & 120 & $2.3 \times 10^{15}$ & 10 & 64 \\
	& 10 & 85 & $1.4 \times 10^{14}$ & 35 & 67 \\
H$_2$CS & 2.8 & 120 & $5.0 \times 10^{16}$ & 10 & 62 \\
OCS & 2.8 & 200 & $3.5 \times 10^{17}$ & 8 & 63 \\
NS & 2.8 & 90 & $1.0 \times 10^{16}$ & 9 & 64 \\
	& 2.8 & 90 & $6.1 \times 10^{15}$ & 8 & 73 \\
CCH & 22 & 50 & $6.0 \times 10^{14}$ & 13 & 64 \\
\enddata

\tablenotetext{a}{Where no value is given, the model is deemed to be purely an effective lineshape fit due to high optical depth in that species.}
\end{deluxetable}

\begin{deluxetable}{c c c c c c}
\tablenum{7}
\tablewidth{0pt}
\tablecaption{LTE parameters for the Sgr B2(N) fullband model (cold absorption components, $T_\textnormal{ex} < 40$ K).}
\tablehead{Molecule & Source size & $T_\textnormal{rot}$ & $N$ & $\Delta v$ & $v_\textnormal{LSR}$ \\
		                  &     $('')$     &           (K)                    & (cm$^{-2}$) &  (km s$^{-1}$)  & (km s$^{-1}$)}
\startdata

CH$_3$OH-$A$ & ext. & 10 & $6.0 \times 10^{15}$ & 12 & 64 \\
	& ext. & 10 & $1.0 \times 10^{16}$ & 12 & 74 \\
	& ext. & 5 & $2.0 \times 10^{15}$ & 12 & 74 \\
	& ext. & 4 & $2.0 \times 10^{14}$ & 15 & 7 \\
	& ext. & 4 & $1.0 \times 10^{14}$ & 6 & 2 \\
CH$_3$OH-$E$ & ext. & 7.5 & $1.45 \times 10^{16}$ & 13 & 63.5 \\
	& ext. & 10.5 & $1.0 \times 10^{16}$ & 12 & 74 \\
	& ext. & 2.7 & $4.0 \times 10^{15}$ & 13 & 63.5 \\
	& ext. & 2.7 & $8.0 \times 10^{15}$ & 12 & 74 \\
	& ext. & 6 & $2.0 \times 10^{14}$ & 15 & 7 \\
	& ext. & 6 & $1.0 \times 10^{14}$ & 6 & 2 \\
$^{13}$CH$_3$OH & ext. & 5 & $1.5 \times 10^{15}$ & 13 & 64 \\
	& ext. & 7 & $7.9 \times 10^{14}$ & 12 & 77 \\
HOCO$^+$ & ext. & 14 & $3.0 \times 10^{13}$ & 11 & 64 \\
	& ext. & 14 & $1.5 \times 10^{13}$ & 10 & 78 \\
OH$^+$ & ext. & 2.7 & $2.7 \times 10^{14}$ & 28 & 64 \\
H$_2$O$^+$ & ext. & 2.7 & $5.7 \times 10^{14}$ & 27 & 64 \\
OH & ext. & 20 & $2.5 \times 10^{17}$ & 10 & 59 \\
	& ext. & 20 & $5.0 \times 10^{17}$ & 6.5 & 73 \\
CH$_2$NH & ext. & 6 & $1.6 \times 10^{14}$ & 11 & 62 \\
	& ext. & 11 & $6.0 \times 10^{13}$ & 6 & 72 \\
	& ext. & 6 & $1.4 \times 10^{14}$ & 15 & 78 \\
NH$_2$ & ext. & 14 & $4.0 \times 10^{15}$ & 12 & 63 \\
	& ext & 18 & $1.2 \times 10^{15}$ & 10 & 74 \\
$^{14}$NH$_3$ & ext. & 20 & $8.6 \times 10^{15}$ & 10 & 64 \\
	& ext. & 8 & $2.0 \times 10^{15}$ & 10 & 64 \\
	& ext. & 4 & $2.0 \times 10^{15}$ & 15 & 64 \\
	& ext. & 20 & $1.0 \times 10^{16}$ & 10 & 73 \\
	& ext. & 8 & $1.2 \times 10^{15}$ & 12 & 77 \\
$^{15}$NH$_3$ & ext. & 11 & $5.0 \times 10^{13}$ & 10 & 59 \\
	& ext. & 35 & $6.0 \times 10^{14}$ & 13 & 59 \\
	& ext. & 3 & $4.0 \times 10^{12}$ & 10 & 74 \\
NH & ext. & 6 & $7.0 \times 10^{15}$ & 15 & 64 \\
	& ext. & 6 & $1.4 \times 10^{15}$ & 7 & 79 \\
	& ext. & 6 & $3.0 \times 10^{14}$ & 7 & 86 \\
N II & ext. & 13 & $5.0 \times 10^{17}$ & 13 & 0 \\
HNCO & ext. & 13.1 & $1.3 \times 10^{15}$ & 13 & 64 \\
	& ext. & 15.5 & $1.0 \times 10^{15}$ & 12 & 81 \\
HNO & ext. & 8.5 & $3.2 \times 10^{13}$ & 9 & 64 \\
H$_2$S & ext. & 30 & $1.4 \times 10^{14}$ & 9 & 64 \\
	& ext. & 6 & $1.0 \times 10^{16}$ & 25 & 64 \\
	& ext. & 30 & $4.4 \times 10^{14}$ & 9 & 73 \\
H$_2$$^{34}$S & ext. & 8 & $3.0 \times 10^{13}$ & 9 & 55 \\
	& ext. & 4.8 & $1.4 \times 10^{15}$ & 9 & 64 \\
	& ext. & 12 & $8.0 \times 10^{13}$ & 6 & 73 \\
	& ext. & 2 & $8.0 \times 10^{16}$ & 18 & 73 \\
H$_2$$^{33}$S & ext. & 8 & $3.0 \times 10^{13}$ & 9 & 55 \\
	& ext. & 2 & $9.0 \times 10^{15}$ & 9 & 64 \\	
	& ext. & 2 & $8.0 \times 10^{15}$ & 12 & 73 \\
SH$^+$ & ext. & 3 & $1.5 \times 10^{14}$ & 19 & 58 \\
	& ext. & 3 & $4.4 \times 10^{13}$ & 12 & 77 \\
CH & ext. & 2.7 & $8.5 \times 10^{13}$ & 10.5 & 52 \\
	& ext. & 2.7 & $2.9 \times 10^{14}$ & 10.5 & 62 \\
	& ext. & 2.7 & $3.0 \times 10^{13}$ & 8 & 78 \\
CH$^+$ & ext. & 4.4 & $1.7 \times 10^{14}$ & 31 & 55 \\
	& ext. & 4.4 & $1.0 \times 10^{14}$ & 10 & 64 \\
	& ext. & 4.4 & $2.8 \times 10^{13}$ & 8 & 76 \\
	& ext. & 4.4 & $4.5 \times 10^{12}$ & 14 & 89 \\
$^{13}$CH$^+$ & ext. & 4.4 & $3.9 \times 10^{12}$ & 31 & 55 \\	
	& ext. & 4.4 & $2.1 \times 10^{12}$ & 8 & 66 \\	
	& ext. & 4.4 & $8.4 \times 10^{11}$ & 8 & 78 \\
C$_3$ & ext. & 32 & $1.6 \times 10^{16}$ & 8 & 65 \\	
	& ext. & 36 & $1.6 \times 10^{16}$ & 10 & 76 \\
HF & ext. & 2.7 & $1.5 \times 10^{13}$ & 9 & 50 \\
	& ext. & 2.7 & $8.2 \times 10^{14}$ & 16 & 68 \\
	& ext. & 2.7 & $4.0 \times 10^{13}$ & 9 & 82 \\
H$^{35}$Cl & ext. & 8.5 & $1.0 \times 10^{14}$ & 20 & 8 \\
	& ext. & 8.5 & $9.0 \times 10^{14}$ & 20 & 66 \\
	& ext. & 8.5 & $2.2 \times 10^{14}$ & 6 & 73 \\
	& ext. & 2.7 & $3.0 \times 10^{14}$ & 20 & 72 \\
H$^{37}$Cl & ext. & 8.5 & $3.0 \times 10^{13}$ & 20 & 8 \\
	& ext. & 8.5 & $3.5 \times 10^{14}$ & 20 & 66 \\
	& ext. & 8.5 & $8.6 \times 10^{13}$ & 6 & 73 \\
H$_2$$^{35}$Cl$^+$ & ext. & 2.7 & $4.1 \times 10^{13}$ & 12 & 2 \\
	& ext. & 2.7 & $9.3 \times 10^{12}$ & 4.5 & 16 \\
H$_2$$^{37}$Cl$^+$ & ext. & 2.7 & $1.1 \times 10^{13}$ & 12 & 2 \\
	& ext. & 2.7 & $2.0 \times 10^{12}$ & 4.5 & 16 \\

\enddata
\end{deluxetable}

\begin{deluxetable}{c c c c c c}
\tablenum{8}
\tablewidth{0pt}
\tablecaption{LTE parameters for the Sgr B2(N) fullband model (hot absorption components, $T_\textnormal{ex} > 40$ K).}
\tablehead{Molecule & Source size & $T_\textnormal{rot}$ & $N$ & $\Delta v$ & $v_\textnormal{LSR}$ \\
		                  &     $('')$     &           (K)                    & (cm$^{-2}$) &  (km s$^{-1}$)  & (km s$^{-1}$)}
\startdata

H$_2$CO & 3.5 & 200 & $1.5 \times 10^{16}$ & 10 & 58 \\
HCO$^+$ & 5 & 55 & $1.2 \times 10^{14}$ & 6 & 73 \\
HNCO & 5.0 & 150 & $6.0 \times 10^{14}$ & 8 & 55 \\
	& 5.0 & 40 & $3.0 \times 10^{14}$ & 5 & 73 \\
HCN & 3.5 & 190 & $2.5 \times 10^{15}$ & 13 & 58 \\
	& 3.5 & 50 & $9.0 \times 10^{14}$ & 6 & 73 \\
H$^{13}$CN & 5.0 & 140 & $2.0 \times 10^{14}$ & 9 & 55 \\
HC$^{15}$N & 5.0 & 140 & $2.0 \times 10^{14}$ & 9 & 55 \\
DCN & 5.0 & 140 & $2.0 \times 10^{13}$ & 9 & 55 \\
HNC & 3.5 & 150 & $4.0 \times 10^{14}$ & 9 & 55 \\
	& 3.5 & 50 & $1.0 \times 10^{14}$ & 6 & 55 \\
HN$^{13}$C& 5.0 & 140 & $1.0 \times 10^{14}$ & 9 & 55 \\
H$_2$S & 3.5 & 350 & $1.5 \times 10^{16}$ & 9 & 55 \\
	& ext. & 100 & $4.0 \times 10^{13}$ & 9 & 64 \\
H$_2$$^{34}$S & 5.0 & 80 & $5.0 \times 10^{14}$ & 9 & 55 \\
H$_2$$^{33}$S & 5.0 & 80 & $4.0 \times 10^{14}$ & 9 & 55 \\
CS & 5.0 & 150 & $3.0 \times 10^{15}$ & 10 & 58 \\

\enddata
\end{deluxetable}

\begin{deluxetable}{c c c c c c}
\tablenum{9}
\tablewidth{0pt}
\tablecaption{Most important reactions relevant to the formation and destruction of HOCO$^+$ and HCO$^+$.}
\tablehead{No. & \multicolumn{3}{c}{Reaction} & Rate (100 K) & Reference \\
                                 & & & & (cm$^3$ s$^{-1}$) & }
\startdata
1 & CO + H$_3$$^+$ & $\rightarrow$ & HCO$^+$ + H$_2$ & $1.6 \times 10^{-9}$ & \cite{rakshit82} \\
2 & CO$_2$ + H$_3$$^+$ & $\rightarrow$ & HOCO$^+$ + H$_2$ & $2.0 \times 10^{-9}$ & \cite{klippenstein10} \\
3 & HOCO$^+$ + CO & $\rightarrow$ & HCO$^+$ + CO$_2$ & $7.8 \times 10^{-10}$ & \cite{prasad80} \\
4 & HCO$^+$ + $e^-$ & $\rightarrow$ & H + CO & $5.2 \times 10^{-7}$ & \cite{brian90} \\
5 & HCO$^+$ + H$_2$O & $\rightarrow$ & H$_3$O$^+$ + CO & $4.3 \times 10^{-9}$ & \cite{adams78} \\
6 & HOCO$^+$ + $e^-$ & $\rightarrow$ & H + CO + O & $2.4 \times 10^{-6}$ & \cite{geppert04}\tablenotemark{a} \\
7 & HOCO$^+$ + H$_2$O & $\rightarrow$ & H$_3$O$^+$ + CO$_2$ & $4.0 \times 10^{-9}$ & \cite{rakshit82} \\
\enddata
\tablenotetext{a}{This reaction has multiple product channels, with the H + CO + O channel being the most significant.  The rate given in the table is the sum of all observed channels.}
\end{deluxetable}

\end{document}